\documentclass[AMA,Times1COL]{WileyNJDv5} %STIX1COL,STIX2COL,STIXSMALL

\articletype{Article Type}%

\received{Date Month Year}
\revised{Date Month Year}
\accepted{Date Month Year}
\journal{Journal}
\volume{00}
\startpage{1}

\usepackage{subcaption}
\usepackage{tikz}
\usepackage{bm}
\usepackage{pgfplots}
\usepgfplotslibrary{fillbetween}
\usetikzlibrary{intersections}
\usepackage{caption}
\usepackage{enumitem}
\newlist{steps}{enumerate}{1}
\setlist[steps, 1]{label = Step \arabic*:}
\newcommand{\rhof}{\rho^f}
\newcommand{\rhos}{\rho^s}
\newcommand{\invrhof}{\frac{1}{\rho^f}}
\newcommand{\invrhos}{\frac{1}{\rho^s}}
\newcommand{\muf}{\mu^f}
\newcommand{\mus}{\mu^s}

\newcommand{\uif}{u_i^f}
\newcommand{\ujf}{u_j^f}
\newcommand{\uis}{u_i^s}
\newcommand{\ujs}{u_j^s}
\newcommand{\phis}{\phi}

\newcommand{\Btil}{{\bm{B}}}
\newcommand{\Btilvar}{{B}}

\newcommand{\der}[2]{\frac{\partial{#1}}{\partial{#2}}}

\newcommand{\onebytwo}{\frac{1}{2}}

\newcommand{\CubeLength}{4}
\newcommand{\CubeAxisLength}{4.5}
\newcommand{\CubeHalfLength}{2}

\DeclareRobustCommand\square{\tikz{\filldraw[] (0,0) rectangle (0.2,0.2);}}
\DeclareRobustCommand\emptysquare{\tikz{\draw[] (0,0) rectangle (0.2,0.2);}}
\DeclareRobustCommand\circle{\tikz{\filldraw[] (0,0) circle (0.1);}}
\DeclareRobustCommand\emptycircle{\tikz{\draw[] (0,0) circle (0.1);}}
\DeclareRobustCommand\emptytriangle{\tikz{\draw[] (0,0) -- (0.2,0) -- (0.1,0.2) -- cycle;}}
\DeclareRobustCommand\emptydiamond{\tikz{\draw[] (0.1,0) -- (0.2,0.1) -- (0.1,0.2) -- (0,0.1) -- (0.1,0);}}

\DeclareRobustCommand\fullthick  {\tikz[baseline=-0.6ex]\draw[ultra thick] (0,0)--(0.5,0);}
\DeclareRobustCommand\dottedthick{\tikz[baseline=-0.6ex]\draw[ultra thick,dotted] (0,0)--(0.54,0);}

\DeclareRobustCommand\dasheddottedthick {\tikz[baseline=-0.6ex]\draw[ultra thick,dash dot] (0,0)--(0.5,0);}
\DeclareRobustCommand\dashedthick{\tikz[baseline=-0.6ex]\draw[ultra thick,dashed] (0,0)--(0.54,0);}
\begin{document}

\title{Monolithic framework to simulate fluid-structure interaction problems using geometric volume-of-fluid method}

\author[1]{Soham Prajapati}

\author[1]{Ali Fakhreddine}

\author[1,2]{Krishnan Mahesh}

\authormark{PRAJAPATI \textsc{et al.}}
\titlemark{Monolithic framework to simulate fluid-structure interaction problems using geometric volume-of-fluid method}

\address[1]{\orgdiv{Department of Aerospace Engineering and Mechanics}, \orgname{University of Minnesota - Twin Cities}, %\orgaddress{\state{Minnesota}, \country{USA}}}
\orgaddress{\city{Minneapolis}, \state{Minnesota}, \country{USA}}}

\address[2]{\orgdiv{Department of Naval Architecture and Marine Engineering}, \orgname{University of Michigan}, \orgaddress{\city{Ann Arbor}, \state{Michigan}, \country{USA}}}

\corres{Corresponding author Krishnan Mahesh, Department of Naval Architecture and Marine Engineering, University of Michigan, Ann Arbor, Michigan, USA \email{krmahesh@umich.edu}}

\fundingInfo{United States Office of Naval Research (ONR), Grant/Award Numbers: N00014-17-1-2676 and N00014-21-1-2455}

\presentaddress{Ali Fakhreddine: Department of Metallurgy, Montanuniversität Leoben, Leoben, Austria}

\abstract[Abstract]{We develop a three-dimensional Eulerian framework to simulate fluid-structure interaction (FSI) problems on a fixed Cartesian grid using the geometric volume-of-fluid (VOF) method. The coupled problem involves incompressible flow and viscous hyperelastic solids. A VOF-based one-continuum formulation is used to describe the unified momentum conservation equations with incompressibility constraints that are solved using the finite volume method (FVM) \cite{mahesh2004numerical}. In the geometric VOF interface-capturing (IC) approach, the piecewise linear interface calculation (PLIC) method is used to reconstruct the interface, and the Lagrangian Explicit (LE) method is used in the directionally split advection procedure. To model the hyperelastic behavior of the solid, we consider linear and nonlinear Mooney-Rivlin material models, where we use the left Cauchy-Green deformation tensor ($\bm{B}$) to account for the solid deformation on an Eulerian grid and the fifth-order weighted essentially non-oscillatory (WENO-Z) \citep{borges2008improved,jiang1996efficient} finite difference reconstruction method is utilized to treat the advection terms involved in the transport equation of $\bm{B}$. Multiple benchmark problems and reversibility tests are considered to verify the accuracy of the approach. Furthermore, to demonstrate the capability of the solver to handle turbulent interactions, we perform direct numerical simulation (DNS) of turbulent channel flow with a deformable compliant bottom wall and a rigid top wall; our observations align well with previous experimental and numerical works. The detailed numerical experiments show that: (i) Despite the discontinuity of the interface across the cell boundaries and stress discontinuity across the interface, a VOF/PLIC-based FSI framework can provide stable and accurate solutions that significantly minimizes numerical artifacts (e.g., flotsam and spurious currents) while maintaining a sharp interface. (ii) The accuracy of a  VOF/PLIC-based FSI approach on coarse grids is comparable to the accuracy of a diffusive IC method-based FSI approach on much finer grids.}


\maketitle

\renewcommand\thefootnote{}

\renewcommand\thefootnote{\fnsymbol{footnote}}
\setcounter{footnote}{1}

% -------- Introduction -----------------
% Main text
\section{Introduction}\label{sec:introduction}

The dynamic interaction of fluid flows with deformable structures is crucial in various sectors, such as the marine, healthcare, and aerospace industries. Accurately modeling such fluid-structure interaction (FSI) problems enables us to perform detailed assessments of complex processes such as the interaction of deformable compliant coatings with turbulent flows \citep{wang2020interaction, esteghamatian2022spatiotemporal, rosti2017numerical,rosti2020low}, the impact of marine biofouling on the turbulent flowfield near the affected surfaces \citep{kaminaris2024direct, monty2016assessment, sarakinos2022investigation}, and the rheology of red blood cells \citep{liu2006rheology, ii2012computational, ii2011implicit, sugiyama2010full}.         

Most of the mesh-based numerical methods used to model FSI problems can be broadly classified into three categories \cite{jain2019conservative}: (i) fully Eulerian methods, where the Eulerian grid represents both subsystems, (ii) mixed Lagrangian-Eulerian methods, where mostly Lagrangian and Eulerian grids represent solid and fluid subsystems, respectively, and (iii) fully Lagrangian methods, where the Lagrangian grid represents both subsystems. For a brief overview of the different types of FSI approaches, we refer the readers to Jain et al.\cite{jain2019conservative}. The mixed Lagrangian-Eulerian methods can be further classified into two categories: (i) partitioned approach, where both systems are solved separately, and (ii) monolithic approach, where both systems are solved simultaneously. The partitioned mixed Lagrangian-Eulerian methods such as arbitrary Lagrangian-Eulerian (ALE) \cite{hu2001direct,hirt1974arbitrary,nitikitpaiboon1993arbitrary,hughes1981lagrangian} and deforming-spatial-domain/stabilized-space-time (DSD/SST) \cite{TEZDUYAR1992339,tezduyar1992new,hughes1996space} procedures are extensively adopted to simulate a variety of FSI problems such as blood flow \cite{torii2007influence,torii2008fluid,taylor1998finite,leuprecht2003blood},
parachute modeling \citep{stein2005fluid,takizawa2011fluid}, and flapping wings \cite{mittal1995parallel,takizawa2011}. However, such partitioned approaches can become expensive for large deformations due to grid reconstruction requirements. To address the computational cost issue, monolithic approaches such as Immersed Boundary (IB) \cite{peskin1972flow,peskin2002immersed,mittalIBMreview2005}, Fictious Domain (FD) \cite{glowinski1999distributed,glowinski2001fictitious,yu2005dlm}, immersed finite-element \cite{liu2006immersed,zhang2007immersed}, immersed interface \cite{li2001iim,LeVeque1994}, and immersed continuum \cite{wang2006immersed} methods were developed. These methods are utilized to simulate a wide range of FSI problems, such as red blood cell dynamics \cite{eggleton1998large,gong2009deformation}, flexible plate vibration \cite{yu2005dlm,kadapa2016fictitious,zhao2008fixed}, and biofouling surface effects \cite{kaminaris2024direct}.

The fully Eulerian methods perform computations on a fixed Eulerian grid and generally utilize an interface capturing (IC) procedure to track the fluid-solid (FS) interface, similar to multiphase flows. These methods are cost-effective for the following reasons: (i) they do not require complex grid generation procedures, (ii) they can leverage the existing frameworks of fluid solvers, and (iii) they have the potential to scale well for large-scale problems. Over the years, multiple IC methods \citep{tryggvason2011direct} have been developed to perform multiphase simulations. Some of the well-known approaches are the volume-of-fluid (VOF) method \cite{hirt1981volume}, level-set (LS) method \cite{lsmethod}, and phase-field (PF) method \cite{pfmethod}. These methods are increasingly used to analyze FSI problems such as modeling red-blood cells \cite{sugiyama2010full,ii2011implicit,ii2012computational}, modeling blood vessel \citep{nagano2010full}, examining droplet impact on flexible cantilever \cite{shin2020interaction}, and interaction of compliant coatings with turbulent flows \cite{rosti2017numerical,rosti2020low,esteghamatian2022spatiotemporal}. These methods have their benefits and limitations. On one hand, VOF methods are well-known for their potential to conserve mass. However, for sharp IC methods like geometric VOF, interface normal estimation is not a straightforward task due to the discontinuous nature of the volume fraction. On the other hand, LS methods can estimate the interface normal more accurately due to the smoothness and differentiability of the LS function. However, LS methods do not inherently conserve mass and, in most cases, requires a reinitialization procedure that is problem-dependent and computationally costly.  

Recently, FSI simulations involving incompressible flows and hyperelastic solids have been performed using the Eulerian method. Sugiyama et al. \cite{sugiyama2011full} used a fifth-order weighted essentially non-oscillatory (WENO) scheme \citep{jiang1996efficient} to treat the advection terms involved in the transport equations of the volume fraction and the left Cauchy-Green deformation tensor, Ii et al. \cite{ii2012computational} used MTHINC (multi-dimensional tangent of hyperbola for interface capturing) method \citep{xiao2005simple,yokoi2007efficient,ii2012interface} to simulate red blood cells, and Esmailzadeh and Passandideh-Fard \cite{esmailzadeh2014numerical} used the piecewise linear interface calculation (PLIC) method to capture interface in 2D problems. Unlike Lagrangian approaches where mesh movement represents solid deformation, the Eulerian approach generally uses deformation tensors to account for solid displacements. Recently, a reference map technique (RMT) \cite{valkov2015eulerian,jain2019conservative} was introduced that utilizes a reference-map vector field to track solid deformation and was used to simulate the collision of elastic solids with rigid walls \cite{jain2019conservative}.     

To analyze fully coupled problems involving turbulent flows, the two-way coupled FSI approach needs to be robust enough to tackle complex interactions that occur on a wide range of length scales and time scales, and the solver needs to scale well with the problem size. To perform a detailed analysis of the involved coupling, sometimes a simplistic one-way coupled FSI approach provides valuable insights. For example, Prajapati et al. \cite{prajapati2023direct} investigated the vibroacoustic response of elastic plates excited by turbulent flows and proposed new scaling laws, and  Anantharamu and Mahesh \cite{anantharamu2021response} analyzed the sources of elastic plate vibrations excited by turbulent flows. However, such simplistic approaches are limited to a range of problem parameters because the dynamic coupling is assumed to be negligible. Recently, Eulerian methods have shown good potential to handle fully coupled large-scale problems such as the interaction of turbulent flows with deformable compliant coatings \cite{rosti2017numerical,esteghamatian2022spatiotemporal,rosti2020low}, where Esteghamatian et al. \cite{esteghamatian2022spatiotemporal} used LS method to capture the fluid-solid interface, and Rosti and Brandt \cite{rosti2017numerical} used the VOF formulation and employed a fifth-order WENO scheme to treat the advection terms involved in the transport equations.         

Geometric VOF methods \citep{hirt1981volume,gueyffier1999volume,aulisa2007interface,weymouth2010conservative,scardovelli2003interface,baraldi2014mass,arrufat2021mass} are well-known procedures that use geometric representations to capture the interface. Two primary advantages of these methods are: (i) they maintain a sharp interface while avoiding interface diffusion, and (ii) they have the potential to achieve mass conservation to machine precision \cite{weymouth2010conservative,baraldi2014mass,arrufat2021mass}, depending on the numerical procedure. These methods involve two fundamental steps: reconstruction and advection. In the reconstruction step, interface shape and location in each computational cell are estimated, and in the advection step, the volume fractions across the cell boundaries are exchanged using the velocity field. One of the well-known reconstruction procedures is the PLIC \citep{debar1974fundamentals,youngs1982time,youngs1984interface,li1995calcul} method that uses the neighboring cells to construct a piecewise linear interface. In the advection step, the volume fraction can be exchanged across cell boundaries using unsplit or directionally split methods. The unsplit geometric VOF methods allow the advection procedure in a single step, however, it involves implementing complex algorithms and has difficulties conserving mass. The split geometric VOF methods, on the other hand, perform the advection procedure in multiple steps, however, the numerics involved are relatively less complicated than the unsplit methods and have a high potential to conserve mass.

The geometric VOF method has many attractive features that make it suitable for 3D complex and turbulent FSI problems involving a wide range of length scales and time scales. However, to the best of our knowledge, such a 3D FSI algorithm involving VOF/PLIC (geometric VOF using PLIC technique) method has not been developed yet. As a result, their potential to investigate such problems remains unexplored.

Designing a robust 3D FSI algorithm involving the VOF/PLIC method is not trivial because the discontinuity of the interface across the cell boundaries and discontinuities across the sharp interface can lead to numerical artifacts such as (i) nonphysical occurrence of solid fragments (flotsam) due to inappropriate numerical discretization and treatment of stress contributions, and (ii) unwanted surface oscillations (especially at equilibrium states) generated due to strong unphysical parasitic currents caused by an imbalance and jumps in the stress distribution across the sharp interface. Furthermore, a 3D interface reconstruction procedure is much more complex than a 2D procedure. This paper develops a 3D Eulerian framework that leverages the benefits of the VOF/PLIC method while avoiding any numerical artifacts typically associated with it.

In this work, we develop a three-dimensional monolithic FSI framework to simulate the interaction of incompressible flow and viscous hyperelastic solids on a fixed cartesian grid. We use the geometric VOF method to capture the fluid-solid interface, where the PLIC method is used to reconstruct the interface and Lagrangian Explicit (LE) \cite{gueyffier1999volume,alame2020numerical,scardovelli2003interface} directionally split advection procedure is used to exchange the volume fraction across cell boundaries. Both linear and nonlinear Mooney-Rivlin hyperelastic material models are considered to capture the mechanical behavior of solids, where the left Cauchy-Green deformation tensor ($\bm{B}$) is used to account for the solid deformation on an Eulerian grid. To treat the advection terms involved in the transport of $\bm{B}$ tensor, we use the fifth-order WENO-Z \cite{borges2008improved,jiang1996efficient} finite difference reconstruction procedure, and to solve the unified momentum conservation equations with incompressibility constraints, we use the finite volume method (FVM) \cite{mahesh2004numerical}. The procedure is designed to efficiently handle large-scale turbulent FSI problems involving a wide range of time scales and length scales. We thoroughly test the accuracy and robustness of the FSI algorithm using a variety of benchmark problems.

In section \ref{sec:methodology}, we discuss the methodology, where section \ref{subsec:gov_eqns} presents the governing equations and section \ref{subsec:numerical_details} provides the numerical details. In section \ref{sec:num_experiments}, we discuss the numerical experiments. Finally, we conclude the paper in section \ref{sec:conclusions}.

% ---------------------------------------

% --------- Methodology -----------------
\section{Methodology}\label{sec:methodology}
% -- Governing equations
\subsection{Governing equations} \label{subsec:gov_eqns}

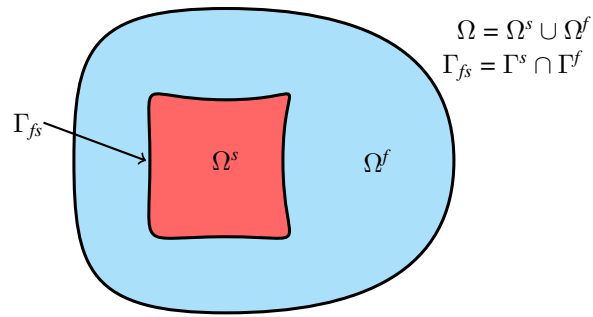
\begin{figure}[h!]
    \centering
            \begin{tikzpicture}
            \draw[color=black!100, fill=cyan!30,opacity=1, line width=0.4mm] plot[smooth cycle, tension=1.3] coordinates {(-2, 0) (0, -2)  (3, 0) (0, 2)};            
            \draw[color=black!100, fill=red!60, line width=0.4mm] plot[smooth cycle, tension=2.5] coordinates {(-1, 0) (0, -1) (0.75,0)  (0, 0.8)}; 
            \node at (0,0) {\color{black}{$\Omega^s$}};
            \node at (2,0) {\color{black}{$\Omega^f$}};
            \node at (3.93,1.75) {\color{black}{$\Omega = \Omega^s\cup\Omega^f$}};
            \node at (3.78,1.25) {\color{black}{$\Gamma_{fs} = \Gamma^s\cap\Gamma^f$}};    
            \draw[black,thick,<-] (-1.05,0) -- (-2.4,0.5);
            \node at (-2.6,0.45) {$\Gamma_{fs}$}; 
            
        \end{tikzpicture}
    \caption{Schematic of the problem.\\\hspace{\textwidth}}
    \label{fig:problem_schematic}
\end{figure}

We consider the interaction of fluid and solid systems as shown in figure \ref{fig:problem_schematic}. It shows the combined fluid-solid domain ($\Omega$), where $\Omega^f$ is the fluid subdomain, $\Omega^s$ is the solid subdomain, and $\Gamma_{fs}$ is the fluid-solid interface.

We solve the momentum conservation equation with an incompressibility constraint in both subdomains, i.e., in the fluid subdomain ($\Omega^f$), the governing equations are
\begin{equation}
\begin{split}
    \der{\uif}{t} + \der{\uif\ujf}{x_j} & = -\invrhof\der{p^f}{x_i} + \invrhof\der{\sigma_{ij}^f}{x_j}, \\
    \der{\uif}{x_i} & = 0,
\end{split}
\label{eq:GOV_fluid}
\end{equation}
and in the solid subdomain ($\Omega^s$),  
\begin{equation}
\begin{split}
    \der{\uis}{t} + \der{\uis\ujs}{x_j} & = -\invrhos\der{p^s}{x_i} + \invrhos\der{\sigma_{ij}^s}{x_j}, \\
    \der{\uis}{x_i} & = 0,
\end{split}
\label{eq:GOV_solid}
\end{equation}
where $\rho$ is density, $p$ is pressure, $u_i$ is the velocity component, and $\sigma_{ij}$ is a component of the stress tensor $\bm{\sigma}$. The subscripts $i = $ 1, 2, and 3 correspond to $x$, $y$, and $z$ directions, respectively. Superscripts $f$ and $s$ refer to fluid and solid quantities, respectively.

At the fluid-solid interface ($\Gamma_{fs}$), the kinematic and dynamic interaction requires continuity of velocity and traction force,  
%\begin{subequations}
    \begin{align}
    u_i^f & = u_i^s, \label{eq:GOV_fs_kin} \\
    \sigma_{ij}^f n_j & = \sigma_{ij}^s n_j, \label{eq:GOV_fs_dyn}    
    \end{align}
    %\label{eq:GOV_fs}
%\end{subequations}
where $n_j$ is the unit normal vector at $\Gamma_{fs}$.

To distinguish subdomains in the Eulerian VOF framework, we use a Heaviside function ($H$) that represents subdomains as 
\begin{equation}
        H(x,y,z) = 
    \begin{cases}
        0,& \text{if } (x,y,z) \in \Omega^f,\\
        1,& \text{if } (x,y,z) \in \Omega^s,
    \end{cases}
\end{equation}
and in the computational domain, we approximate it in each computational cell as
\begin{equation}
    \phis = \frac{1}{V_{c}}\int_{\Omega_{c}} H(x,y,z)\text{d}V,
\end{equation}
where $\phis$ is the solid volume fraction, $\Omega_{c}$ represents the volume integral over the cell volume, and $V_{c} = \int_{\Omega_{c}} \text{d}V$. By definition, $\phis = 0$ in the fluid domain, $\phis = 1$ in the solid domain, and $\phis \in (0,1)$ at the fluid-solid interface. 

To solve the governing equations simultaneously, we use the one-continuum formulation \citep{tryggvason2007immersed}, where monolithic quantities are defined over the whole domain ($\Omega = \Omega^f\cup\Omega^s$). We use a volume-averaging technique to define the monolithic velocity ($u_i$) as 
\begin{equation}
    u_i = \phis u_i^s + (1-\phis)u_i^f,    
\end{equation}
and it inherently satisfies the kinematic interaction condition (\ref{eq:GOV_fs_kin}) at the fluid-solid interface. The unified representation of the governing equations (\ref{eq:GOV_fluid} and \ref{eq:GOV_solid}) is  
\begin{equation}
\begin{split}
    \der{u_i}{t} + \der{u_i u_j}{x_j} & = -\frac{1}{\rho}\der{p}{x_i} + \frac{1}{\rho}\der{\sigma_{ij}}{x_j}, \\
    \der{u_i}{x_i} & = 0,    
\end{split}    
\label{eq:GOV}
\end{equation}
where $\rho$, $p$, and $\sigma_{ij}$ are defined over the whole domain. The density and stress distribution over the entire domain are defined as
\begin{align}
    \rho & = \phis \rhos + (1-\phis)\rhof, \\
    \sigma_{ij} & = \phis \sigma_{ij}^s + (1-\phis)\sigma_{ij}^f,
\end{align}
and by using such a mixture model for stress, we essentially satisfy the dynamic interaction condition (\ref{eq:GOV_fs_dyn}) at the fluid-solid interface. 

In the fluid domain, we consider Newtonian fluid, and therefore, the fluid stress is prescribed as 
\begin{equation}
    \sigma_{ij}^f = 2\muf D_{ij},
\end{equation}
where $\muf$ is the fluid dynamic viscosity and $D_{ij}$ is a component of the strain-rate tensor $\bm{D}$,
\begin{equation}
    D_{ij} = \frac{1}{2}\left(\der{u_i}{x_j} + \der{u_j}{x_i}\right).
\end{equation}

In the solid domain, we consider viscous hyperelastic material, and therefore, the solid stress is prescribed as 
\begin{equation}
    \sigma_{ij}^s = 2\mus D_{ij} + \sigma_{ij}^{hy},    
\end{equation}
where $2\mus D_{ij}$ is the viscous contribution and $\sigma_{ij}^{hy}$ is the hyperelastic stress contribution. We consider the mechanical behavior of linear and nonlinear hyperelastic materials that satisfy the incompressible Mooney-Rivlin law \cite{mooney1940theory,rivlin1948large} and are often used to model biological and soft rubber-like materials. Specifically, we consider neo-Hookean and Saint Venant-Kirchhoff materials that are subsets of linear and nonlinear Mooney-Rivlin materials, respectively \cite{ii2011implicit,sugiyama2011full}.

For the neo-Hookean material, the hyperelastic stress is prescribed as \cite{ii2011implicit,sugiyama2011full}
\begin{equation}
\sigma^{hy}_{ij} = G^s(B_{ij}-\delta_{ij}), 
\end{equation}
where $G^s$ is the modulus of transverse elasticity, $B_{ij}$ is a component of the left Cauchy-Green deformation tensor $\bm{B}$, and $\delta_{ij}$ is the Kronecker delta function. 

For the Saint Venant-Kirchhoff material, the hyperelastic stress is modeled as \cite{ii2011implicit,sugiyama2011full} 
\begin{equation}
\sigma^{hy}_{ij} = G^s(B_{ik}B_{kj}-B_{ij}).
\end{equation}

As the upper convected time derivative of $\bm{B}$ is identically zero \cite{bonet1997nonlinear,sugiyama2011full},  we can update $\bm{B}$ as
\begin{equation}
    \der{B_{ij}}{t} + u_k \der{B_{ij}}{x_k} = B_{ik}\der{u_j}{x_k} + B_{kj}\der{u_i}{x_k}.
    \label{eq:B}
\end{equation}

As the fluid and solid subdomains are assumed to be insoluble, the material derivative of $H$ is zero, %i.e., 
\begin{equation}
    \frac{\partial H}{\partial t} + u_i\frac{\partial H}{\partial x_i} = 0,
\end{equation}
and the volume fraction advection in geometric VOF corresponds to this transport equation.

\subsection{Numerical details} \label{subsec:numerical_details}
% -- Basic Introduction
We use an Eulerian VOF model similar to Sugiyama et al. \cite{sugiyama2011full} to simulate the interaction of incompressible flow and incompressible viscous hyperelastic structures on a fixed Cartesian grid. The fluid-solid interface is captured using the geometric VOF procedure that ensures a sharp interface. We first reconstruct the interface in each computational cell using piecewise linear interface calculation (PLIC) method \citep{debar1974fundamentals,youngs1982time,youngs1984interface,li1995calcul} and then perform a directionally split advection procedure to exchange volume fraction across cell boundaries using Lagrangian Explicit (LE) method \cite{gueyffier1999volume,alame2020numerical,scardovelli2003interface}. The transport equation of $\bm{B}$ tensor (\ref{eq:B}) is solved using the finite difference method (FDM) utilizing the fifth-order weighted essentially non-oscillatory (WENO-Z) \citep{jiang1996efficient,borges2008improved} finite difference reconstruction procedure, as the fifth-order WENO scheme \cite{jiang1996efficient} was shown to work well by Sugiyama et al. \cite{sugiyama2011full} However, the unified momentum conservation equations with incompressibility constraints (\ref{eq:GOV}) are solved using a finite volume method (FVM) \cite{mahesh2004numerical} that was developed for incompressible flows with a focus on discrete kinetic energy conservation in the inviscid limit and is robust at high Reynolds numbers. The FSI solver is built on the framework of our in-house FVM solver that has been extensively used in the past to investigate a variety of problems such as flow over superhydrophobic surfaces \cite{li2017feature,alame2019wall}, random roughness surfaces \cite{ma2021direct}, and roughness elements \cite{ma2022global,ma2023boundary}. It has also been used to analyze sources of turbulent wall-pressure fluctuations \cite{anantharamu2020analysis} and provide turbulent wall-pressure forcing to excite elastic plate vibrations \cite{anantharamu2021response, prajapati2023direct}.  

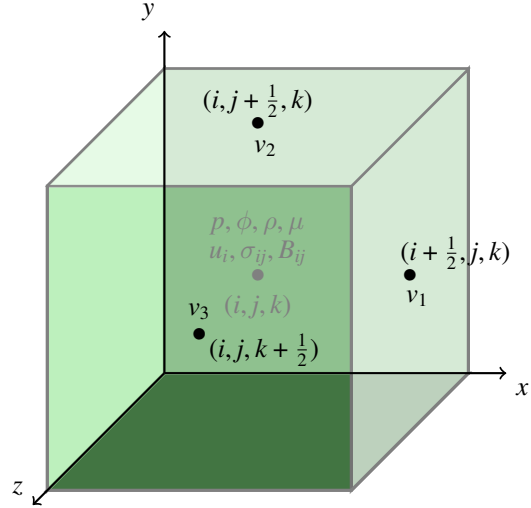
\begin{figure}[h!]
    \centering
    \begin{tikzpicture}

        \coordinate (p0) at (0,0,0);
        \coordinate (p1) at (0,\CubeLength,0);
        \coordinate (p2) at (0,\CubeLength,\CubeLength);
        \coordinate (p3) at (0,0,\CubeLength);
        \coordinate (p4) at (\CubeLength,0,0);
        \coordinate (p5) at (\CubeLength,\CubeLength,0);
        \coordinate (p6) at (\CubeLength,\CubeLength,\CubeLength);
        \coordinate (p7) at (\CubeLength,0,\CubeLength);
        \coordinate (a04) at (\CubeAxisLength,0,0);        
        \coordinate (a03) at (0,0,\CubeAxisLength);
        \coordinate (a01) at (0,\CubeAxisLength,0);        
        \coordinate (CubeCenter) at (\CubeHalfLength,\CubeHalfLength,\CubeHalfLength);
        \coordinate (XFaceCenter) at (\CubeLength,\CubeHalfLength,\CubeHalfLength);
        \coordinate (YFaceCenter) at (\CubeHalfLength,\CubeLength,\CubeHalfLength);
        \coordinate (ZFaceCenter) at (\CubeHalfLength,\CubeHalfLength,\CubeLength);
        \coordinate (CubeApex) at (\CubeLength,\CubeLength,\CubeLength);
        
        \draw[gray, very thick,fill=black!90] (p0) -- (p3) -- (p7) -- (p4) -- cycle;
        \draw[gray, very thick,fill=black!40] (p0) -- (p1) -- (p5) -- (p4) -- cycle;
        \draw[gray, very thick,fill=black!10] (p0) -- (p1) -- (p2) -- (p3) -- cycle;
        \draw[gray, very thick,fill=green!10,opacity=0.8] (p4) -- (p5) -- (p6) -- (p7) -- cycle;
        \draw[gray, very thick,fill=green!50,opacity=0.4] (p3) -- (p2) -- (p6) -- (p7) -- cycle;
        \draw[gray, thick,fill=green!10,opacity=0.8] (p1) -- (p2) -- (p6) -- (p5) -- cycle;

        \draw[black,thick,->] (p0) -- (a04) node[anchor=north west] {$x$};
        \draw[black,thick,->] (p0) -- (a01) node[anchor=south east] {$y$};
        \draw[black,thick,->] (p0) -- (a03) node[anchor=south east] {$z$};

        \foreach \Point in {(CubeCenter)}{
        \node at \Point {\Large{\color{gray}{\textbullet}}};}
        \foreach \Point in {(XFaceCenter)}{
        \node at \Point {\Large{\color{black}{\textbullet}}};}
        \foreach \Point in {(YFaceCenter)}{
        \node at \Point {\Large{\color{black}{\textbullet}}};}        
        \foreach \Point in {(ZFaceCenter)}{
        \node at \Point {\Large{\color{black}{\textbullet}}};}

        \node[gray,thick] at ([shift=({0,-0.35,0})]CubeCenter) {\color{gray}{$(i,j,k)$}};
        \node[black,thick] at ([shift=({0.60,0.35,0})]XFaceCenter) {\color{black}{$(i+\onebytwo,j,k)$}};
        \node[black,thick] at ([shift=({0,0.35,0})]YFaceCenter) {\color{black}{$(i,j+\onebytwo,k)$}};
        \node[black,thick] at ([shift=({0.85,-0.15,0})]ZFaceCenter) {\color{black}{$(i,j,k+\onebytwo)$}};

        \node[gray,thick] at ([shift=({0,0.35,0})]CubeCenter) {\color{gray}{$u_i, \sigma_{ij}, \Btilvar_{ij}$}};
        \node[gray,thick] at ([shift=({0,0.75,0})]CubeCenter) {\color{gray}{$p, \phis, \rho, \mu$}};        
        \node[black,thick] at ([shift=({0.10,-0.25,0})]XFaceCenter) {\color{black}{$v_{1}$}};
        \node[black,thick] at ([shift=({0.10,-0.27,0})]YFaceCenter) {\color{black}{$v_{2}$}};
        \node[black,thick] at ([shift=({0,0.35,0})]ZFaceCenter) {\color{black}{$v_{3}$}};
    \end{tikzpicture}
    \caption{Location of variables for a three-dimensional unit computational cell. {\color{gray}{\circle}} cell center; {\color{black}{\circle}} face center.\\\hspace{\textwidth}}
    \label{fig:3Dvarstorage}
\end{figure}

The problem is discretized such that the velocity ($u_i$), stress ($\sigma_{ij}$), left Cauchy-Green deformation tensor ($\Btil$), pressure ($p$), solid volume fraction ($\phis$), density ($\rho$), and viscosity ($\mu$) are defined at the cell center; the face-normal velocities ($v_{n}$) are defined at the center of the cell faces. Figure \ref{fig:3Dvarstorage} shows the location of variables for a three-dimensional unit computational cell.       

\subsubsection{Time advancement and spatial discretizations} \label{subsubsec:discretization}

To describe the initial shape of the solid structure, we prescribe the corresponding $\phis$ distribution on the fixed Cartesian grid. In addition, we set $\bm{B}=\bm{I}$ in the whole domain, corresponding to the initial unstressed state of the solid system, where $\bm{I}$ is the unit tensor. Such a simplistic definition facilitates us in handling multicomponent solid systems and complex geometries.   

We update the variables at $\{n+1\}^{\mathrm{th}}$ timestep from the previous timesteps ($\{n\}, \{n-1\}$) in four steps:   

%\begin{enumerate}[label=(\roman*)]
\begin{steps}
   \item Update $\phis$ using geometric VOF method. \\
   The details are discussed in section \ref{subsubsec:geo_VOF}.\\
   \item Update $\bm{B}$ tensor. \\
   We use the second-order Adams-Bashforth scheme to update $\Btil$, 
    \begin{equation}
    \begin{split}
    \Btilvar_{ij}^{n+1} = \Btilvar_{ij}^{n} & + \Delta t \left[ \frac{3}{2}\left(\Btilvar_{ik}\der{u_j}{x_k} + \Btilvar_{kj}\der{u_i}{x_k} \right)^{n} - \frac{1}{2}\left(\Btilvar_{ik}\der{u_j}{x_k} + \Btilvar_{kj}\der{u_i}{x_k} \right)^{n-1} \right] \\
    & - \Delta t \left[ \frac{3}{2}\left(u_k\der{ \Btilvar_{ij}}{x_k}\right)^{n} - \frac{1}{2}\left(u_k\der{ \Btilvar_{ij}}{x_k}\right)^{n-1} \right],    
    \end{split}
    \label{eq:Bupdate}
    \end{equation}
    where the time step $\Delta t$ is set such that the CFL numbers based on advection speed and shear wave speed are less than 0.5.\\
   \item Fractional step - predict velocity while omitting the pressure contribution. \\
    We use an implicit Crank-Nicolson time advancement scheme to predict the cell-centered velocity,
    \begin{equation}
        \frac{u_i^* - u_i^n}{\Delta t} = \frac{1}{2}\left[ \left( \frac{1}{\rho}\der{\sigma_{ij}}{x_j} - \der{u_i u_j}{x_j} \right)^{n} + \left( \frac{1}{\rho}\der{\sigma_{ij}}{x_j} - \der{u_i u_j}{x_j} \right)^{n+1} \right],
    \end{equation}
    where we linearize the nonlinear convection terms and set $u_i^{n+1}$ as $u_i^{*}$. We perform symmetric interpolation ($\mathrm{O}(\Delta_x^2)$) to obtain the face-normal velocities from the cell-centered velocities,
    \begin{equation}
        v_n = \left( \frac{u_{cv1} + u_{cv2}}{2} \right),
    \end{equation}
    where $u_{cv1}$ and $u_{cv2}$ are the cell-centered velocity components of control volumes (CVs) $cv1$ and $cv2$ that share the considered face. Also, $u_{cv1}$ and $u_{cv2}$ are aligned in the face-normal direction. \\
    \item Fractional step - correct velocity using the pressure distribution yielded from the incompressibility constraint. \\
    We project the corrector step,
    \begin{equation}
        \frac{u_i^{n+1} - u_i^*}{\Delta t} = - \frac{1}{\rho} \frac{\partial p^{n+1}}{\partial x_i},    
    \end{equation}
    on the face-normal, 
    \begin{equation}
        \frac{v_n^{n+1} - v_n^*}{\Delta t} = - \frac{1}{\rho} \frac{\partial p^{n+1}}{\partial n},    
    \end{equation}
    and solve the Poisson equation derived using the incompressibility constraint to obtain the pressure distribution,
    \begin{equation}
        \sum_{k = 1}^{N_{faces}} \frac{\Delta t}{\rho} \frac{\partial p^{n+1}}{\partial n} A_{k} = \sum_{k = 1}^{N_{faces}} v_n^* A_{k},
    \end{equation}
    where $k = 1, 2, \dots, N_{faces}$ correspond to the faces of a CV and $A_{k}$ is the face area. The Poisson equation is solved using the multigrid preconditioned conjugate gradient (CG) method utilizing the Trilinos libraries (Sandia National Labs). Finally, we correct the predicted velocity using the pressure gradient,
    \begin{equation}
        \frac{u_i^{n+1} - u_i^*}{\Delta t} = - \frac{1}{\rho} \frac{\partial p^{n+1}}{\partial x_i}.
    \end{equation} 

\end{steps}   
%\end{enumerate}

In the above time advancement equations, we compute the derivatives (except for the advection terms involved in equation \ref{eq:Bupdate}) by computing the derivative at faces and using symmetric interpolation as required. The derivative terms at the face center are defined as
\begin{equation}
    \frac{\partial (\cdot)}{\partial n} = \frac{(\cdot)_{cv2} - (\cdot)_{cv1}}{\Delta_f}, 
\end{equation}
where the face-normal $n$ points from $cv1$ to $cv2$ and $\Delta_f = (x_{i,cv2}-x_{i,cv1})$. For the advection term involved in equation \ref{eq:Bupdate}, we use the fifth-order WENO-Z finite difference reconstruction procedure (discussed in section \ref{subsubsec:WENO}).

\subsubsection{Geometric VOF (VOF/PLIC) interface-capturing method}\label{subsubsec:geo_VOF}

The geometric VOF method involves two fundamental steps: reconstruction and advection.\\

\noindent \underline{\textit{Reconstruction step}} \\

In the reconstruction step, we find the interface location in each computational cell using the PLIC technique \citep{debar1974fundamentals,youngs1982time,youngs1984interface,li1995calcul}. It involves two steps: (i) estimating the normal direction of the interface and (ii) geometrically shifting the interface location along the normal direction to attain the required volume under the interface.   

We estimate the normal ($\mathbf{m}$) of the local interface as $\mathbf{m} = m_i\mathbf{e}_i =  -\nabla_{\delta_{x_i}}\phis$ using Young's finite difference method \cite{youngs1984interface,li1995calcul}, where the finite difference approximation of the normals at the cell corners are used to compute the normals at the cell centers.
For 2D problems with uniform cells, the normal components at the cell corner are estimated as,
\begin{align}
    m_{x:i+\onebytwo,j+\onebytwo} & = -\frac{1}{2\Delta_x}\left( \phis_{i+1,j} - \phis_{i,j} + \phis_{i+1,j+1} - \phis_{i,j+1} \right), \\
    m_{y:i+\onebytwo,j+\onebytwo} & = -\frac{1}{2\Delta_y}\left( \phis_{i,j+1} - \phis_{i,j} + \phis_{i+1,j+1} - \phis_{i+1,j} \right),    
\end{align}
where $\Delta_x$ and $\Delta_y$ are the grid spacing in $x$ and $y$ directions, respectively. Then, we compute the components at the cell center as,
\begin{align}
    m_{x/y:i,j} & = \frac{1}{4}\left( m_{x/y:i-\onebytwo,j-\onebytwo} + m_{x/y:i+\onebytwo,j-\onebytwo} + m_{x/y:i+\onebytwo,j+\onebytwo} + m_{x/y:i-\onebytwo,j+\onebytwo} \right).
\end{align}
Similarly, we estimate normals for 3D problems and represent the interface as,
\begin{equation}
    m_{x:i,j,k} x + m_{y:i,j,k} y + m_{z:i,j,k} z = \beta_{i,j,k},
\end{equation}
and we use the analytical relations \cite{scardovelli2000analytical} to adjust $\beta_{i,j,k}$ such that the local solid volume under the interface is equal to $\phis_{i,j,k} V_{c}$, where $V_c$ is the cell volume.\\

\vspace{16pt}
\noindent \underline{\textit{Advection step}}\\

In the advection step, we use the directionally split advection procedure where Lagrangian Explicit (LE) \citep{gueyffier1999volume,alame2020numerical,scardovelli2003interface} geometrical linear-mapping method is used to exchange the reference volume fraction across the cell boundaries. Each timestep involves three directionally split advection substeps for 3D problems, where $\phis^{(n,l+1)}$ is estimated using $\phis^{(n,l)}$, where $l = 1, 2,3$ corresponds to the directional sweeps, $\phis^{(n,1)}=\phis^n$, and $\phis^{(n,4)}=\phis^{n+1}$. Before each advection substep, we reconstruct the interface to be consistent with $\phis^{(n,l)}$. As a result, each timestep involves three reconstruction steps and three advection steps. To avoid preferential direction, the directional sweeps are rotated cyclically. 

To advect the reconstructed interface in $x_d$ direction, we use a linearized velocity field to move the endpoints of the interface segment in each cell. We rescale space and time using the local grid spacing ($\Delta x_d$) and time step ($\Delta t$). It yields the following non-dimensional variables: $x^*_d=x_d/\Delta x_d$, $t^*=t/\Delta t$, and $u^*_d=u_d\Delta t/\Delta x_d$, where the non-dimensional velocity is the local CFL number. Such rescaling enables the interface capturing algorithm to efficiently handle cuboidal cells, as all the cells are mapped to a unit cube. For brevity, superscript `*' is dropped in the following discussion.   

The general LE steps are summarized below.

\begin{figure}
    \centering
    \begin{subfigure}[t]{0.90\textwidth}
        \centering
        \begin{tikzpicture}
        % main grid
        \draw[step=3,black,ultra thin] (0,0) grid (9,3);          
        % lines
        \draw [thick, red] (0,1.5) -- (3,2.3);
        \draw [thick, red] (6,2) -- (9,1.4);  
        \draw [ultra thick, red] (3,2) -- (6,2.2);
        % arrows
        \draw [thick, black, ->] (3,2.3) -- (4,2.3);
        \draw [thick, black, <-] (5.5,2) -- (6,2);  
        % center rectangle
        \draw[black,ultra thick] (3,0) rectangle (6,3);

        % labels
        \draw[fill=black] (1.5,1.5) circle[radius=1.5pt];
        \draw[fill=black] (4.5,1.5) circle[radius=1.5pt];
        \draw[fill=black] (7.5,1.5) circle[radius=1.5pt];
        \node[] at (1.5,1.2) {$i-1,j$};
        \node[] at (4.5,1.2) {$i,j$};
        \node[] at (7.5,1.2) {$i+1,j$};
        \node[] at (2.2,2.4) {$u_{i-1/2,j}>0$};        
        \node[] at (6.8,2.4) {$u_{i+1/2,j}<0$};
        \node[] at (-0.8,2.4) {$u_{i-3/2,j}=0$};
        \node[] at (9.8,2.4) {$u_{i+3/2,j}=0$};
        
        \end{tikzpicture}
        \caption{}
    \end{subfigure}
    \\~\\
    \begin{subfigure}[t]{0.90\textwidth}
        \centering
        \begin{tikzpicture}
        % main grid
        \draw[step=3,black,ultra thin] (0,0) grid (9,3);
        % thick line (center) 
        \draw [ultra thick, red] (3,2) -- (6,2.2);
        % hiding boxes (center)
        \draw[fill=white,opacity=1] (3,0) rectangle (4,3); 
        \draw[fill=white,opacity=1] (5.5,0) rectangle (6,3);
        % thick line (left, right)
        \draw [ultra thick, red] (0,1.5) -- (4,2.3);
        \draw [ultra thick, red] (5.5,2) -- (9,1.4);
        % hiding boxes (left, right)
        \draw[fill=white,opacity=1] (0,0) rectangle (3,3);  
        \draw[fill=white,opacity=1] (6,0) rectangle (9,3);
        % lines (left, right)
        \draw [thick, red] (0,1.5) -- (4,2.3);
        \draw [thick, red] (5.5,2) -- (9,1.4);
        % thick box border (center)
        \draw[black,ultra thick] (4,0) rectangle (5.5,3); 
        \draw[black,ultra thick] (3,0) rectangle (6,3);
        \node[] at (3.55,1.2) {$\Phi_1$};
        \node[] at (4.8,1.2) {$\Phi_2$};
        \node[] at (5.77,1.2) {$\Phi_3$};          
        %\draw [thin, black, ->] (3,2.3) -- (4,2.3);
        %\draw [thin, black, <-] (5.5,2) -- (6,2);             
        \end{tikzpicture}
        \caption{}
    \end{subfigure}   
\caption{(a) Reconstructed interface with horizontal velocities and (b) volume formation ($\Phi_i$) for the central cell by the Lagrangian advection in $x$ direction.\\\hspace{\textwidth}}
\label{fig:geovof_schematic}
\end{figure}
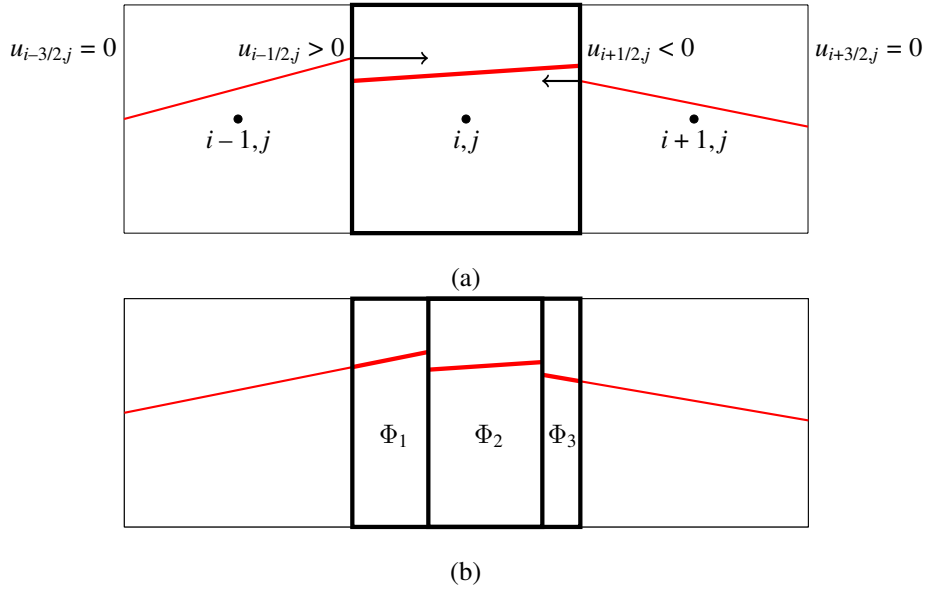

%\begin{enumerate}[label=(\roman*)]
\begin{steps}
   \item Translate the reconstructed segments using the equation $x_d(t^{n+1}) = (1+u_{d+}-u_{d-})x_d + u_{d-}$, where the left face is considered as the origin for $x_d$, $u_{d-}$ is the left face velocity, and $u_{d+}$ is the right face velocity. This expression is obtained by integrating the equation of motion using an explicit first-order scheme ($dx_d/dt=u_d(x_d)$) and linearly interpolating velocity ($u_d(\mathbf{x})=u_{d-}(1-x_d) + u_{d+}x_d$). 
   \item Update the volume fraction by accounting for the contributions from the neighboring cells, i.e., $\phis_{i,j,k}^{(n,l+1)} = \Phi_1^d + \Phi_2^d + \Phi_3^d$, where $\Phi_2^d$ corresponds to the contribution from the cell itself, and $\Phi_1^d$ and $\Phi_3^d$ corresponds to the contribution from the left and right cells, respectively. Figure \ref{fig:geovof_schematic} shows a simple representation of these volume contributions for a 2D cell.
   \item Scale all the measures of space using the `compression' factor $(1+u_{d+}-u_{d-})$ in the cell region, translate the cells by $u_{d-}$, and the final update can be expressed as $\phis_{i,j,k}^{(n,l+1)} = \phis_{i,j,k}^{(n,l)}(1+u_{d+}-u_{d-})- F_{d+} + F_{d-}$, where $F_{d-}$ corresponds to the flux of $u_dH$ at the left face and $F_{d+}$ corresponds to the flux at the right face.
    
\end{steps}
%\end{enumerate}    

To avoid the unwanted effects of the arithmetic floating-point round-off error, we perform the clipping procedure after each directional sweep. We set $\phis_{i,j,k}=0$ if $\phis_{i,j,k}<\epsilon_c$ and $\phis_{i,j,k}=1$ if $\phis_{i,j,k}>1-\epsilon_c$, where the clipping tolerance $\epsilon_c$ is set to $10^{-8}$ for all the problems. For more details, we refer the reader to Alam{\'e} \cite{alame2020numerical}. To prevent the exponential growth of $\bm{B}$ near the interface due to the shearing motion in the fluid domain, we set $\bm{B} = \bm{I}$ for $\phis_{i,j,k}<\epsilon_c$.

\subsubsection{Fifth-order WENO-Z finite difference reconstruction}\label{subsubsec:WENO}

The fifth-order WENO-Z finite difference reconstruction procedure \citep{jiang1996efficient,borges2008improved} is used to treat the advection terms involved in the transport equation of $\bm{B}$ tensor (\ref{eq:Bupdate}), as the fifth-order WENO scheme \cite{jiang1996efficient} was shown to perform well by Sugiyama et al. \cite{sugiyama2011full} We discuss the reconstruction procedure to treat the advection term involved in the one-dimensional transport equation of a scalar $f$,   
\begin{equation}
    %\der{f}{t} + \der{L}{x} = 0,
    \der{f}{t} + u\der{f}{x} = 0,    
    \label{eq:phis}
\end{equation}
and a similar procedure is used to treat the advection terms involved in the 3D transport equation of the $\bm{B}$ tensor.

We consider the compact upwind/downwind form \citep{pawar2019cfd} of the above transport equation, 
\begin{equation}
    \der{f_i}{t} + u_{i}^p\left( \frac{\tilde{f}^+_{i+\frac{1}{2}} - \tilde{f}^+_{i-\frac{1}{2}}}{\Delta_x} \right) + u_{i}^m\left( \frac{\tilde{f}^-_{i+\frac{1}{2}} - \tilde{f}^-_{i-\frac{1}{2}}}{\Delta_x} \right) = 0,   
\end{equation}
where $f_i$ is the cell-centered value of $f$ on a uniform grid with $\Delta_x$ spacing, $u_{i}^p = \max (u_i,0)$, $u_{i}^m = \min (u_i,0)$, $\tilde{f}$ is a high-order reconstruction of $f$, and the superscripts $+$ and $-$ reflect correspondence to positive ($u_{i}^p$) and negative ($u_{i}^m$) cell-center velocities, respectively. 

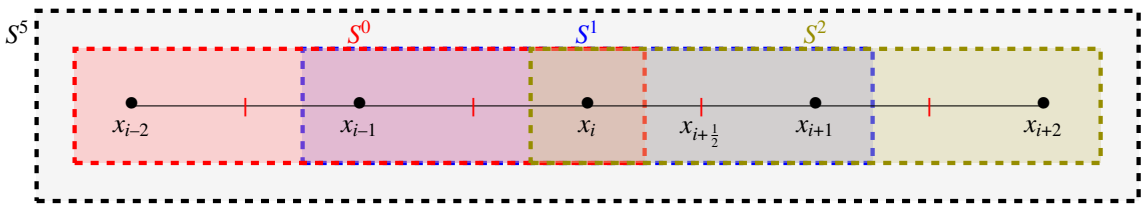
\begin{figure}[h!]
    \centering
    \begin{tikzpicture}[node distance=3cm]

    \filldraw[fill=black!40!white, draw=none, opacity=0.1] (-1.25,-1.25) -- (13.25,-1.25) -- (13.25,1.25) -- (-1.25,1.25) -- (-1.25,-1.25);
    \draw[ultra thick, dashed, draw=black, opacity=1] (-1.25,-1.25) -- (13.25,-1.25) -- (13.25,1.25) -- (-1.25,1.25) -- (-1.25,-1.25);

    \filldraw[fill=blue!40!white, draw=none, opacity=0.4] (2.25,-0.75) -- (9.75,-0.75) -- (9.75,0.75) -- (2.25,0.75) -- (2.25,-0.75);
    \draw[ultra thick, dashed, draw=blue, opacity=1] (2.25,-0.75) -- (9.75,-0.75) -- (9.75,0.75) -- (2.25,0.75) -- (2.25,-0.75); 
    
    \filldraw[fill=red!40!white, draw=none, opacity=0.4] (-0.75,-0.75) -- (6.75,-0.75) -- (6.75,0.75) -- (-0.75,0.75) -- (-0.75,-0.75);
    \draw[ultra thick, dashed, draw=red, opacity=1] (-0.75,-0.75) -- (6.75,-0.75) -- (6.75,0.75) -- (-0.75,0.75) -- (-0.75,-0.75);
    
    \filldraw[fill=olive!40!white, draw=none, opacity=0.4] (5.25,-0.75) -- (12.75,-0.75) -- (12.75,0.75) -- (5.25,0.75) -- (5.25,-0.75);    
    \draw[ultra thick, dashed, draw=olive, opacity=1] (5.25,-0.75) -- (12.75,-0.75) -- (12.75,0.75) -- (5.25,0.75) -- (5.25,-0.75);

    \draw[anchor=center,black,thin] (0,0) -- (12,0);

    % cell-center points
    \node (c1)  [shift=({0,-0.02})]            {{\Large{\color{black}{\textbullet}}}};
    \node (c2)  [right of=c1] {{\Large{\color{black}{\textbullet}}}};
    \node (c3)  [right of=c2] {{\Large{\color{black}{\textbullet}}}};  
    \node (c4)  [right of=c3] {{\Large{\color{black}{\textbullet}}}};    
    \node (c5)  [right of=c4] {{\Large{\color{black}{\textbullet}}}};

    % face-center points
    %\node (f1) [shift=({1.5cm,-0.02})]   {{\Large{\color{red}{$\circ$}}}};
    \node (f1) [shift=({1.5cm,-0.02})]   {{\color{red}{$|$}}};    
    \node (f2) [right of=f1]   {{\color{red}{$|$}}};
    \node (f3) [right of=f2]   {{\color{red}{$|$}}};    
    \node (f4) [right of=f3]   {{\color{red}{$|$}}};

    % grid labels
    \node (g1) [shift=({0,-0.3})]   {$x_{i-2}$};
    \node (g2) [right of=g1]        {$x_{i-1}$};
    \node (g3) [right of=g2]        {$x_{i}$};
    \node (g3_f) [right of=g3] [shift=({-1.5cm,-0.07})] {$x_{i+\frac{1}{2}}$};    
    \node (g4) [right of=g3]        {$x_{i+1}$};
    \node (g5) [right of=g4]        {$x_{i+2}$};
    % --------

    % -------- Stencil names
    \node (s5) [shift=({-1.5,0.975})] {\color{black}{$S^5$}};
    \node (s0) [shift=({3,0.975})] {\color{red}{$S^0$}};
    \node (s1) [right of=s0] {\color{blue}{$S^1$}};
    \node (s2) [right of=s1] {\color{olive}{$S^2$}};    
    % --------
    
    \end{tikzpicture}
    \caption{Stencils involved in the fifth-order WENO reconstruction of $g_{i+\frac{1}{2}}$.\\\hspace{\textwidth}}
    \label{fig:WENO_stencils}
\end{figure}

The reconstruction procedure for $\Tilde{f}_{i+\frac{1}{2}}^+$ (referred to as $g_{i+\frac{1}{2}}$) is discussed, and the procedure for $\Tilde{f}_{i+\frac{1}{2}}^-$ is similar with symmetry about $x_{i+\frac{1}{2}}$. We use a 5-point stencil ($S^5 = \{x_{i-2},x_{i-1},x_{i},x_{i+1},x_{i+2}\}$) to compute $g_{i+\frac{1}{2}}$ and it comprises of three 3-point stencils ($S^0=\{x_{i-2},x_{i-1},x_{i}\},S^1=\{x_{i-1},x_{i},x_{i+1}\},S^2=\{x_{i},x_{i+1},x_{i+2}\}$), as shown in figure \ref{fig:WENO_stencils}. We compute $g_{i+\onebytwo}$ as
\begin{equation}
    g_{i+\onebytwo} = \sum_{j=0}^2\omega_j g_{i+\onebytwo}^j, 
\end{equation}
where  $g_{i+\onebytwo}$ is computed as a convex combination of interpolated values $g_{i+\onebytwo}^j$ on 3-point stencils,
\begin{equation}
    g_{i+\onebytwo}^j = \sum_{k=0}^2\alpha_{jk} f_{i+j+k-2}, 
\end{equation}
the weights $\omega_j$ corresponds to 3-point stencils $S^j$, and the coefficients $\alpha_{jk}$ for all 3-point stencils are given by Jiang and Shu \cite{jiang1996efficient} as shown in table \ref{tab:WENO_coeffs}.

\begin{table}[]
    \centering
    \begin{tabular}{|c|c|c|c|}
         \hline
         $j$ & $k=0$ & $k=1$ & $k=2$ \\
         \hline
         0  & $1/3$ & $-7/6$ & $11/6$ \\
         1  & $-1/6$ & $5/6$ & $1/3$ \\
         2  & $1/3$ & $5/6$ & $-1/6$ \\    
         \hline
    \end{tabular}
    \caption{Coefficients $\alpha_{jk}$ for fifth-order WENO reconstruction \cite{jiang1996efficient}.\\\hspace{\textwidth}}
    \label{tab:WENO_coeffs}
\end{table} 

The weights $\omega_j$ are defined using smoothness indicators $IS_j$ that estimates the smoothness of $f$ over each stencil $S_j$. We consider the non-oscillatory weights suggested by  Borges et al. \cite{borges2008improved}, where  Borges et al. \cite{borges2008improved} developed an improved version of the classical WENO-JS \cite{jiang1996efficient} scheme to decrease dissipation and increase resolution. The non-oscillatory weights and smoothness indicators are given below.

Borges et al. \cite{borges2008improved} used the 5-point stencil $S^5$ to define the smoothness indicators, 
\begin{equation}
    IS_j^Z = \left( \frac{IS_j^{JS} + \epsilon}{IS_j^{JS} + IS_5^Z + \epsilon} \right), \quad j = 0, 1, 2,
\end{equation}
where the smoothness indicators $IS_j^{JS}$ are defined by Jiang and Shu \cite{jiang1996efficient} as
\begin{align}
   IS_0^{JS} & = \frac{13}{12}(f_{i-2} - 2f_{i-1} + f_i)^2 + \frac{1}{4}(f_{i-2} - 4f_{i-1} + 3f_i)^2, \\
   IS_1^{JS} & = \frac{13}{12}(f_{i-1} - 2f_{i} + f_{i+1})^2 + \frac{1}{4}(f_{i-1} - f_{i+1} )^2, \\
   IS_2^{JS} & = \frac{13}{12}(f_{i} - 2f_{i+1} + f_{i+2})^2 + \frac{1}{4}(3f_{i} - 4f_{i+1} + f_{i+2})^2,   
\end{align}
the smoothness indicator $IS_5^Z$ is defined by Borges et al. \cite{borges2008improved} as  
\begin{equation}
    IS_5^Z = \left| IS_0^{JS}-IS_2^{JS} \right| ,
\end{equation}
and the corresponding weights $\omega_j^Z$ are defined as
\begin{equation}
    \omega_j^Z = \frac{\beta_j^Z}{\beta_0^Z +\beta_1^Z + \beta_2^Z}, \quad \beta_j^Z = \frac{\gamma_j^{JS}}{IS_j^{Z}} =  \gamma_j^{JS} \left( 1 +  \frac{IS_5^Z}{IS_j^{JS}+\epsilon} \right),
\end{equation}
where $\gamma_0^{JS} = 1/10$, $\gamma_1^{JS} = 6/10$, $\gamma_2^{JS} = 3/10$ are the coefficients proposed by Jiang and Shu \cite{jiang1996efficient}, and $\epsilon = 10^{-40}$ is a small constant used to avoid division by zero.

% ---------------------------------------

% ------- Numerical experiments ---------
\section{Numerical experiments} \label{sec:num_experiments}

To assess the performance of the FSI solver, we consider multiple benchmark problems: rigid body motion of a slotted disk \cite{zalesak1979fully}, compliant wall in a lid-driven cavity \cite{zhao2008fixed,wang2010interpolation,dunne2006eulerian}, circular disk in a lid-driven cavity \cite{sugiyama2011full, zhao2008fixed}, disk in a shear flow \cite{ii2011implicit}, and sphere in a lid-driven cavity \cite{valizadeh2025monolithic,mao20243d}. Furthermore, we use the framework to perform direct numerical simulation (DNS) of turbulent channel flow with a deformable compliant bottom wall and a rigid top wall.  

\subsection{Rigid body motion of a slotted disk}

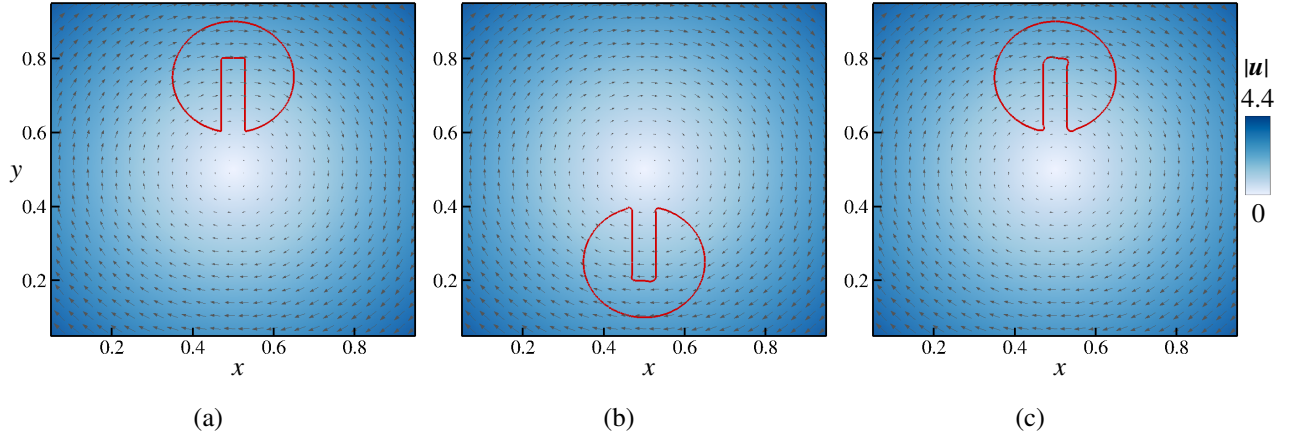
\begin{figure}[h]
\centering
% ----------
\begin{subfigure}[t]{0.30\textwidth}
\centering
\begin{tikzpicture}
    \node[anchor=north west] at (0,0) {\includegraphics[height=1\linewidth,trim={60 50 10 10},clip]
{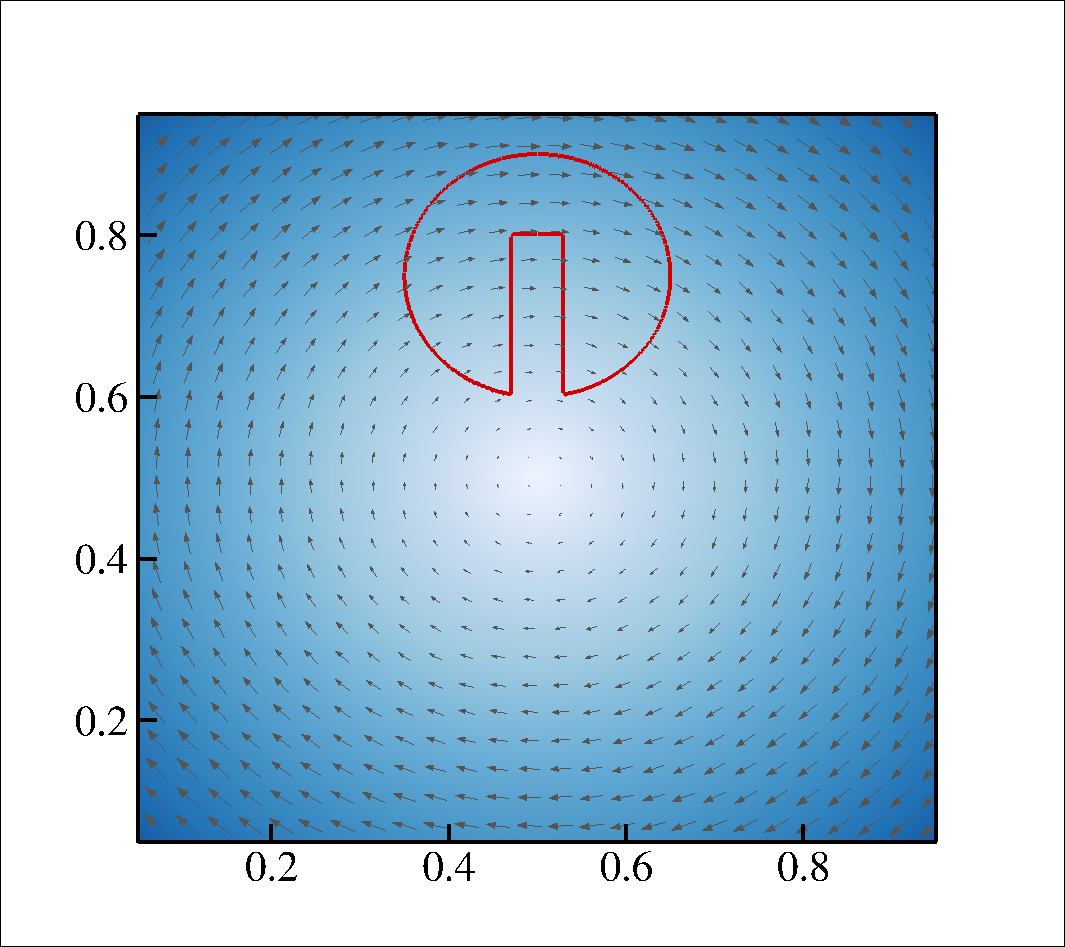}};
\end{tikzpicture}
\put(-175,72){{$y$}}
\put(-92,-2){{$x$}}
\caption{}
\label{subfig:ZD_a}
\end{subfigure}
% ----------
\begin{subfigure}[t]{0.30\textwidth}
\centering
\begin{tikzpicture}
    \node[anchor=north west] at (0,0) {\includegraphics[height=1\linewidth,trim={60 50 10 10},clip]
{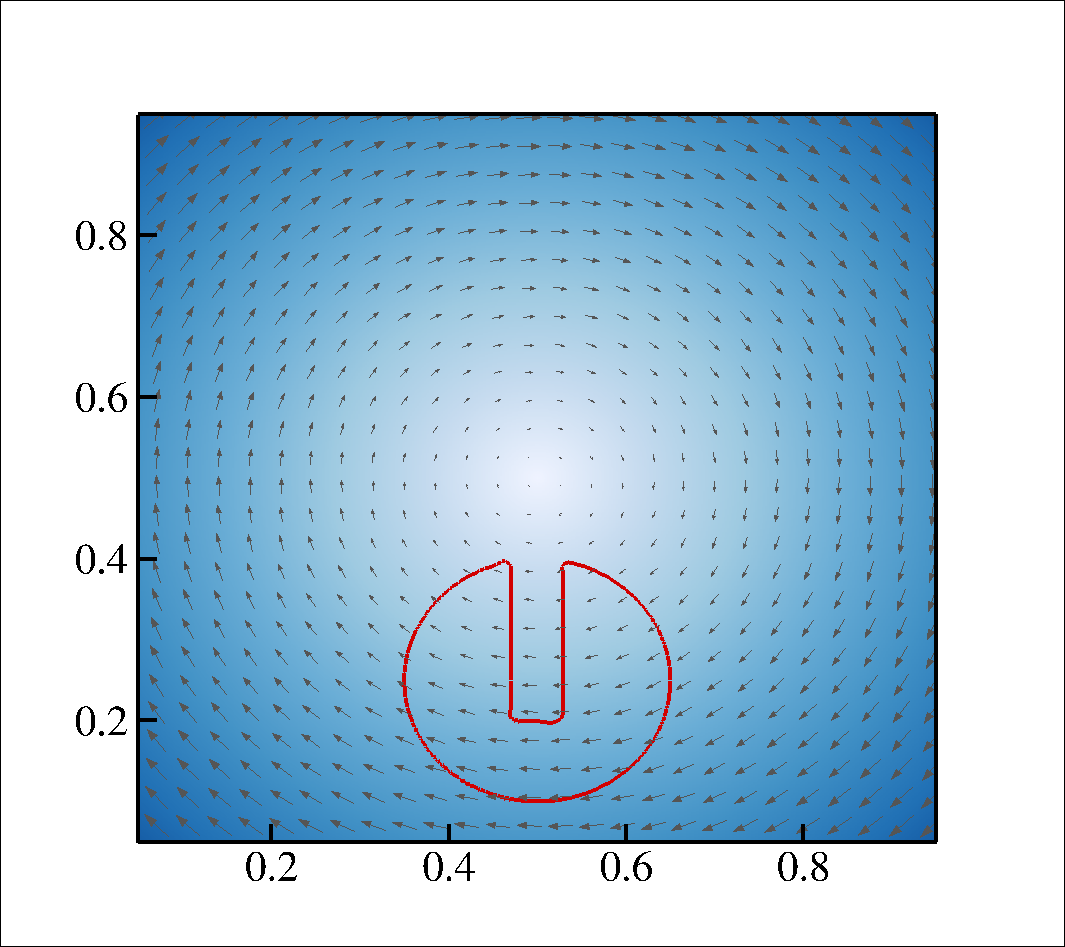}};
\end{tikzpicture}
\put(-92,-2){{$x$}}
\caption{}
\label{subfig:ZD_b}
\end{subfigure}
% ----------
\begin{subfigure}[t]{0.30\textwidth}
\centering
\begin{tikzpicture}
    \node[anchor=north west] at (0,0) {\includegraphics[height=1\linewidth,trim={60 50 10 10},clip]
{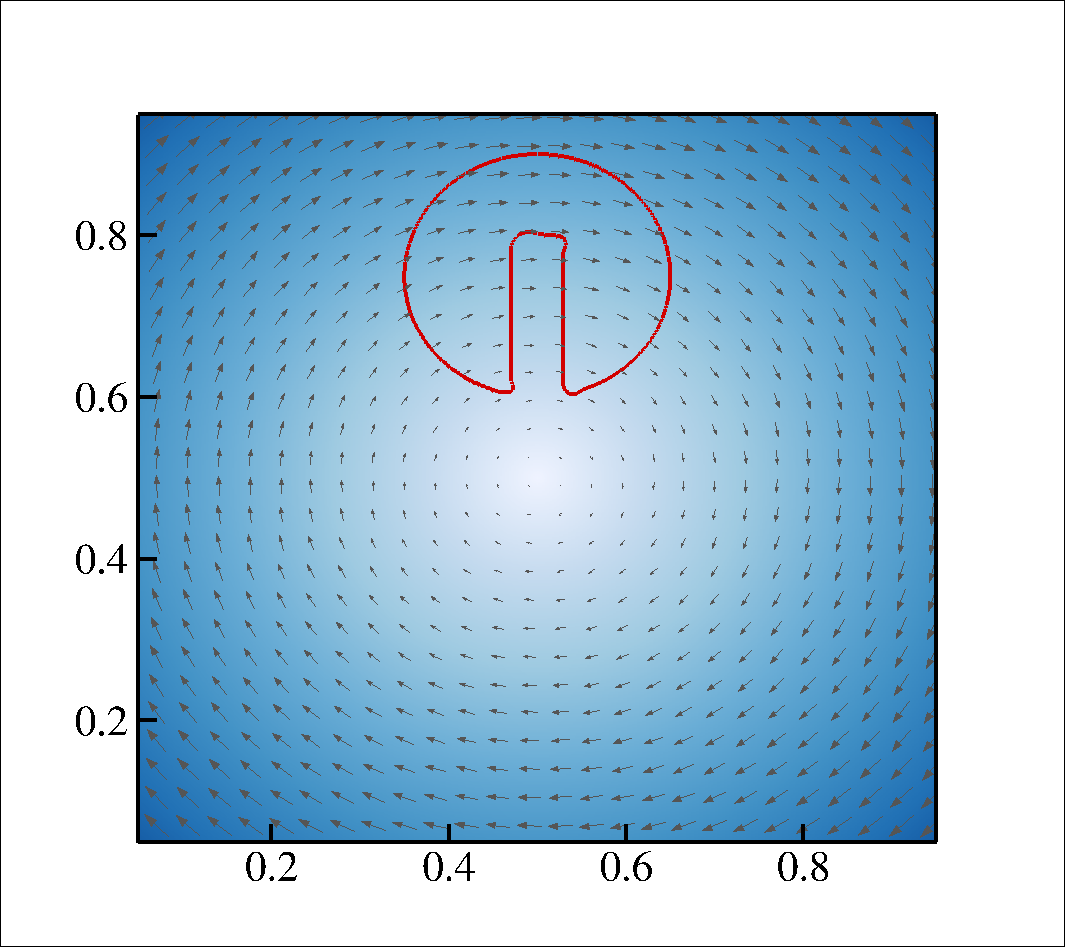}};
    \node [shift=({5.65,-2.75}),rotate=-90] {\includegraphics[width=0.2\textwidth,
        height=0.07\textwidth,trim={0 0 0 0},clip]
    {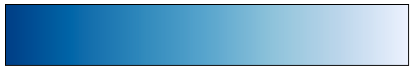}};
\end{tikzpicture}
\put(-92,-2){{$x$}}
\put(-18,55){{$0$}}
\put(-22,98){{$4.4$}}
\put(-21,110){{$\left|\bm{u}\right|$}}
\caption{}
\label{subfig:ZD_c}
\end{subfigure}
% ----------
\caption{Rigid slotted disk in a prescribed velocity field at (a) $t=0$, (b) $t=0.5$, and (c) $t=1$. The red curve shows the slotted disk shape ($\phis=0.5$), the filled contours show velocity magnitude, and the arrows represent velocity vectors. The unit square domain is discretized using $N_x\times N_y=512\times512$ cells.\\\hspace{\textwidth}}
\label{fig:Zalesak_snapshots}
\end{figure}

To validate the VOF/PLIC method, we simulate the rigid body motion of a slotted disk \cite{zalesak1979fully} in a prescribed velocity field. Initially, the slotted disk is centered at $(0.5,0.75)$ with a radius of $r = 0.15$ and a slot that is $l = 0.2$ deep and $w = 0.06$ wide. The computational domain is a unit square ($L_x\times L_y = 1\times 1$). The disk undergoes rigid body motion in the prescribed velocity field, 
\begin{equation}
    u(y) = 2\pi(y-y_c), \quad v(x) = -2\pi(x-x_c),
\end{equation}
where $(x_c,y_c)=(0.5,0.5)$ is the center of rotation. At $t = 1$, the disk completes one cycle of rotation in the velocity field, as shown in figure \ref{fig:Zalesak_snapshots}.

% =============================================================
% =======================================================

\begin{figure}[h]
\centering
% ----------
\begin{tikzpicture}
    \node[anchor=north west] at (0,0) {\includegraphics[width=0.40\linewidth,trim={60 50 100 110},clip]
{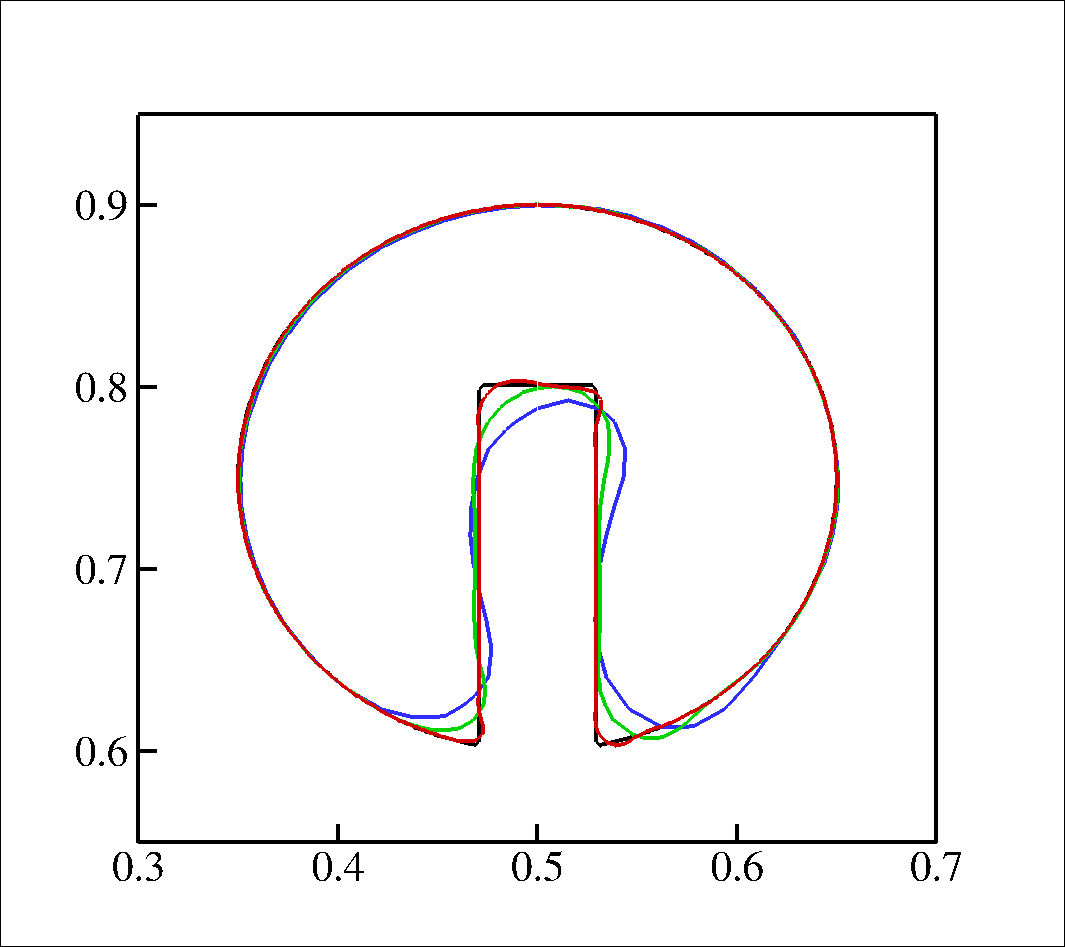}};
\end{tikzpicture}
\put(-206,92){{$y$}}
\put(-100.5,-2){{$x$}}
\caption{Rigid slotted disk shape. {\color{black}\fullthick} $512\times512$ ($t=0$); {\color{blue}\fullthick} $64\times 64$ ($t=1$); {\color{green}\fullthick} $128\times 128$ ($t=1$); {\color{red}\fullthick} $512\times 512$ ($t=1$).\\\hspace{\textwidth}}
\label{fig:Zalesak_shape}
\end{figure}

The performance of the VOF/PLIC method improves with grid resolution. Figure \ref{fig:Zalesak_shape} compares the disk shape ($\phis=0.5$) after one cycle ($t = 1$) for different grid sizes ($N_x \times N_y$, where $N_x$ and $N_y$ are the number of computational cells in $x$ and $y$ directions, respectively) with the initial shape ($t=0$) on $512 \times 512$ grid. Increasing the grid size decreases the difference between the initial and final shape of the slotted disk, especially near the corners.  

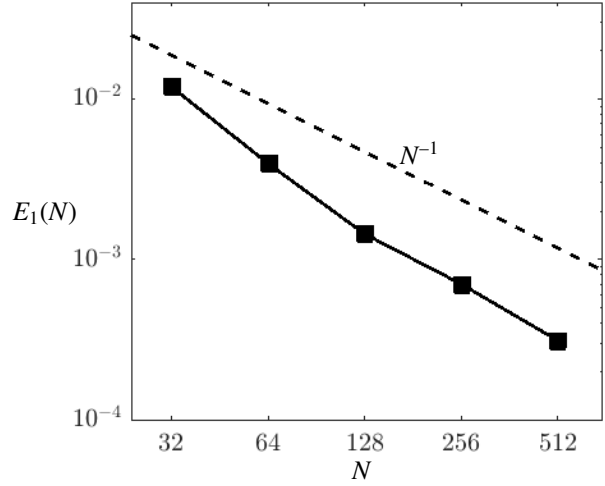
\begin{figure}[h]
\centering
% ----------
\begin{subfigure}[t]{0.45\textwidth}
    \centering
    \begin{tikzpicture}
    \node[anchor=north west] at (0,0) {\includegraphics[width=1\linewidth,trim={0 0 0 0},clip] {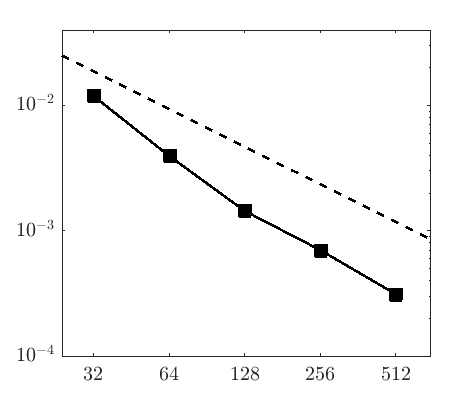}};
    \end{tikzpicture}
    \put(-245,99){{$E_1(N)$}}
    \put(-118,2){{$N$}}
    \put(-100,120){{$N^{-1}$}}
%\caption{}
\label{}
\end{subfigure}
\caption{$L_1$ shape error ($E_1(N)$) for rigid slotted disk.\\\hspace{\textwidth}}
\label{fig:Zalesak_error_OOC}
\end{figure}

We compute $L_1$ shape error ($E_1$) to quantify the accuracy of the VOF/PLIC method,
\begin{equation}
    E_1(N=N_x) = \frac{1}{N_xN_y}\sum_{i=1}^{N_x}\sum_{j=1}^{N_y} \left| \phis_{i,j}(t=1) - \phis_{i,j}(t=0) \right|.
\end{equation}
For all the grids considered, $N_x = N_y = N$. Figure \ref{fig:Zalesak_error_OOC} shows the variation of $L_1(N)$ with $N$, where the error roughly decreases as $N^{-1}$ for higher grid resolutions.

\subsection{Compliant wall in a lid-driven cavity}

We simulate the motion of an incompressible neo-Hookean compliant wall at the bottom of a lid-driven cavity \cite{zhao2008fixed,wang2010interpolation,dunne2006eulerian}. The cavity size is $L_x \times L_y = 2 \times 2$ and the grid size is $N_x \times N_y = 512 \times 512$. Initially, the system is at rest, and the unstressed compliant wall has a flat interface. The thickness of the compliant wall is $h^s = 0.5$. The solid material properties are $G^s = 0.2$, $\mus = 0.2$, and $\rhos = 1$. The fluid properties are $\muf = 0.2$ and $\rhof = 1$. The cavity is driven by the top wall that moves with a non-uniform velocity distribution,
\begin{equation}
        V_w(x) = 0.5
    \begin{cases}
        \sin^2(\pi x/0.6),& \text{if } x \in [0,0.3],\\
        1,& \text{if } x \in (0.3,1.7),\\
        \sin^2(\pi (x-2)/0.6),& \text{if } x \in [1.7,2],
    \end{cases}    
\end{equation}
and the other walls are stationary. Similar to Zhao et al. \cite{zhao2008fixed}, the momentum convection is ignored. 

\begin{figure}[h]
\centering
% ----------
\begin{subfigure}[t]{0.40\textwidth}
\centering
\begin{tikzpicture}
    \node[anchor=north west] at (0,0) {\includegraphics[width=1\linewidth,trim={60 60 100 100},clip]
{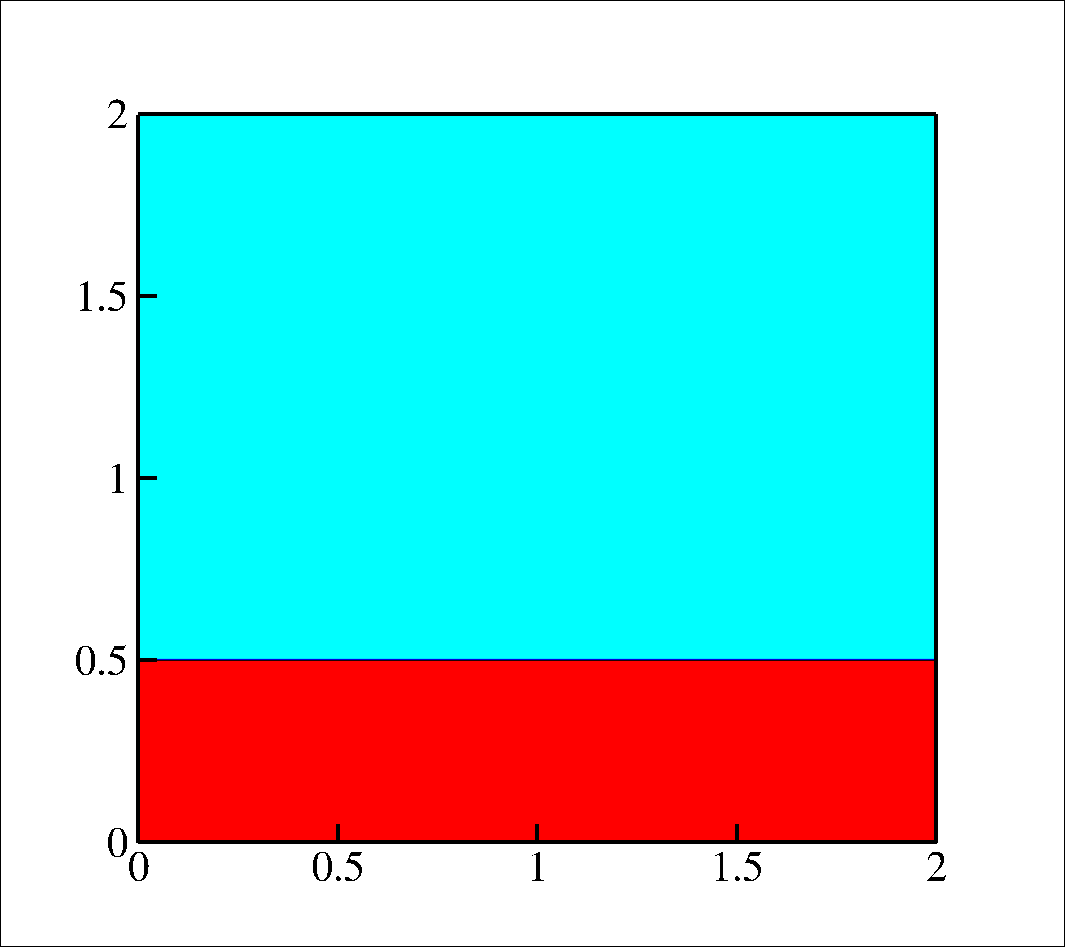}};
\end{tikzpicture}
\put(-101,-2){{$x$}}
\put(-200,92){{$y$}}
\caption{}
\label{subfig:LDC_compwall_t0}
\end{subfigure}
% ----------
\begin{subfigure}[t]{0.40\textwidth}
\centering
\begin{tikzpicture}
    \node[anchor=north west] at (0,0) {\includegraphics[width=1\linewidth,trim={60 60 100 100},clip]
    {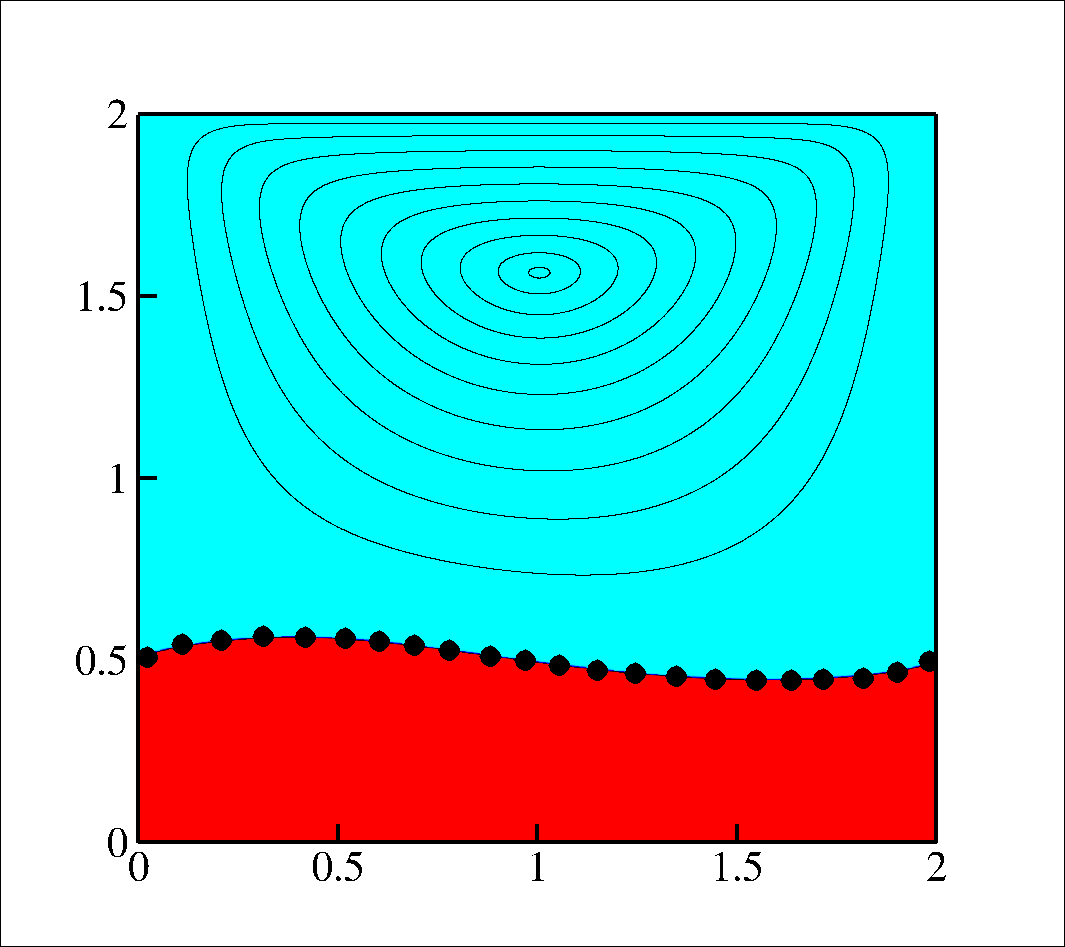}};
    \node[anchor=north west,rotate=90] at (7.0,-3.7) {\includegraphics[width=0.2\textwidth,
        height=0.07\textwidth,trim={0 0 0 0},clip]
    {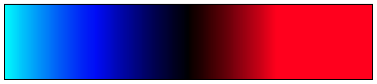}};
\end{tikzpicture}
\put(-12,124){{$1$}}
\put(-12,70){{$0$}}
%\put(-12,80){{$\phis$}}
\put(-13.5,134){{$\phis$}}
\put(-112,-2){{$x$}}
%\put(-188,85){{$y$}}
\caption{}
\label{subfig:LDC_compwall_t8}
\end{subfigure}
\caption{Deformable compliant wall in a lid-driven cavity at (a) $t=0$ and (b) steady state ($t=8$). Filled contours show $\phis$. In (b), thin black lines represent streamlines, and black circles ({\color{black}{\circle}}) indicate deformed compliant surface as reported by Zhao et al. \cite{zhao2008fixed}\\\hspace{\textwidth}}
\label{fig:VAL_compwall}
\end{figure}
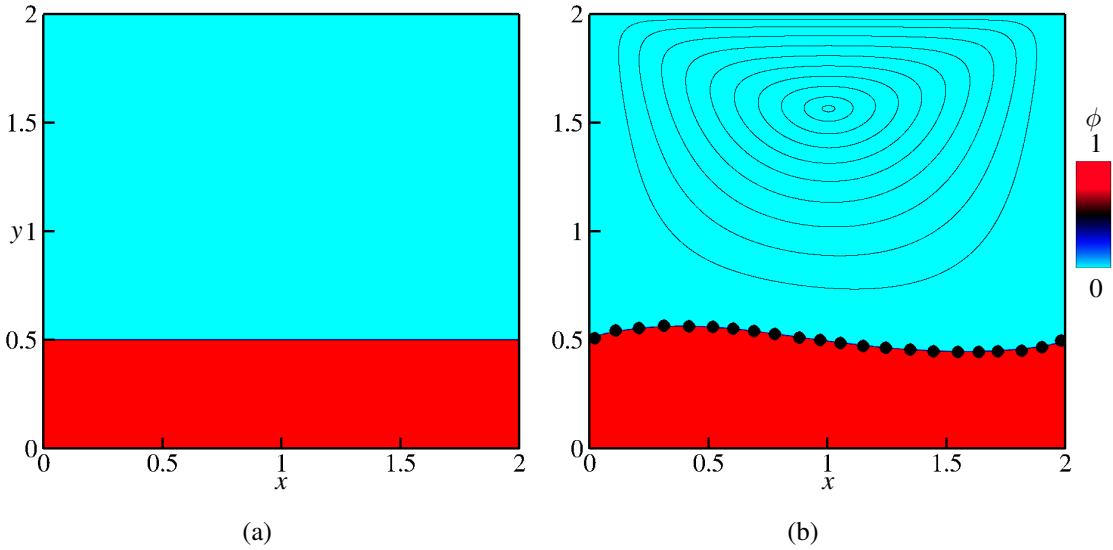

The VOF/PLIC method ensures that the fluid-solid interface remains sharp throughout the time history, and our compliant wall deformation obtained on an Eulerian grid is in excellent agreement with that of Zhao et al. \cite{zhao2008fixed} obtained using a Lagrangian solid mesh. Figure \ref{fig:VAL_compwall} shows $\phis$ distribution inside the lid-driven cavity at $t=0$ and steady state ($t=8$). The fluid-solid interface at steady state is as sharp as at $t=0$. At steady state, the interface shape agrees well with the result of Zhao et al. \cite{zhao2008fixed}, and the streamlines are centered along the x-direction, similar to past works\cite{zhao2008fixed,wang2010interpolation,esmailzadeh2014numerical}.

%------------------------------------------------------------------
%------------------------------------------------------------------

\subsection{Disk in a lid-driven cavity}

We simulate the motion of an incompressible neo-Hookean solid disk in a lid-driven cavity \cite{zhao2008fixed,sugiyama2011full}. The cavity dimensions are $L_x \times L_y = 1 \times 1$ and the grid size is $N_x \times N_y = 128 \times 128$. The system is at rest initially. The top wall starts moving with a constant velocity $V_w = 1$ in $x$ direction, and the remaining boundaries of the cavity stay stationary. The unstressed solid has a circular shape centered at $(x_c, y_c) = (0.6,0.5)$ with a radius of $r = 0.2$. The density and viscosity of the fluid and solid domains are the same, i.e., $\rhos = \rhof = 1$ and $\mus = \muf = 0.01$. We consider two disks: soft disk with $G^s = 0.1$ and stiff disk with $G^s = 10$.      

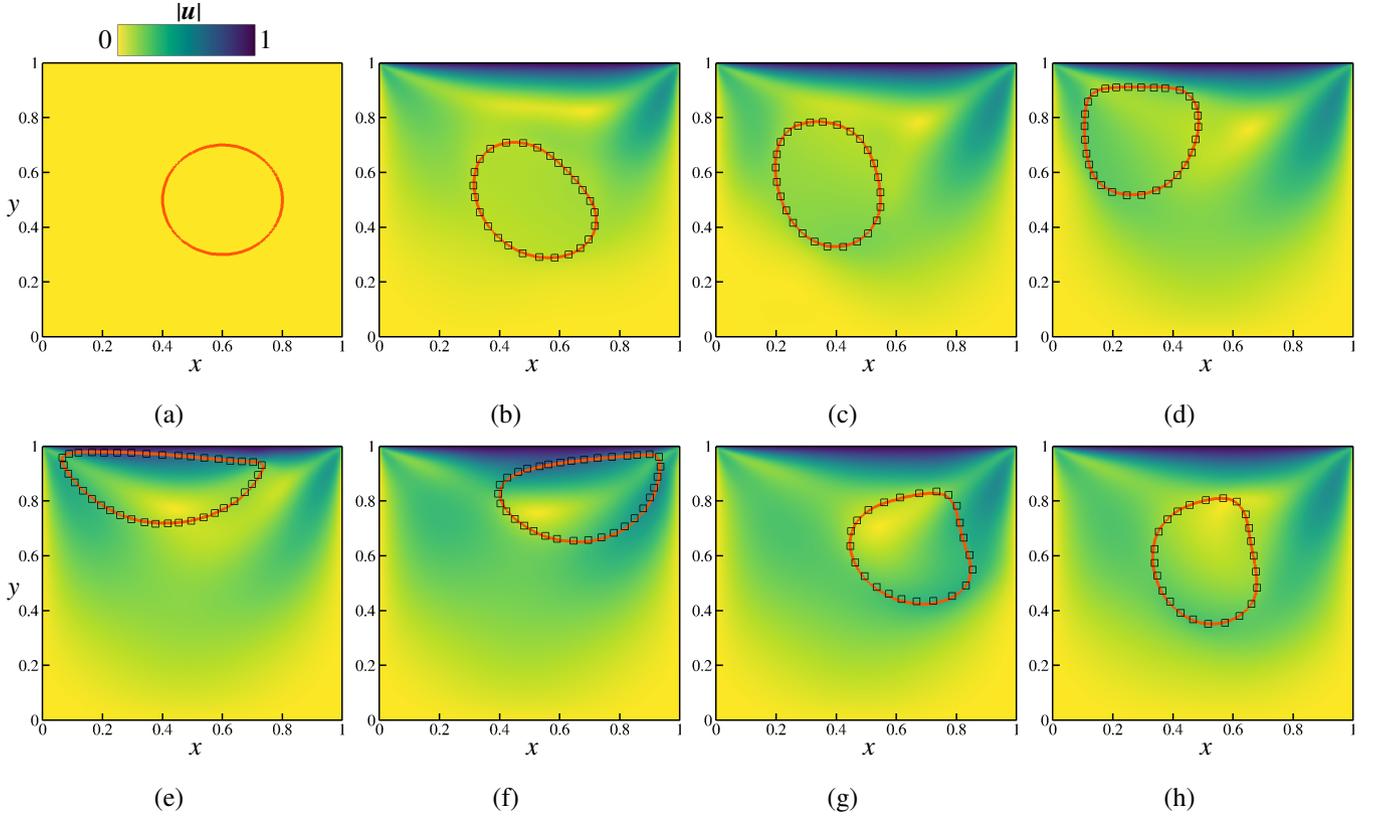
\begin{figure}[h]
\flushleft
% ----------
\begin{subfigure}[t]{0.245\textwidth}
\centering
\begin{tikzpicture}
    \node[anchor=north west] at (0,0) {\includegraphics[width=1\linewidth,trim={60 50 125 100},clip]
{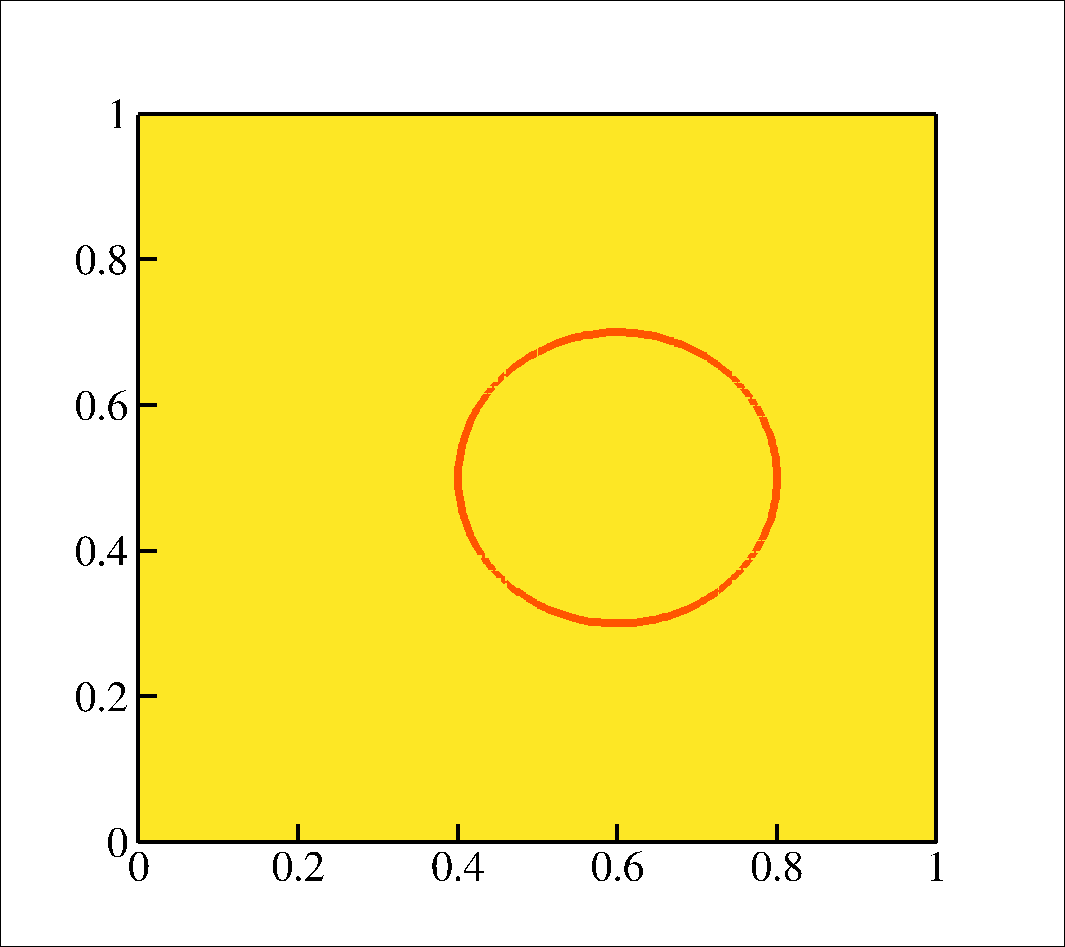}};
    \node[anchor=north west] at (1.36,0.45) {\includegraphics[height=0.1\linewidth,trim={0 0 0 0},clip]
{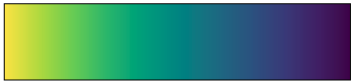}};
\end{tikzpicture}
\put(-129,58){{$y$}}
\put(-95,119){{$0$}}
\put(-35,119){{$1$}}
\put(-66,130){{$\left| \bm{u} \right|$}}
\put(-61,-1.75){{$x$}}
\caption{}
\label{subfig:VAL_disk_in_cavity_a}
\end{subfigure}
% ----------
\begin{subfigure}[t]{0.245\textwidth}
\centering
\begin{tikzpicture}
    \node[anchor=north west] at (0,0) {\includegraphics[width=1\linewidth,trim={60 50 125 100},clip]
{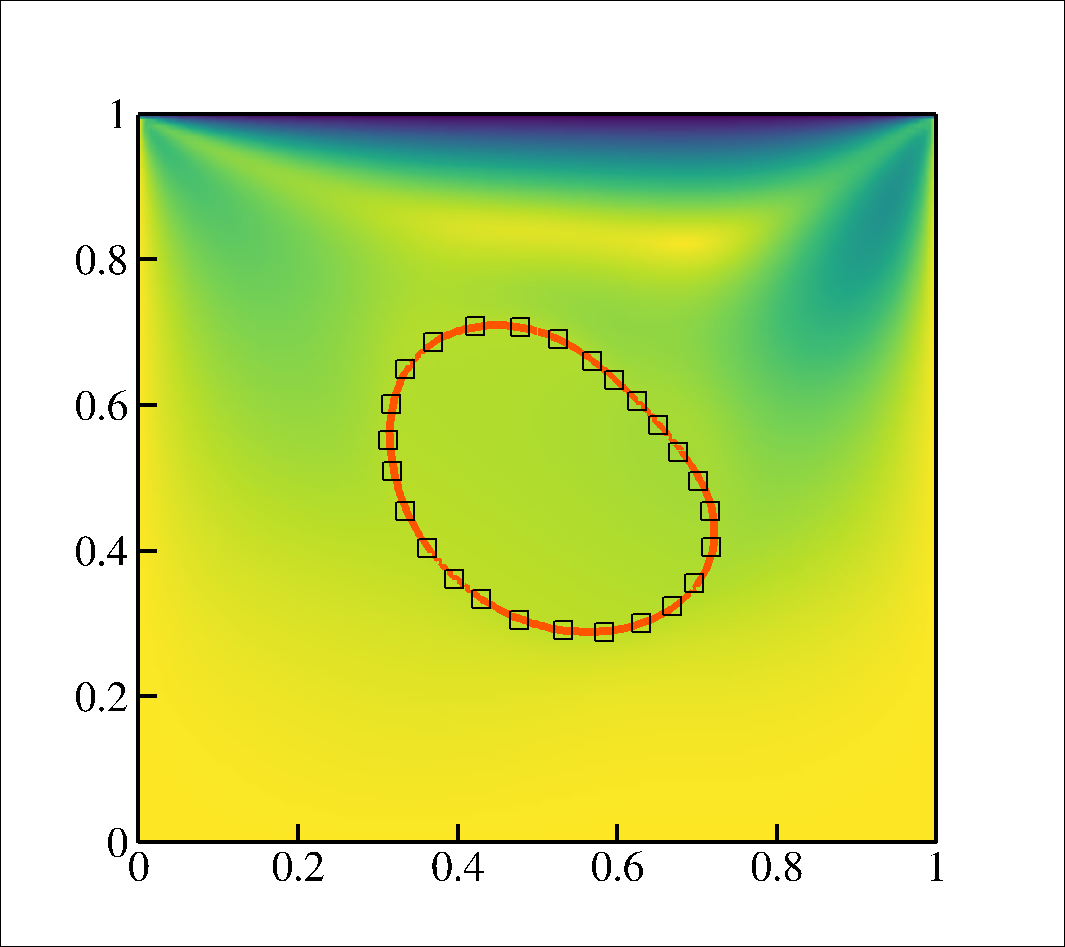}};
\end{tikzpicture}
\put(-61,-1.75){{$x$}}
\caption{}
\label{subfig:VAL_disk_in_cavity_b}
\end{subfigure}
% ----------
\begin{subfigure}[t]{0.245\textwidth}
\centering
\begin{tikzpicture}
    \node[anchor=north west] at (0,0) {\includegraphics[width=1\linewidth,trim={60 50 125 100},clip]
{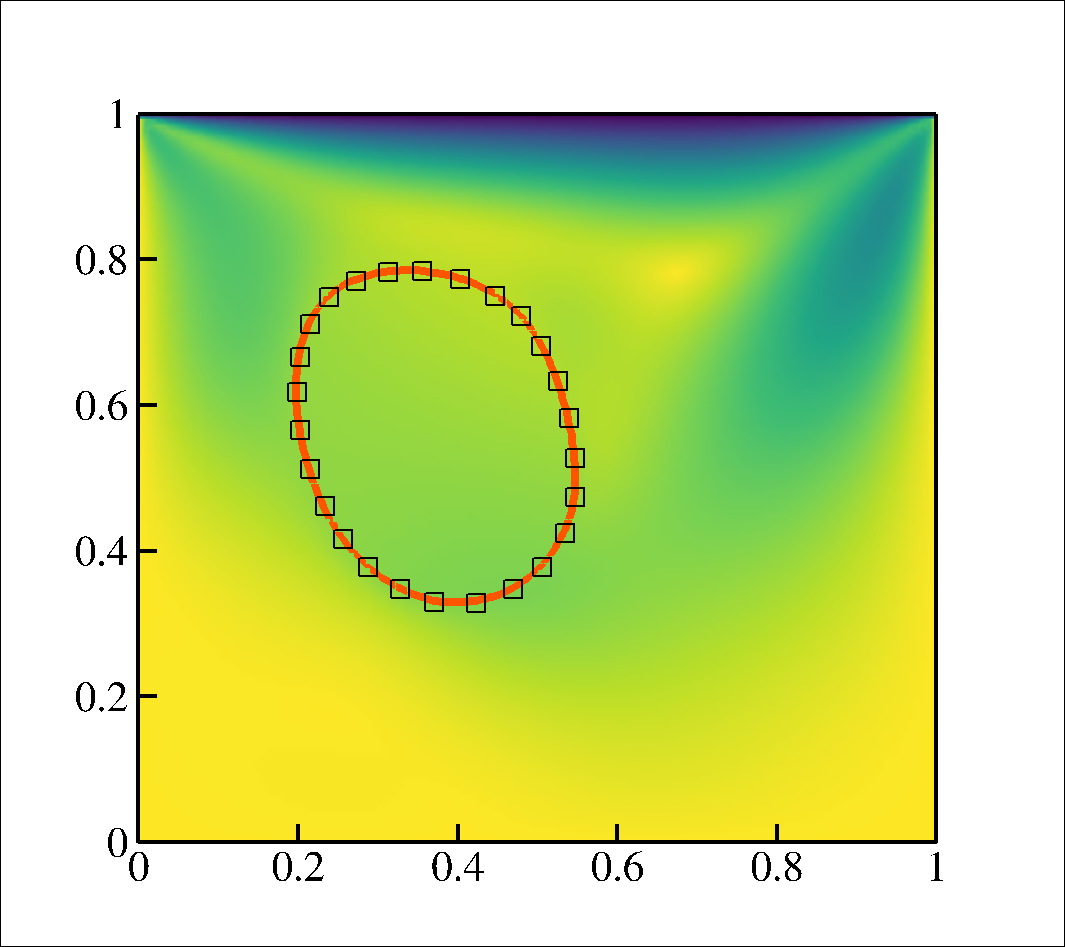}};
\end{tikzpicture}
\put(-61,-1.75){{$x$}}
\caption{}
\label{subfig:VAL_disk_in_cavity_c}
\end{subfigure}
% ----------
\begin{subfigure}[t]{0.245\textwidth}
\centering
\begin{tikzpicture}
    \node[anchor=north west] at (0,0) {\includegraphics[width=1\linewidth,trim={60 50 125 100},clip]
{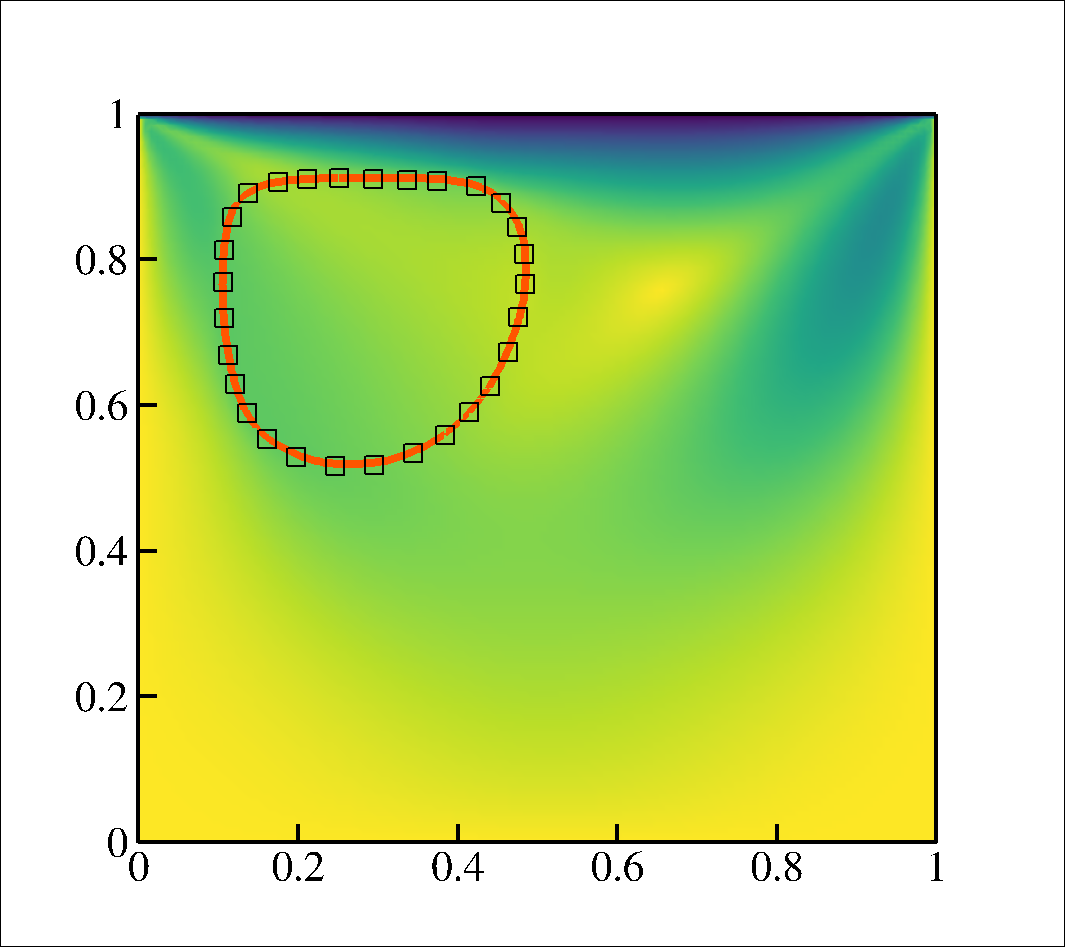}};
\end{tikzpicture}
\put(-61,-1.75){{$x$}}
% \put(-165,67.5){{$y$}}
% \put(-87,-2){{$x$}}
\caption{}
\label{subfig:VAL_disk_in_cavity_d}
\end{subfigure}
% ----------
\\
\begin{subfigure}[t]{0.245\textwidth}
\centering
\begin{tikzpicture}
    \node[anchor=north west] at (0,0) {\includegraphics[width=1\linewidth,trim={60 50 125 100},clip]
{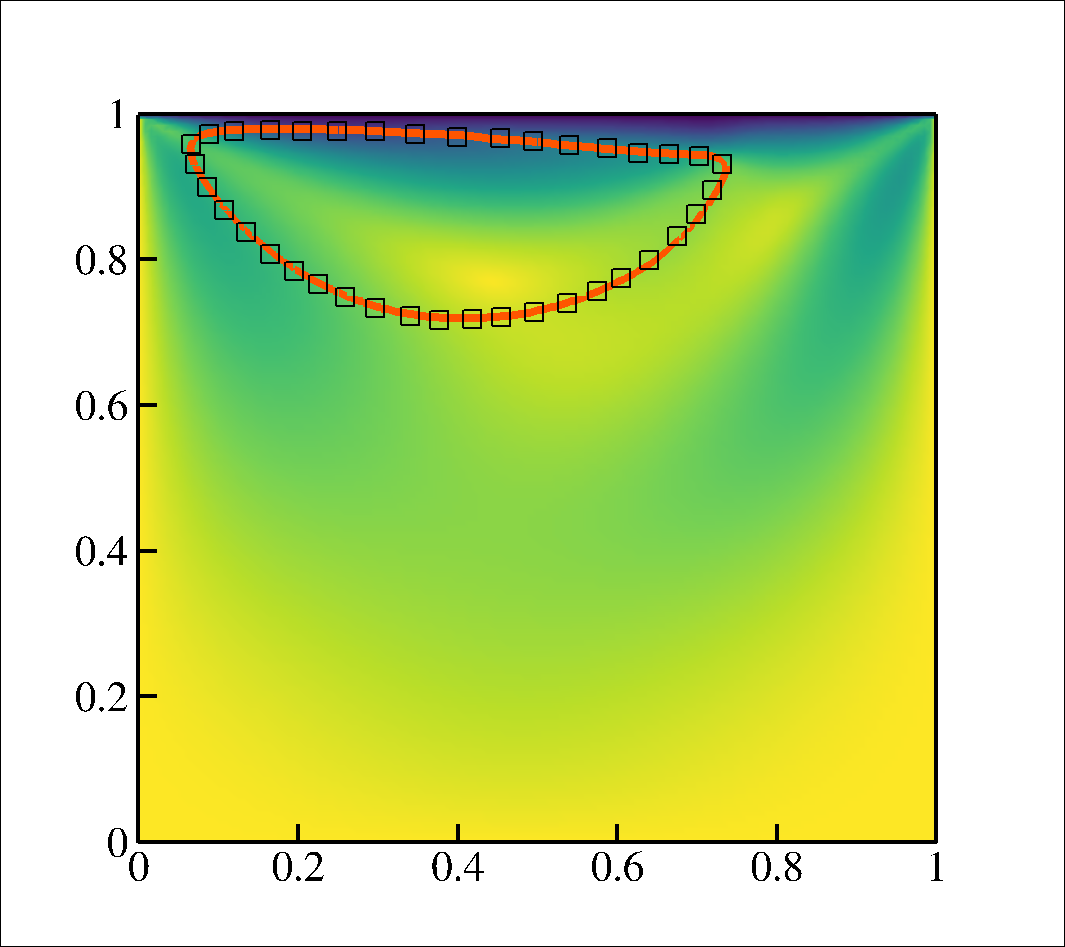}};
\end{tikzpicture}
\put(-129,58){{$y$}}
\put(-61,-1.75){{$x$}}
\caption{}
\label{subfig:VAL_disk_in_cavity_e}
\end{subfigure}
% ----------
\begin{subfigure}[t]{0.245\textwidth}
\centering
\begin{tikzpicture}
    \node[anchor=north west] at (0,0) {\includegraphics[width=1\linewidth,trim={60 50 125 100},clip]
{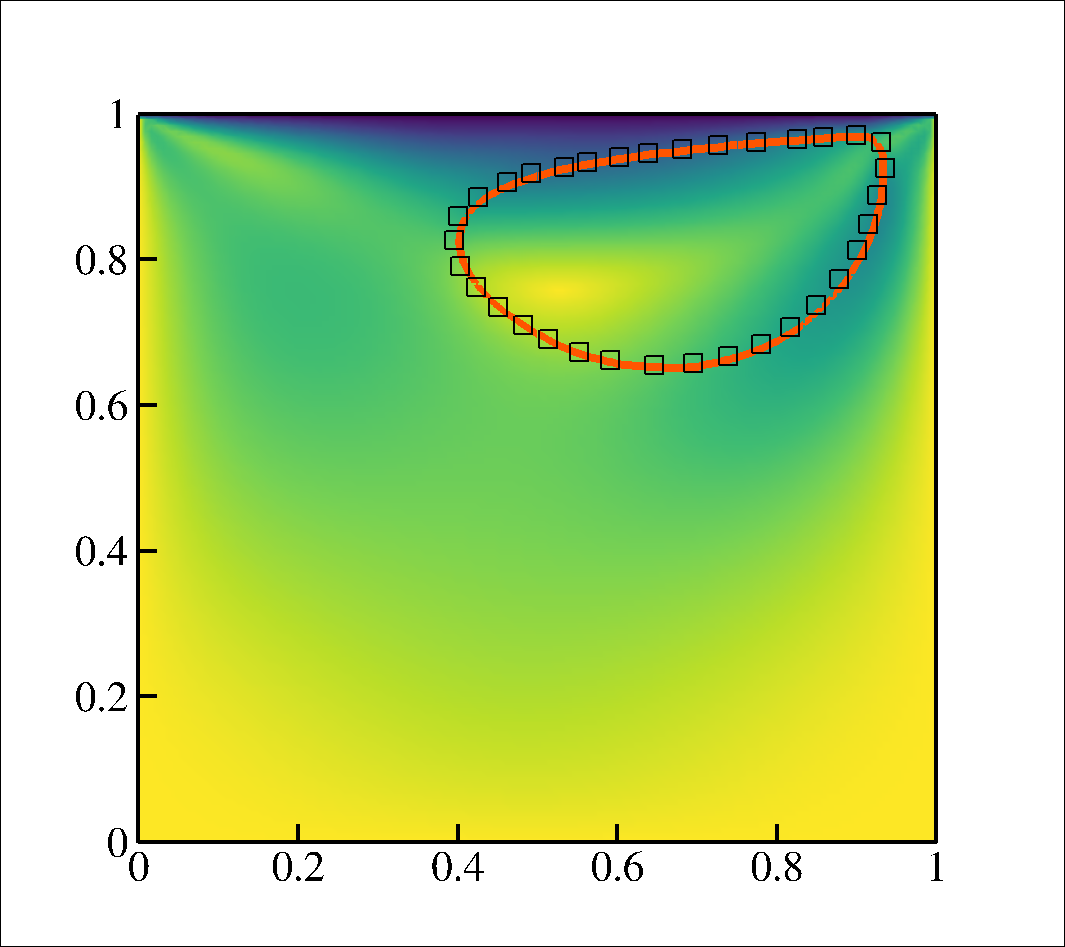}};
\end{tikzpicture}
\put(-61,-1.75){{$x$}}
\caption{}
\label{subfig:VAL_disk_in_cavity_f}
\end{subfigure}
% ----------
\begin{subfigure}[t]{0.245\textwidth}
\centering
\begin{tikzpicture}
    \node[anchor=north west] at (0,0) {\includegraphics[width=1\linewidth,trim={60 50 125 100},clip]
{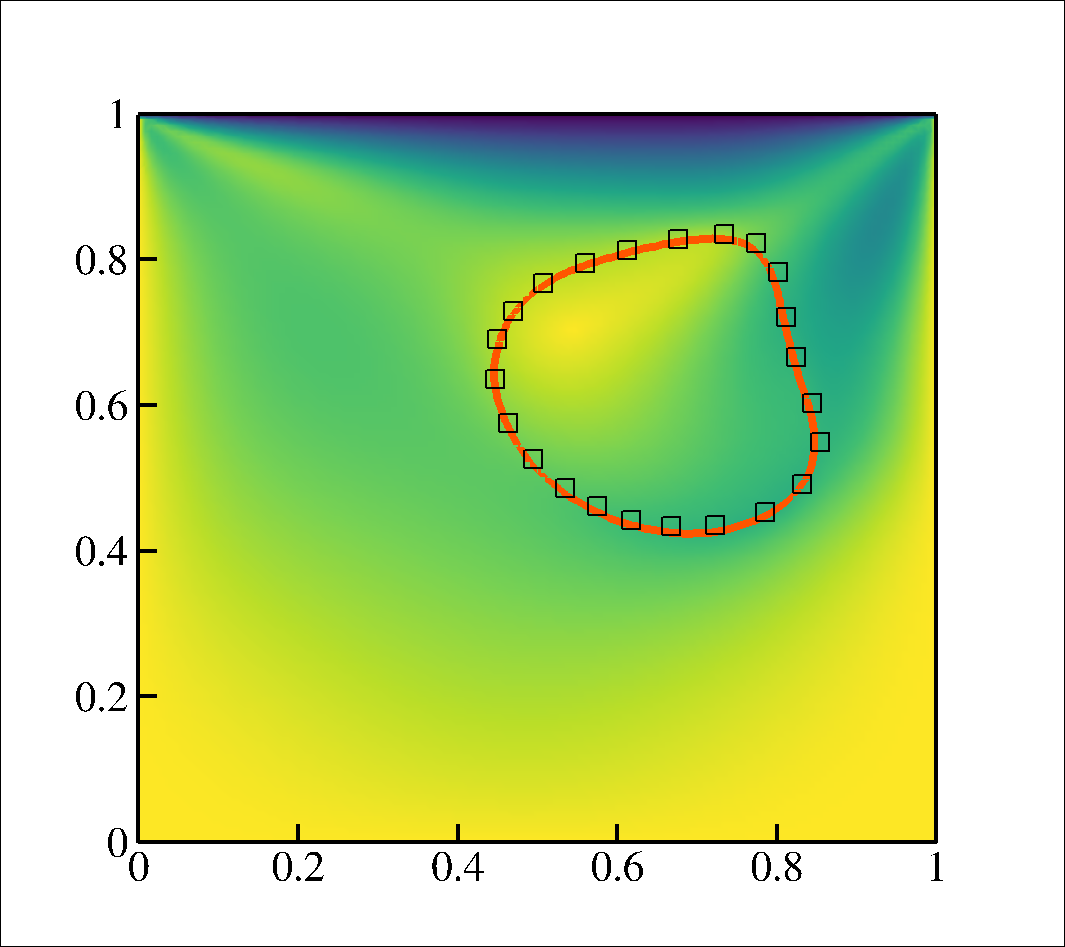}};
\end{tikzpicture}
\put(-61,-1.75){{$x$}}
\caption{}
\label{subfig:VAL_disk_in_cavity_g}
\end{subfigure}
% ----------
\begin{subfigure}[t]{0.245\textwidth}
\centering
\begin{tikzpicture}
    \node[anchor=north west] at (0,0) {\includegraphics[width=1\linewidth,trim={60 50 125 100},clip]
{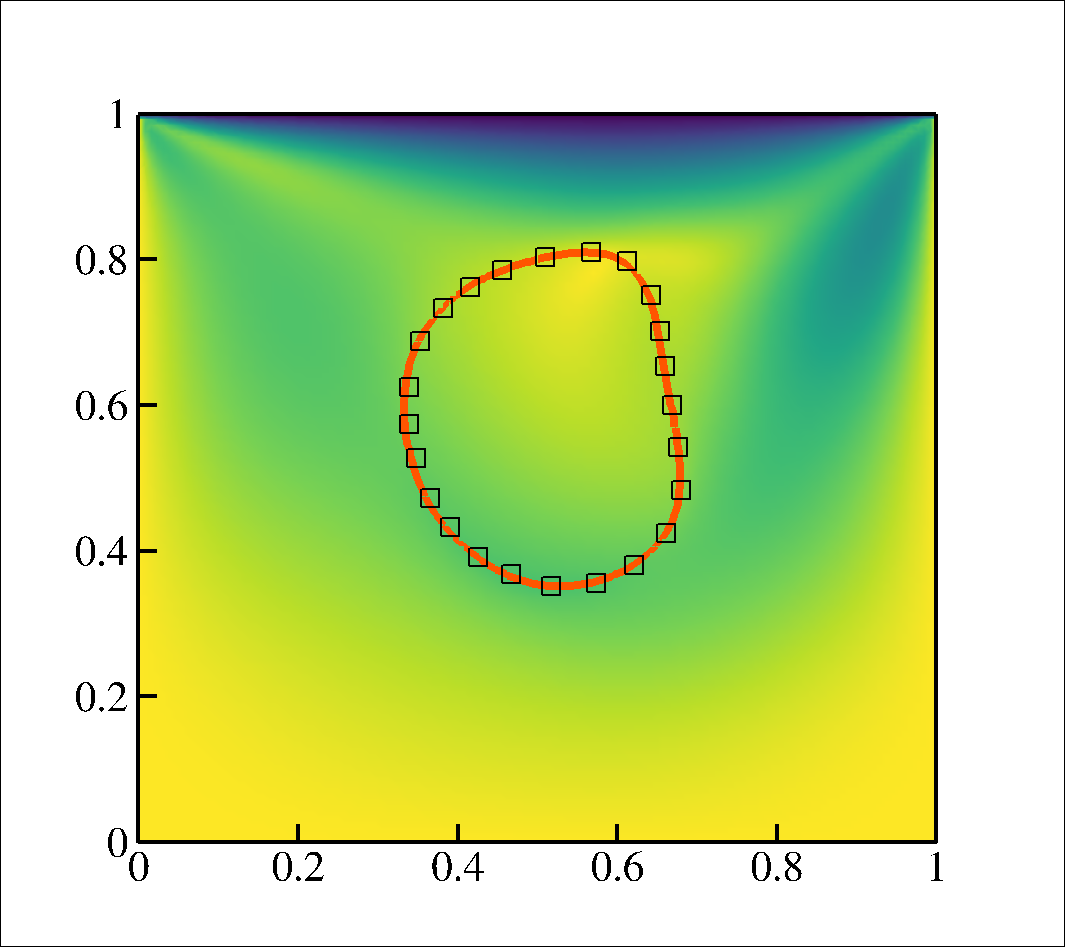}};
\end{tikzpicture}
\put(-61,-1.75){{$x$}}
\caption{}
\label{subfig:VAL_disk_in_cavity_h}
\end{subfigure}
\caption{Soft solid ($G^s=0.1$) in a lid-driven cavity at different time instants. (a-h) $t = 0, 1.17, 2.34, 3.52, 4.69, 5.86, 7.03, 8.20$; {\color{red}\fullthick} VOF/PLIC (solid shape on $128 \times 128$ grid); {{\color{black}{$\Box$}}} Sugiyama et al.\cite{sugiyama2011full} (solid shape on $1024 \times 1024$ grid). The filled contours show flow speed.\\\hspace{\textwidth}}
\label{fig:VAL_LDC_disk_G0p1}
\end{figure}

\begin{figure}[h]
\flushleft
% ----------
\begin{subfigure}[t]{0.245\textwidth}
\centering
\begin{tikzpicture}
    \node[anchor=north west] at (0,0) {\includegraphics[width=1\linewidth,trim={60 50 125 100},clip]
{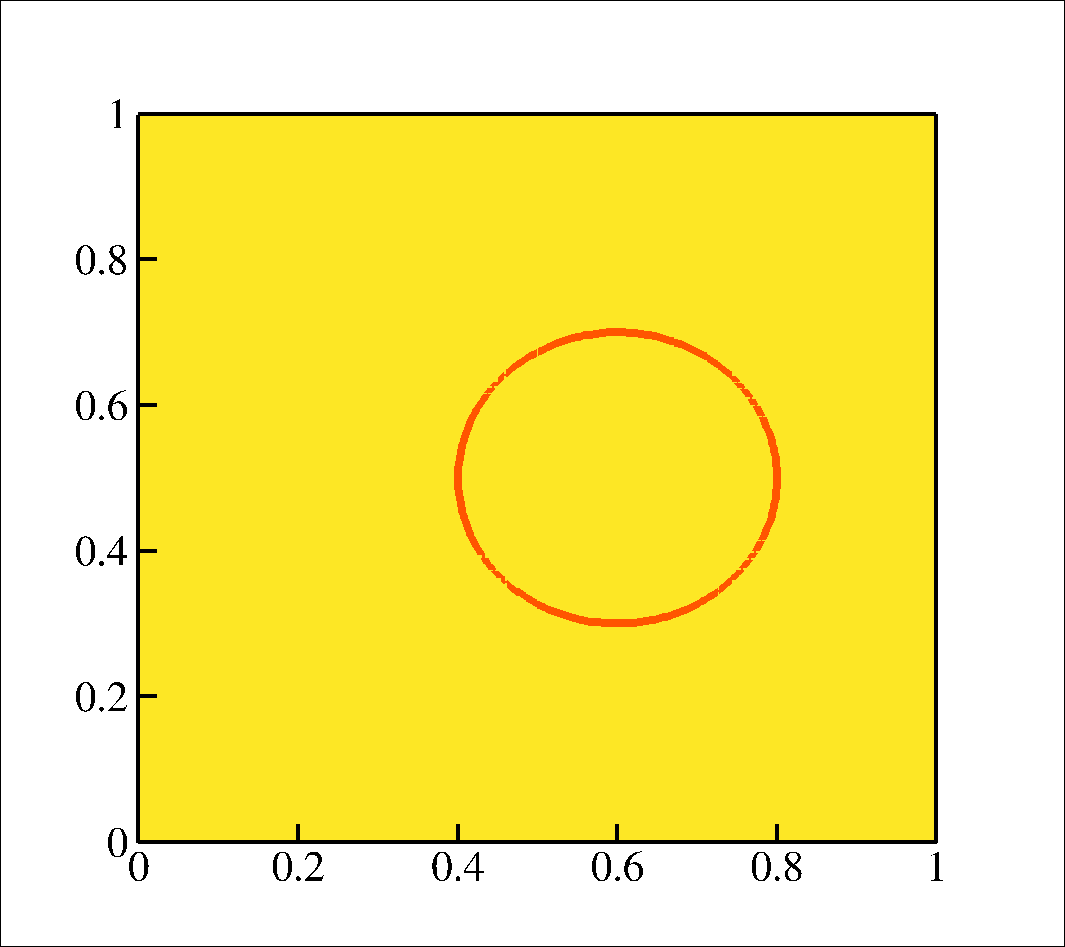}};
    \node[anchor=north west] at (1.36,0.45) {\includegraphics[height=0.1\linewidth,trim={0 0 0 0},clip]
{figures/VOF_method_validations/LDC_disk/Colormap_Sequential_Viridis.png}};
\end{tikzpicture}
\put(-61,-1.75){{$x$}}
\put(-129,58){{$y$}}
\put(-95,119){{$0$}}
\put(-35,119){{$1$}}
\put(-66,130){{$\left| \bm{u} \right|$}}
\caption{}
\label{subfig:VAL_disk_in_cavity_a}
\end{subfigure}
% ----------
\begin{subfigure}[t]{0.245\textwidth}
\centering
\begin{tikzpicture}
    \node[anchor=north west] at (0,0) {\includegraphics[width=1\linewidth,trim={60 50 125 100},clip]
{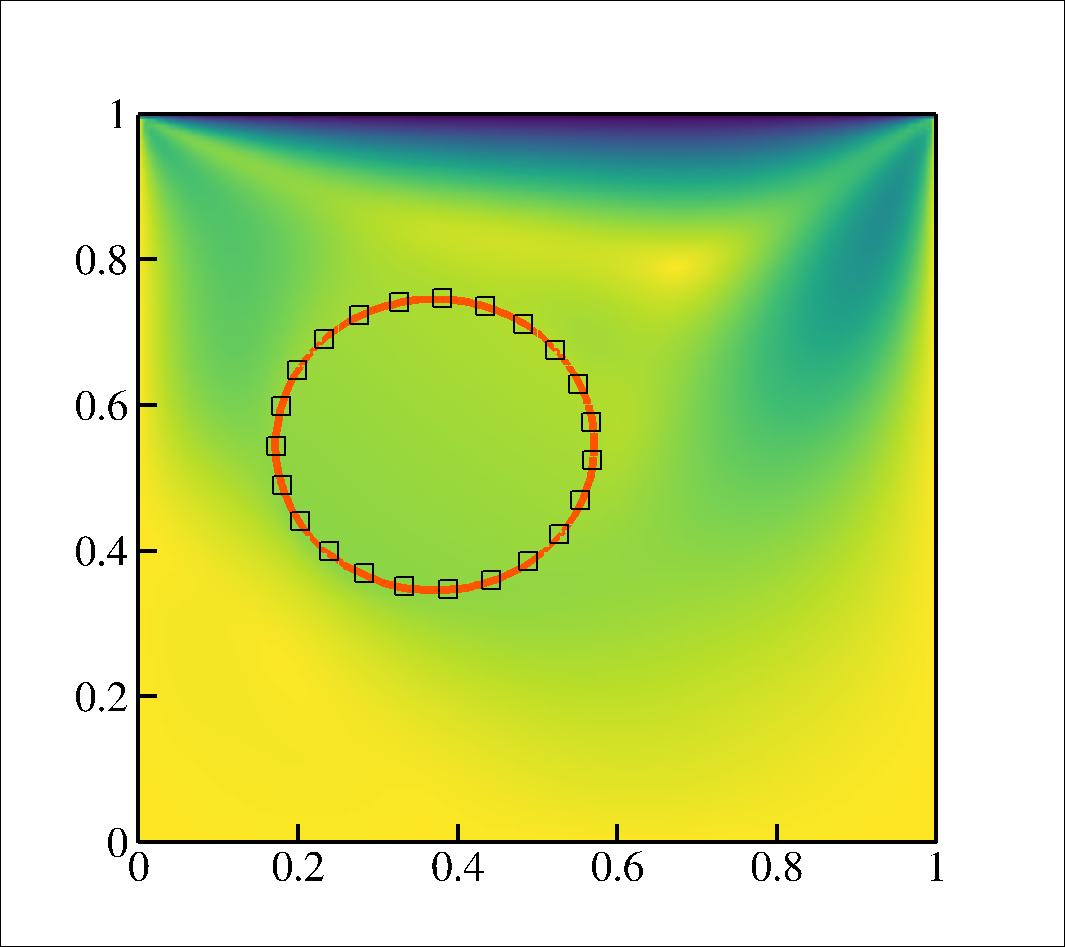}};
\end{tikzpicture}
\put(-61,-1.75){{$x$}}
\caption{}
\label{subfig:VAL_disk_in_cavity_c}
\end{subfigure}
% ----------
\begin{subfigure}[t]{0.245\textwidth}
\centering
\begin{tikzpicture}
    \node[anchor=north west] at (0,0) {\includegraphics[width=1\linewidth,trim={60 50 125 100},clip]
{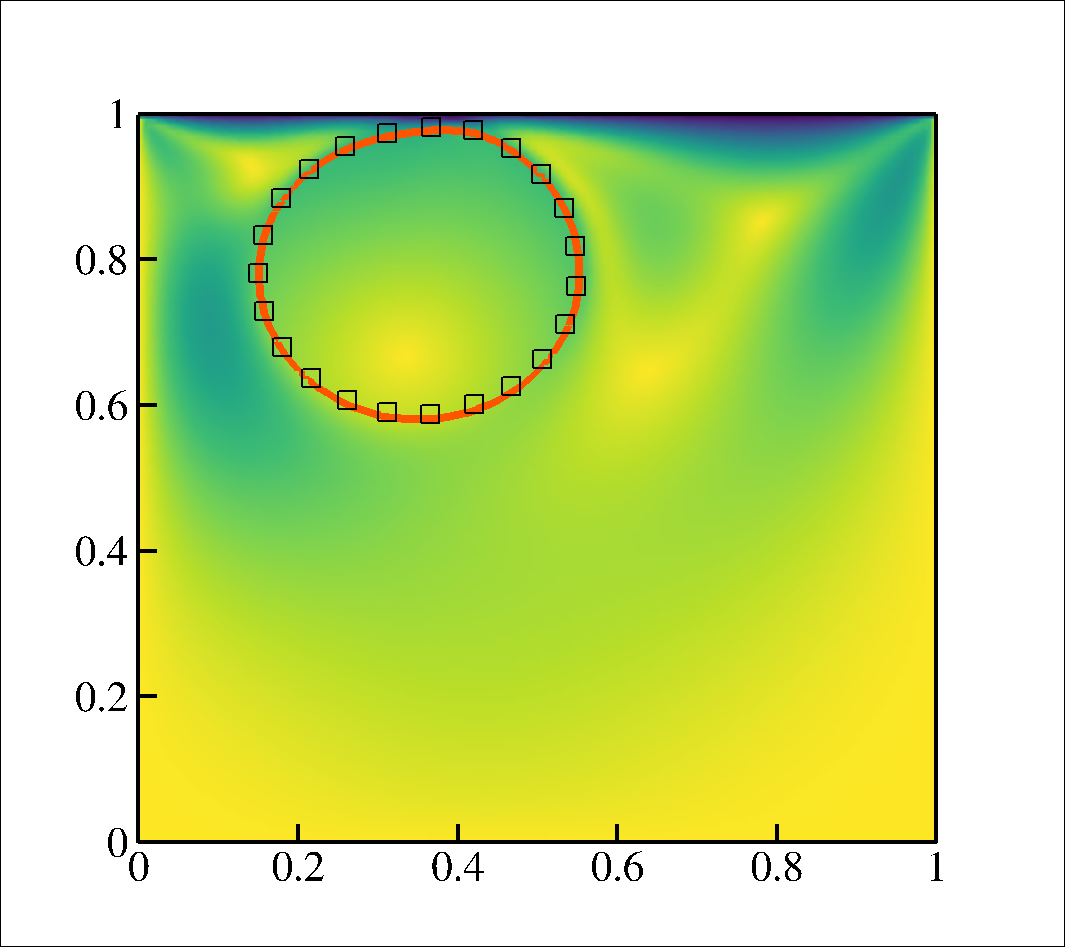}};
\end{tikzpicture}
\put(-61,-1.75){{$x$}}
%\put(-87,-2){{$x$}}
\caption{}
\label{subfig:VAL_disk_in_cavity_e}
\end{subfigure}
% ----------
\begin{subfigure}[t]{0.245\textwidth}
\centering
\begin{tikzpicture}
    \node[anchor=north west] at (0,0) {\includegraphics[width=1\linewidth,trim={60 50 125 100},clip]
{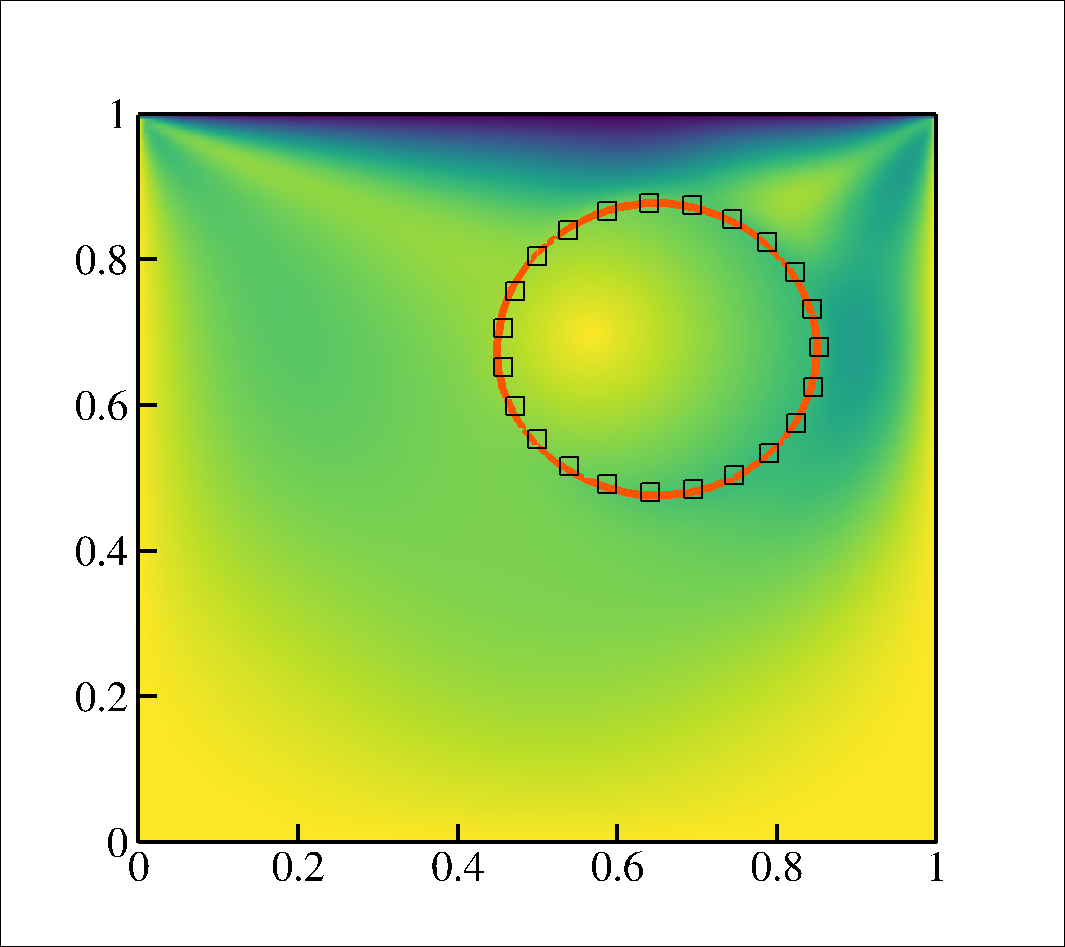}};
\end{tikzpicture}
\put(-61,-1.75){{$x$}}
%\put(-165,67.5){{$y$}}
%\put(-87,-2){{$x$}}
\caption{}
\label{subfig:VAL_disk_in_cavity_g}
\end{subfigure}
\caption{Stiff solid ($G^s=10$) in a lid-driven cavity at different time instants. (a-d) $t = 0, 2.34, 4.69, 7.03$; {\color{red}\fullthick} VOF/PLIC (solid shape); {{\color{black}{$\Box$}}} Zhao et al. \cite{zhao2008fixed} (solid shape). The filled contours show flow speed.\\\hspace{\textwidth}}
\label{fig:VAL_LDC_disk_G10}
\end{figure}

At all time instants, the deformed solid shapes are in excellent agreement with the references. Figure \ref{fig:VAL_LDC_disk_G0p1} shows the soft disk shape and cavity flow at different time instants obtained using the VOF/PLIC-based FSI framework (referred to as VOF/PLIC in figures for brevity), along with results of Sugiyama et al. \cite{sugiyama2011full} obtained on $1024 \times 1024$ Eulerian grid. The shape and location of the soft solid agree well with the reference. Figure \ref{fig:VAL_LDC_disk_G10} shows the stiff disk shape and cavity flow at different time instants, along with results of Zhao et al. \cite{zhao2008fixed} obtained using a Lagrangian solid mesh. For all instances, the shape and location of the stiff solid also agree well with the reference.

Once the lid starts moving, the solid moves toward the top lid due to the flow circulation, where the soft solid undergoes significant deformation, while the stiff solid behaves like a nearly rigid body. However, due to the lubrication effect \citep{skotheim2005soft} between the solid and the wall, the solid does not touch the top lid.

The VOF/PLIC-based FSI framework does not generate unphysical solid fragments even when the solid undergoes significant deformation or has a high pinching tendency. The soft disk experiences high stretching near the top lid ($t=4.69$ in figure \ref{fig:VAL_LDC_disk_G0p1}) before it slides along the lid. As a result, one of the corners (right) has the highest tendency to undergo pinching which could lead to non-physical solid fragments ejecting from the main body in the following time. The time instants $t=4.69, 5.86, 7.03,$ and $8.20$ in figure \ref{fig:VAL_LDC_disk_G0p1} verify that no such mass fragments are generated.

% ======================================
% ======================================

% ======================================
% ======================================

To further verify the results and show grid convergence, we compute the solid centroid trajectory \cite{sugiyama2011full},
\begin{equation}
    \bm{x}_c(t) = \frac{\sum_{i=1}^{N_x}\sum_{j=1}^{N_y}\bm{x}_{i,j}\phis_{i,j}(t)\Delta_x \Delta_y}{\sum_{i=1}^{N_x}\sum_{j=1}^{N_y}\phis_{i,j}(t)\Delta_x \Delta_y},
\end{equation}
for the soft solid.

\begin{figure}[h]
\centering
% ----------
\begin{tikzpicture}
    \node[anchor=north west] at (0,0) {\includegraphics[width=0.45\linewidth,trim={10 10 10 10},clip]
{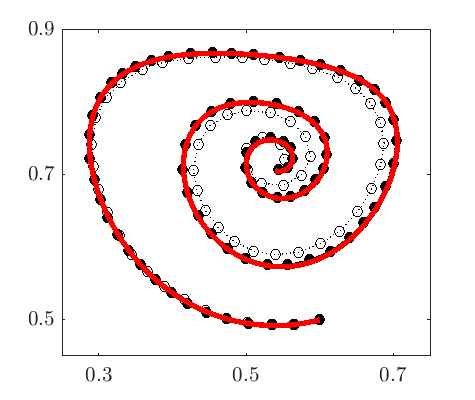}};
\end{tikzpicture}
\put(-230,93){{$y$}}
\put(-114,-2){{$x$}}
%\put(-92,94){{$-1$}}
\caption{ Soft solid centroid trajectory for $t \in [0,20]$. {\color{red}\fullthick} VOF/PLIC ($128 \times 128$); {\color{black}{\circle}} Sugiyama et al. \cite{sugiyama2011full} ($1024 \times 1024$); {\color{black}{\emptycircle}} Sugiyama et al. \cite{sugiyama2011full} ($128 \times 128$).\\\hspace{\textwidth}}
\label{fig:VAL_LDC_disk_G0p1_centroid}
\end{figure}

The accuracy of our VOF/PLIC-based FSI approach on coarse grids is comparable to the accuracy of a diffusive IC method-based FSI approach on much finer grids. Figure \ref{fig:VAL_LDC_disk_G0p1_centroid} shows the soft solid centroid trajectory for $t \in [0,20]$. Our centroid trajectory obtained on $128\times128$ Eulerian grid matches well with that of Sugiyama et al. \cite{sugiyama2011full} obtained on $1024 \times 1024$ Eulerian grid using a fifth-order WENO scheme to advect the interface. Furthermore, as discussed above, at all time instances (figure \ref{fig:VAL_LDC_disk_G0p1}), the deformed soft solid shapes on $128\times128$ grid are in excellent agreement with those of Sugiyama et al. \cite{sugiyama2011full} on $1024\times1024$ grid. It reflects the advantage of using the VOF/PLIC method to capture the fluid-solid interface that maintains a sharp interface, whereas WENO diffuses it.     

% ======================================
% ======================================

\begin{figure}[h]
\centering
% ----------
\begin{subfigure}[t]{0.45\textwidth}
    \centering
    \begin{tikzpicture}
    \node[anchor=north west] at (0,0) {\includegraphics[width=1\linewidth,trim={10 10 10 10},clip]
    {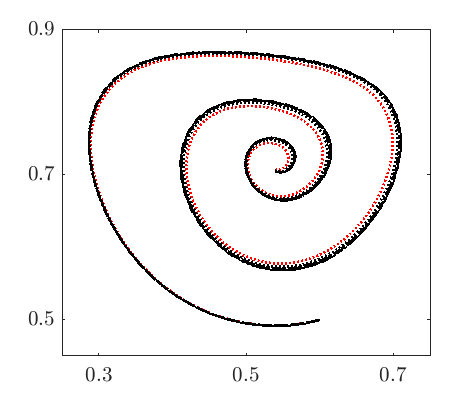}};
    \end{tikzpicture}
    \put(-230,93){{$y$}}
    \put(-114,-2){{$x$}}
    \caption{}
    \label{fig:LDC_disk_G0p1_centroid_grid_conv}
\end{subfigure}
% ----------
\begin{subfigure}[t]{0.45\textwidth}
    \centering
    \begin{tikzpicture}
    \node[anchor=north west] at (0,0) {\includegraphics[width=1\linewidth,trim={10 10 10 10},clip]
    {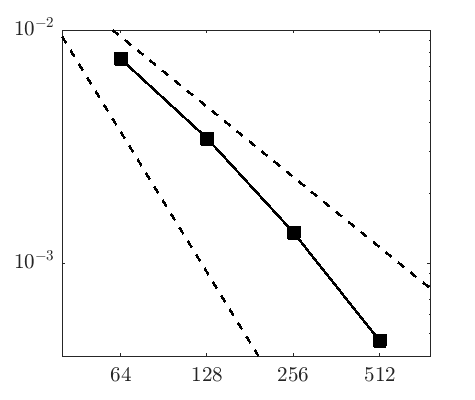}};
    \end{tikzpicture}
    \put(-232,93){{$E_2(N)$}}
    \put(-114,-2){{$N$}}
    \put(-100,120){{$N^{-1}$}}
    \put(-154,100){{$N^{-2}$}}
    \caption{}
    \label{fig:LDC_disk_G0p1_centroid_L2_err}
\end{subfigure}
\caption{ (a) Soft solid centroid trajectory for $t \in [0,20]$ and (b) $L_2$ error ($E_2(N)$) for various grid resolutions. In (a), {{\color{red}{\dottedthick}}} $64 \times 64$; {{\color{black}{\dottedthick}}} $128 \times 128$; {{\color{black}{\dasheddottedthick}}} $256 \times 256$; {{\color{black}{\dashedthick}}} $512 \times 512$; {{\color{black}{\fullthick}}} $1024 \times 1024$.\\\hspace{\textwidth}}
\label{}
\end{figure}

To analyze grid convergence, figure \ref{fig:LDC_disk_G0p1_centroid_grid_conv} shows the soft solid centroid trajectory for various grid sizes ($64 \times 64$, $128 \times 128$, $256 \times 256$, $512 \times 512$, $1024 \times 1024$). For grid sizes greater than $128 \times 128$, there are negligible differences in the centroid trajectory, indicating grid convergence at $128 \times 128$ grid size.

To evaluate the order of convergence, we compute $L_2$ error defined as
\begin{equation}
    E_2(N=N_x) = \left[\frac{1}{T}\int_0^T \left| \bm{x}_c(t,N_x)-\bm{x}_c(t,N_x=1024) \right| ^2\mathrm{d}t \right]^{0.5}.
\end{equation}
Figure \ref{fig:LDC_disk_G0p1_centroid_L2_err} shows $E_2$ monotonically decreases with grid size and the order of convergence is approximately first-order ($N^{-1}$) for smaller grid sizes.   

\begin{figure}[h]
\centering
% ----------
\begin{tikzpicture}
    \node[anchor=north west] at (0,0) {\includegraphics[width=0.45\linewidth,trim={10 10 10 10},clip]
{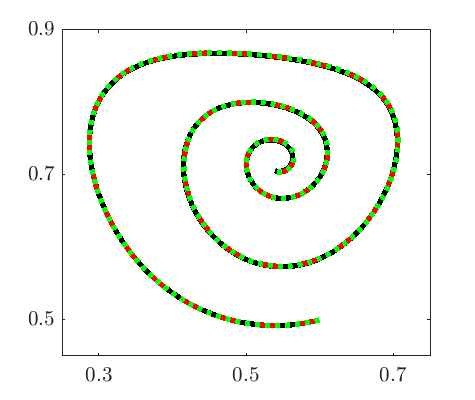}};
\end{tikzpicture}
    \put(-230,93){{$y$}}
    \put(-114,-2){{$x$}}
\caption{Soft solid centroid trajectory for various cell aspect ratios (AR). {\color{black}{\fullthick}} $AR=1$ ($128 \times 128$); {\color{red}{\dashedthick}} $AR=2$ ($256 \times 128$); {\color{green}{\dottedthick}} $AR=4$ ($512 \times 128$).\\\hspace{\textwidth}}
\label{fig:VAL_LDC_disk_G0p1_centroid_ARs}
\end{figure}

The generic implementation of the FSI solver is valid for non-unity cell aspect ratios $AR = \Delta_y/ \Delta_x \ne 1$ as well. To test the validity, we compare the motion of soft solid on $128 \times 128$, $256 \times 128$, and $512 \times 128$ grids that have cell aspect ratios $AR =$ $1$, $2$, and $4$, respectively. Figure \ref{fig:VAL_LDC_disk_G0p1_centroid_ARs} compares the soft solid centroid trajectory for all the grids. For $AR = $ $2$ and $4$, the centroid trajectory agrees well with the validated $AR = 1$ centroid trajectory.     

% ======================================
% ======================================

% ======================================
% ======================================

\subsection{Reversibility test of a circular disk in shear flow}

After releasing the stresses exerted on the hyperelastic solids, they generally return to their initial configuration \citep{sugiyama2011full}. The reversibility test examines this behavior by applying stress to the solid and releasing it. When describing solids on a Lagrangian grid, such reversibility is readily satisfied as the grid tracks the movement of each node relative to their original configuration. However, reversibility is not trivial on a fixed Eulerian grid because the solid deformation is not explicitly traced through mesh motion. While using a VOF/PLIC procedure that exhibits discontinuities across the interface and cell boundaries, verifying the reversibility and stability is of utmost necessity. Otherwise, the strong spurious currents generated due to the stress imbalance might produce an oscillatory interface at equilibrium, resulting in inconsistencies between the initial and final configurations, and an unstable interface.    

To perform a reversibility test and assess the accuracy of the solver while handling nonlinear hyperelastic materials, we consider the evolution of a Saint Venant-Kirchhoff circular disk in shear flow, similar to Ii et al. \cite{ii2011implicit} The computational domain size is $L_x \times L_y = 4 \times 4$ with domain extents $x \in [-L_x/2,L_x/2], y \in [-L_y/2,L_y/2]$, and the grid size is $N_x \times N_y = 256 \times 256$. Initially, the system is at rest and the unstressed circular solid disk is centered at $(x_c,y_c) = (0,0)$ with diameter $d=1$. The top and bottom walls start moving in opposite directions along the $x$ axis with velocities $V_w$ and $-V_w$, respectively. No-slip boundary conditions are imposed on the top and bottom walls, and periodicity is considered in the $x$ direction. The solid is considered to be non-viscous ($\mus = 0$), and the fluid and solid densities are $\rhof = \rhos = 1$.       

The problem parameters are determined according to the required shear rate $\dot{\gamma}=2V_w/L_y$, Reynolds number $Re = \rhof \dot{\gamma} d^2/\muf$, and the capillary number $Ca = \muf\dot{\gamma}/G^s$. We set $\dot{\gamma} = 1$ and $Re = 0.1$ by considering $V_w = 2$ and $\muf = 10$. We consider three values of $Ca$, $Ca = 0.1, 0.3, 0.5$, by varying $G^s = 100, 33.33, 20$.

The simulations are conducted for $t \in [0,20]$, where the walls move in opposite directions for $t \in [0,5]$ and stop moving after $t=5$. For all three values of $Ca$, the disk reaches a steady state by $t=5$, and the time taken to attain a steady state (referred to as the deformed steady state) increases with $Ca$. After the walls stop moving, far before $t = 20$, the disks return to their initial circular shape (referred to as the final steady state).

% =============================================================
\begin{figure}[h]
\centering
% ----------
\begin{subfigure}[t]{0.30\textwidth}
\centering
\begin{tikzpicture}
    \node[anchor=north west] at (0,0) {\includegraphics[height=1\linewidth,trim={60 50 10 10},clip]
{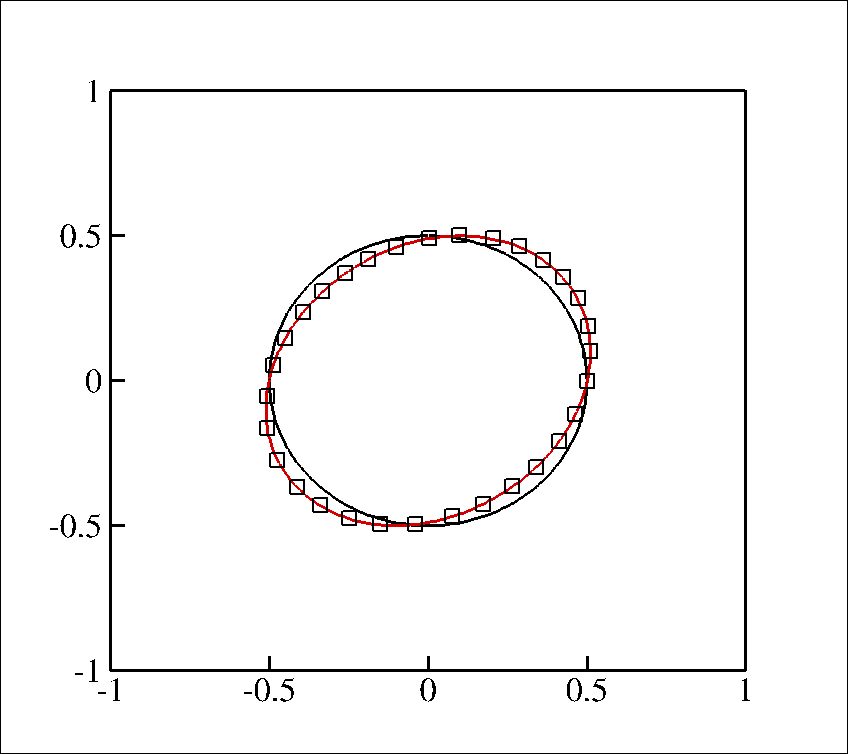}};
\end{tikzpicture}
\put(-175,71){{$y$}}
\put(-95,-2){{$x$}}
\put(-157,125){$Ca=0.1$}
\put(-160,20){{{\color{black}\fullthick}}}
\put(-142,20){$t=0$}
\put(-115,20){{{\color{red}\fullthick}}}
\put(-97,20){$t=5$}
\put(-70,20){{{\color{black}\emptysquare}}}
\put(-61,20){Ii et al.\citep{ii2011implicit}}
\caption{}
\label{subfig:sv_a}
\end{subfigure}
% ----------
\begin{subfigure}[t]{0.30\textwidth}
\centering
\begin{tikzpicture}
    \node[anchor=north west] at (0,0) {\includegraphics[height=1\linewidth,trim={60 50 10 10},clip]
{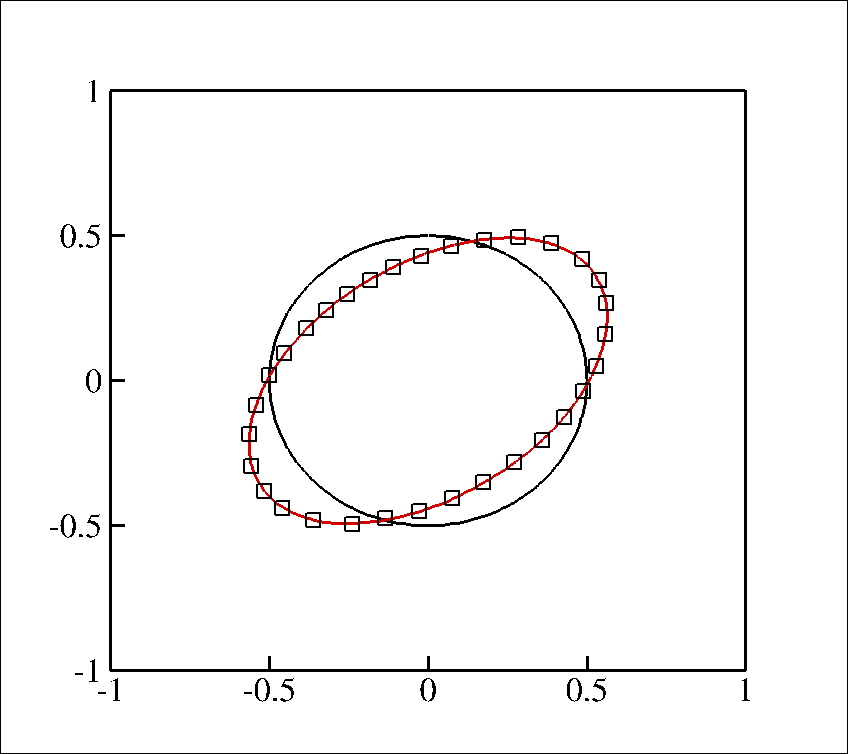}};
\end{tikzpicture}
\put(-95,-2){{$x$}}
\put(-157,125){$Ca=0.3$}
\put(-160,20){{{\color{black}\fullthick}}}
\put(-142,20){$t=0$}
\put(-115,20){{{\color{red}\fullthick}}}
\put(-97,20){$t=5$}
\put(-70,20){{{\color{black}\emptysquare}}}
\put(-61,20){Ii et al.\citep{ii2011implicit}}
\caption{}
\label{subfig:sv_b}
\end{subfigure}
% ----------
\begin{subfigure}[t]{0.30\textwidth}
\centering
\begin{tikzpicture}
    \node[anchor=north west] at (0,0) {\includegraphics[height=1\linewidth,trim={60 50 10 10},clip]
{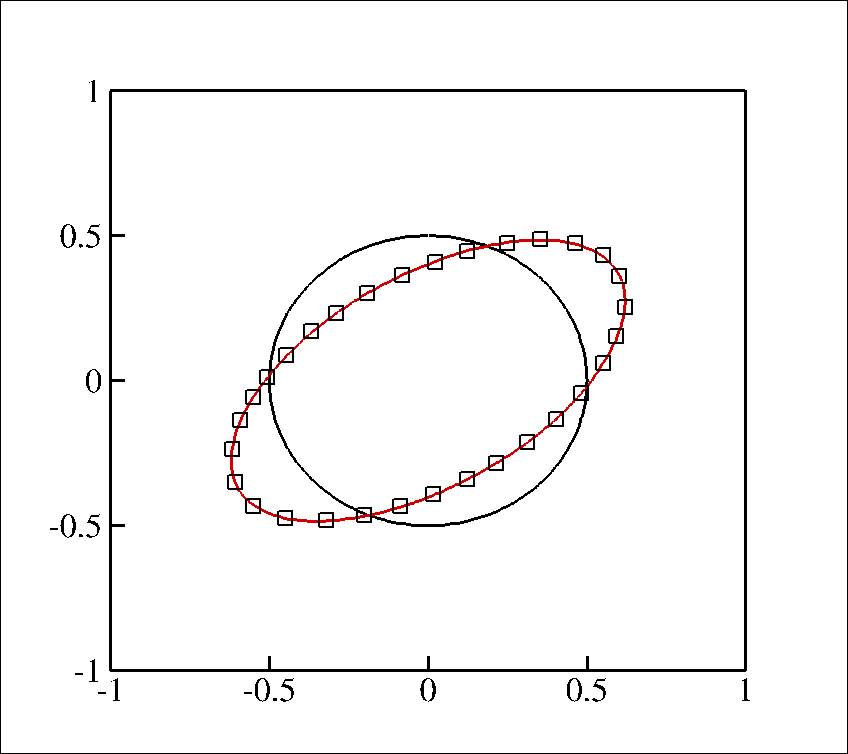}};
\end{tikzpicture}
\put(-95,-2){{$x$}}
\put(-157,125){$Ca=0.5$}
\put(-160,20){{{\color{black}\fullthick}}}
\put(-142,20){$t=0$}
\put(-115,20){{{\color{red}\fullthick}}}
\put(-97,20){$t=5$}
\put(-70,20){{{\color{black}\emptysquare}}}
\put(-61,20){Ii et al.\citep{ii2011implicit}}
\caption{}
\label{subfig:sv_c}
\end{subfigure}
% ----------
\\
% ----------
\begin{subfigure}[t]{0.30\textwidth}
\centering
\begin{tikzpicture}
    \node[anchor=north west] at (0,0) {\includegraphics[height=1\linewidth,trim={60 50 10 10},clip]
{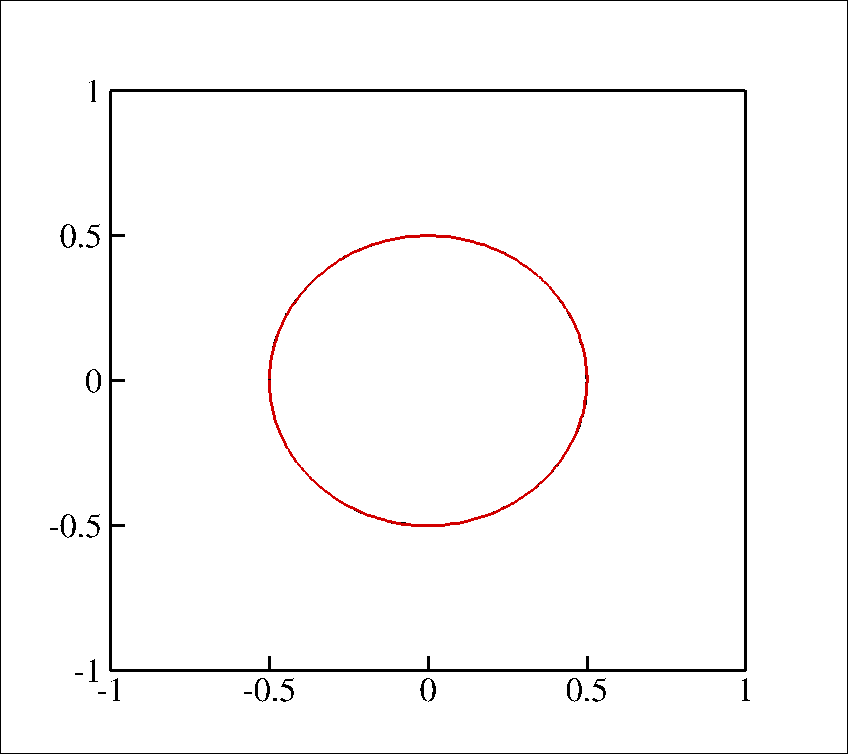}};
\end{tikzpicture}
\put(-175,71){{$y$}}
\put(-95,-2){{$x$}}
\put(-157,125){$Ca=0.1$}
\put(-160,20){{{\color{black}\fullthick}}}
\put(-142,20){$t=0$}
\put(-115,20){{{\color{red}\fullthick}}}
\put(-97,20){$t=20$}
\caption{}
\label{subfig:sv_a}
\end{subfigure}
% ----------
\begin{subfigure}[t]{0.30\textwidth}
\centering
\begin{tikzpicture}
    \node[anchor=north west] at (0,0) {\includegraphics[height=1\linewidth,trim={60 50 10 10},clip]
{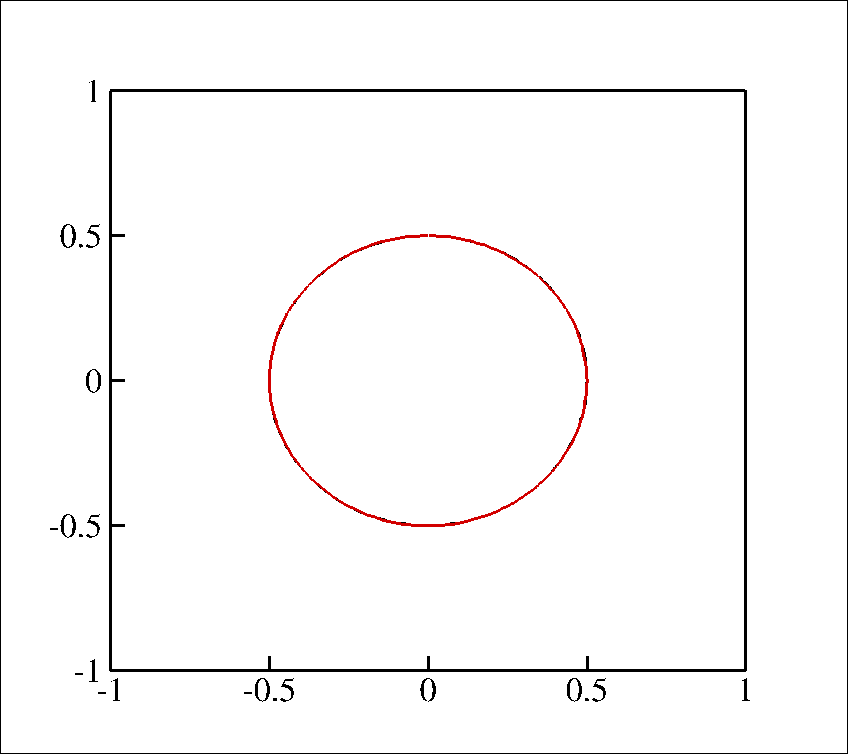}};
\end{tikzpicture}
\put(-95,-2){{$x$}}
\put(-157,125){$Ca=0.3$}
\put(-160,20){{{\color{black}\fullthick}}}
\put(-142,20){$t=0$}
\put(-115,20){{{\color{red}\fullthick}}}
\put(-97,20){$t=20$}
\caption{}
\label{subfig:sv_b}
\end{subfigure}
% ----------
\begin{subfigure}[t]{0.30\textwidth}
\centering
\begin{tikzpicture}
    \node[anchor=north west] at (0,0) {\includegraphics[height=1\linewidth,trim={60 50 10 10},clip]
{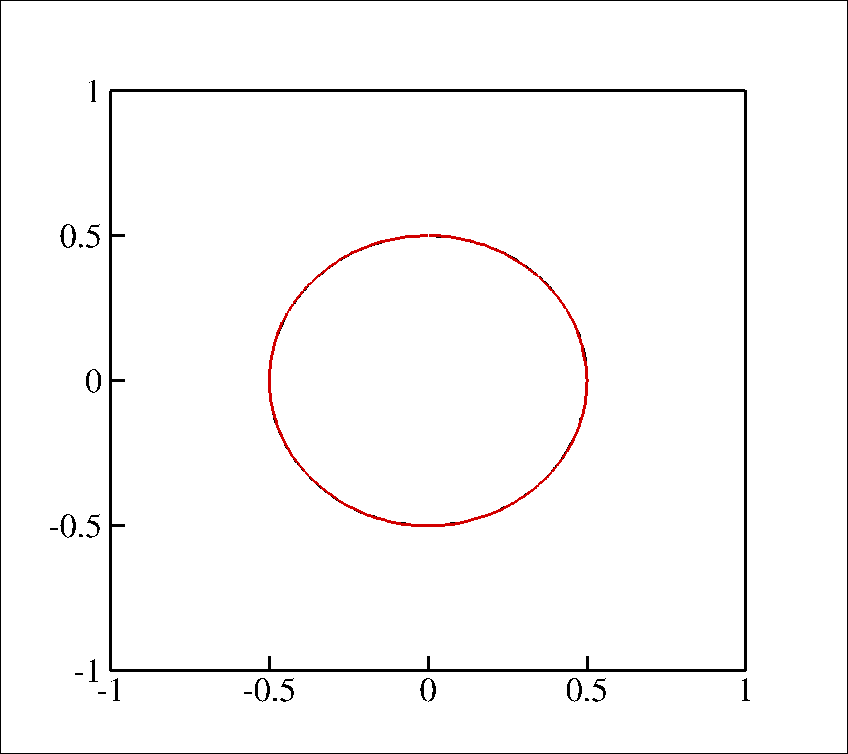}};
\end{tikzpicture}
\put(-95,-2){{$x$}}
\put(-157,125){$Ca=0.5$}
\put(-160,20){{{\color{black}\fullthick}}}
\put(-142,20){$t=0$}
\put(-115,20){{{\color{red}\fullthick}}}
\put(-97,20){$t=20$}
\caption{}
\label{subfig:sv_c}
\end{subfigure}
% ----------
\caption{Saint Venant-Kirchhoff circular disk in shear flow.\\\hspace{\textwidth}}
\label{fig:MR_disk}
\end{figure}
% =======================================================

At the deformed steady state,  the disk attains a smooth and stable elliptical shape without unwanted oscillations at the fluid-solid interface, and the disk shapes are in excellent agreement with the steady-state shapes reported by Ii et al. \cite{ii2011implicit} Figure \ref{fig:MR_disk} shows the current disk shape at $t = 5$ along with the results of Ii et al. \cite{ii2011implicit} The circular disk deforms to an elliptical shape whose major axis increases with $Ca$. The variation of the elliptical shape with $Ca$ and the steady-state shapes are consistent with the results of Ii et al. \cite{ii2011implicit} Furthermore, the deformed steady state has a smooth interface without any wiggles at the fluid-solid interface, reflecting stability.  

At the final steady state, the disk retains its initial smooth circular shape without unwanted oscillations at the fluid-solid interface. Figure \ref{fig:MR_disk} compares the final steady state with the initial state. The final state has a smooth interface without any wiggles on the solid surface. As a result, the initial and final configurations coincide very well.

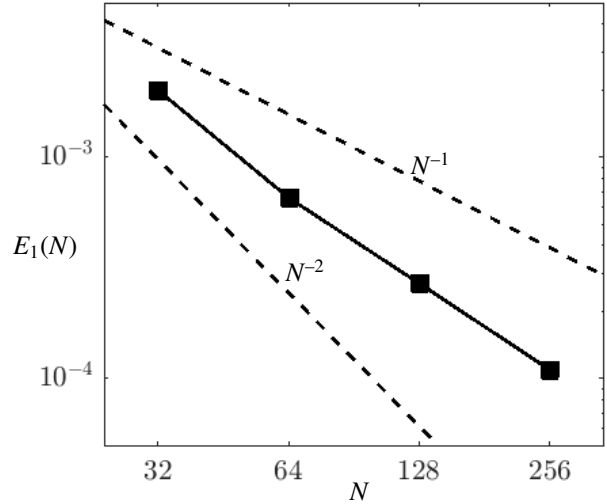
\begin{figure}[h]
\centering
% ----------
\begin{subfigure}[t]{0.45\textwidth}
    \centering
    \begin{tikzpicture}
    \node[anchor=north west] at (0,0) {\includegraphics[width=1\linewidth,trim={10 10 10 10},clip]
    {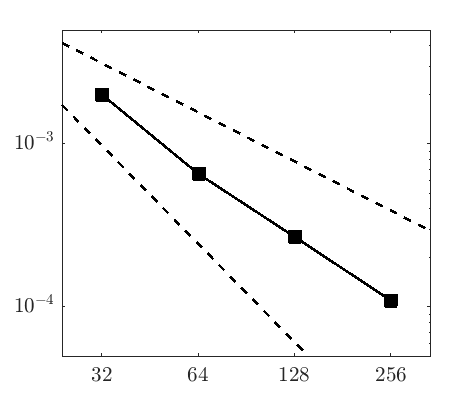}};
    \end{tikzpicture}
    \put(-240,90){{$E_1(N)$}}
    \put(-114,-2){{$N$}}
    \put(-91,120){{$N^{-1}$}}
    \put(-138,80){{$N^{-2}$}}
\end{subfigure}
\caption{ $L_1$ shape error ($E_1(N)$) for Saint Venant-Kirchhoff circular disk in shear flow with $Ca = 0.5$.\\\hspace{\textwidth}}
\label{fig:MR_L1_error}
\end{figure}

To estimate the order of convergence, we compute the $L_1$ shape error at $t = 20$ for $Ca=0.5$ on four different grids ($32 \times 32$, $64 \times 64$, $128 \times 128$, $256 \times 256$). The $L_1$ shape error ($E_1$) is computed as
\begin{equation}
    E_1(N=N_x) = \frac{1}{N_xN_y}\sum_{i=1}^{N_x}\sum_{j=1}^{N_y}\left|\phis_{i,j}(t=20) - \phis_{i,j}(t=0)\right|.
\end{equation} 
Figure \ref{fig:MR_L1_error} shows the variation of $E_1(N)$ with the grid size. The error decreases with grid size approximately as $N^{-1}$ for higher grid resolutions.

% ======================================
% ======================================

\subsection{Sphere in a lid-driven cavity}

We consider the motion of a deformable neo-Hookean sphere in a 3D lid-driven cavity flow. The problem setup is similar to Valizadeh et al. \citep{valizadeh2025monolithic} and Mao et al. \citep{mao20243d} The dimensions of the 3D cavity are $L_x \times L_y \times L_z = 1 \times 1 \times 1$, with no-slip boundary conditions on the walls. Initially, the system is at rest, and the unstressed sphere is centered at $(x_c,y_c,z_c)=(0.6,0.5,0.5)$ with radius $r=0.2$. The top wall ($y=L_y$) starts moving in the $x$ direction with steady velocity $V_w = 1$, while the remaining walls stay stationary. The fluid properties are $\rhof = 1$ and $\muf = 0.02$, and the solid properties are $\rhos = 1$, $\mus = 0.02$, and $G^s = 0.5$. The grid size is $N_x \times N_y \times N_z = 64 \times 64 \times 64$ and we consider the motion for $t \in [0,10]$.  

\begin{figure}[h]
\flushleft
% ----------
\begin{subfigure}[t]{0.245\textwidth}
\centering
\begin{tikzpicture}
    \node[anchor=north west] at (0,0) {\includegraphics[width=1\linewidth,trim={60 50 125 100},clip]
{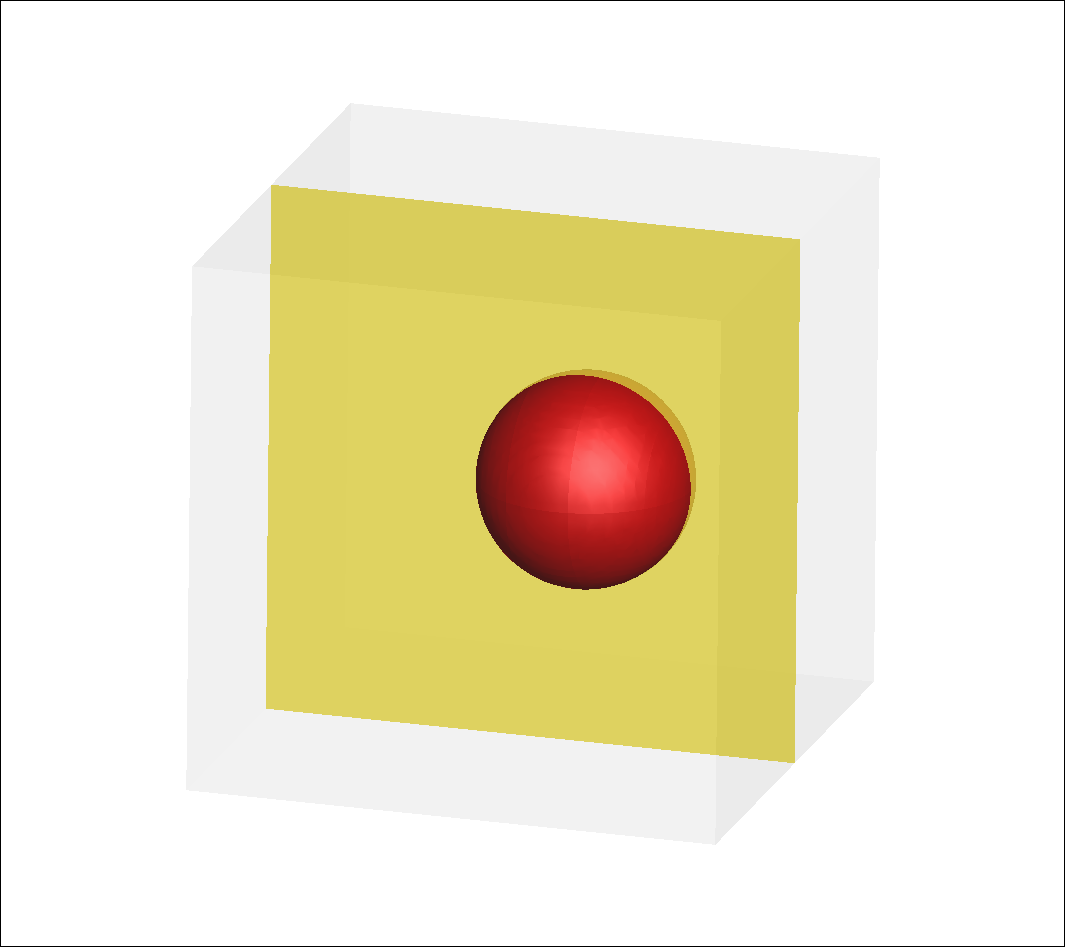}};
    \node[anchor=north west] at (1.45,0.5) {\includegraphics[height=0.1\linewidth,trim={0 0 0 0},clip]
{figures/VOF_method_validations/LDC_disk/Colormap_Sequential_Viridis.png}};
    \draw[black,thick,->] (1.15,-3.14) -- (1.65,-3.18);
    \draw[black,thick,->] (1.15,-3.14) -- (1.15,-2.65);
    \draw[black,thick,->] (1.15,-3.14) -- (0.88,-3.4);    
\end{tikzpicture}
\put(-93,120){{$0$}}
\put(-32,120){{$1$}}
\put(-63,131){{$\left| \bm{u} \right|$}}
\put(-85,22){{$x$}}
\put(-97,46){{$y$}}
\put(-107,17){{$z$}}
\begin{tikzpicture}
%\node at (-2.45,0.40) {$\Gamma_{fs}$};
\end{tikzpicture}
\caption{}
\label{subfig:VAL_disk_in_cavity_a}
\end{subfigure}
% ----------
\begin{subfigure}[t]{0.245\textwidth}
\centering
\begin{tikzpicture}
    \node[anchor=north west] at (0,0) {\includegraphics[width=1\linewidth,trim={60 50 125 100},clip]
{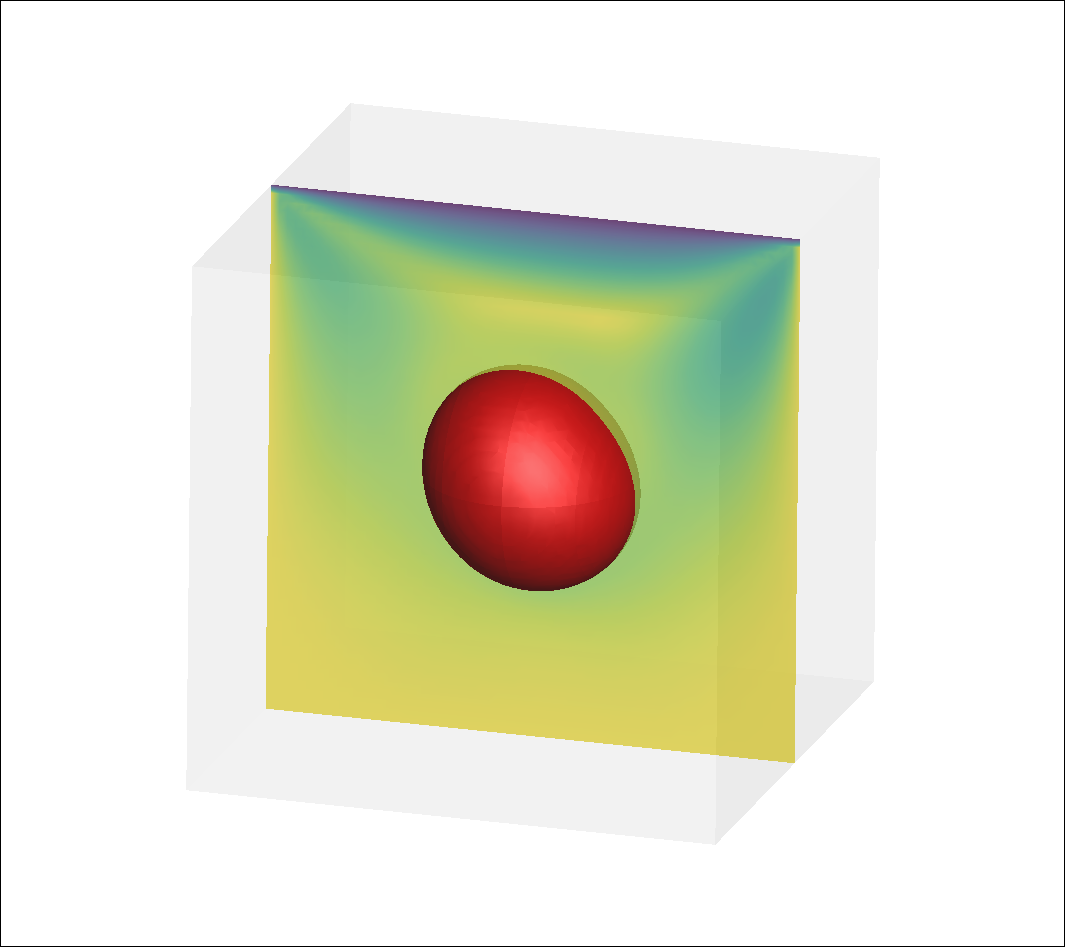}};
\end{tikzpicture}
\caption{}
\label{subfig:VAL_disk_in_cavity_b}
\end{subfigure}
% ----------
\begin{subfigure}[t]{0.245\textwidth}
\centering
\begin{tikzpicture}
    \node[anchor=north west] at (0,0) {\includegraphics[width=1\linewidth,trim={60 50 125 100},clip]
{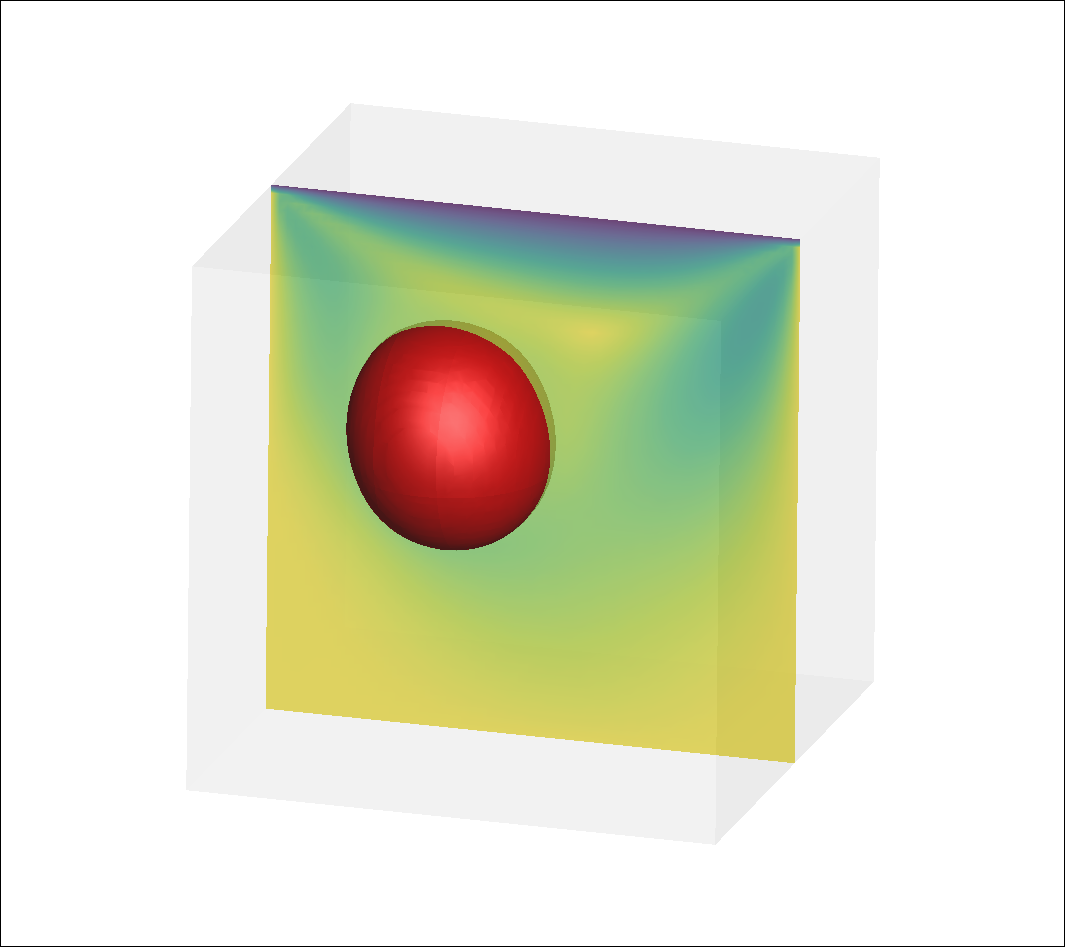}};
\end{tikzpicture}
\caption{}
\label{subfig:VAL_disk_in_cavity_c}
\end{subfigure}
% ----------
\begin{subfigure}[t]{0.245\textwidth}
\centering
\begin{tikzpicture}
    \node[anchor=north west] at (0,0) {\includegraphics[width=1\linewidth,trim={60 50 125 100},clip]
{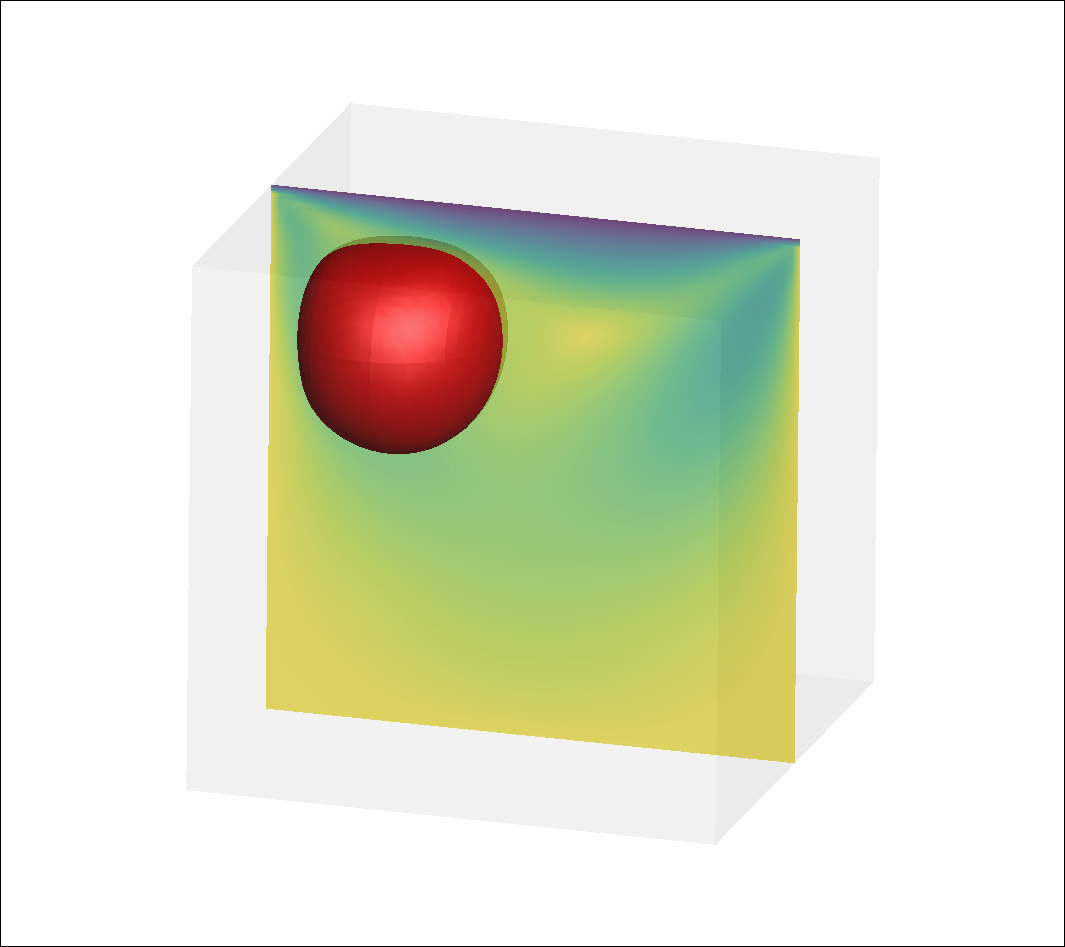}};
\end{tikzpicture}
% \put(-165,67.5){{$y$}}
% \put(-87,-2){{$x$}}
\caption{}
\label{subfig:VAL_disk_in_cavity_d}
\end{subfigure}
% ----------
\\
\begin{subfigure}[t]{0.245\textwidth}
\centering
\begin{tikzpicture}
    \node[anchor=north west] at (0,0) {\includegraphics[width=1\linewidth,trim={60 50 125 100},clip]
{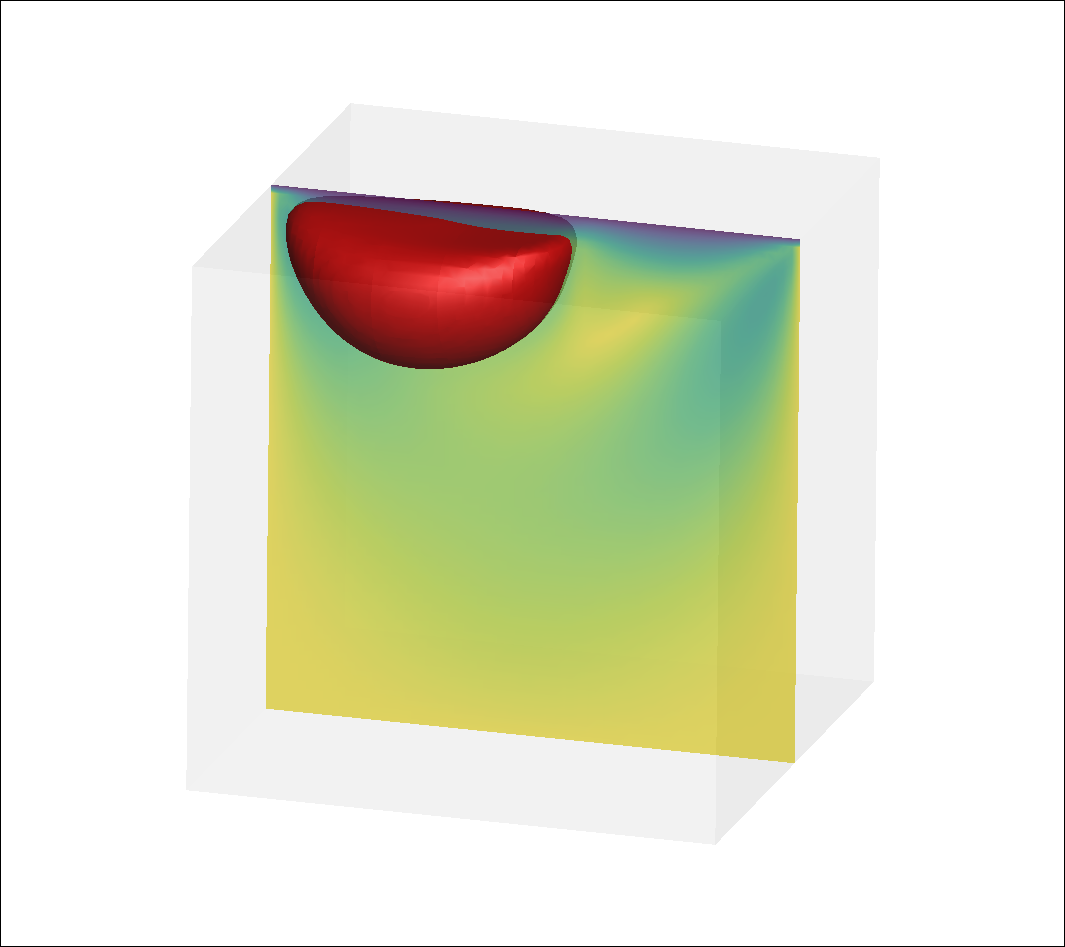}};
\end{tikzpicture}
%\put(-87,-2){{$x$}}
\caption{}
\label{subfig:VAL_disk_in_cavity_e}
\end{subfigure}
% ----------
\begin{subfigure}[t]{0.245\textwidth}
\centering
\begin{tikzpicture}
    \node[anchor=north west] at (0,0) {\includegraphics[width=1\linewidth,trim={60 50 125 100},clip]
{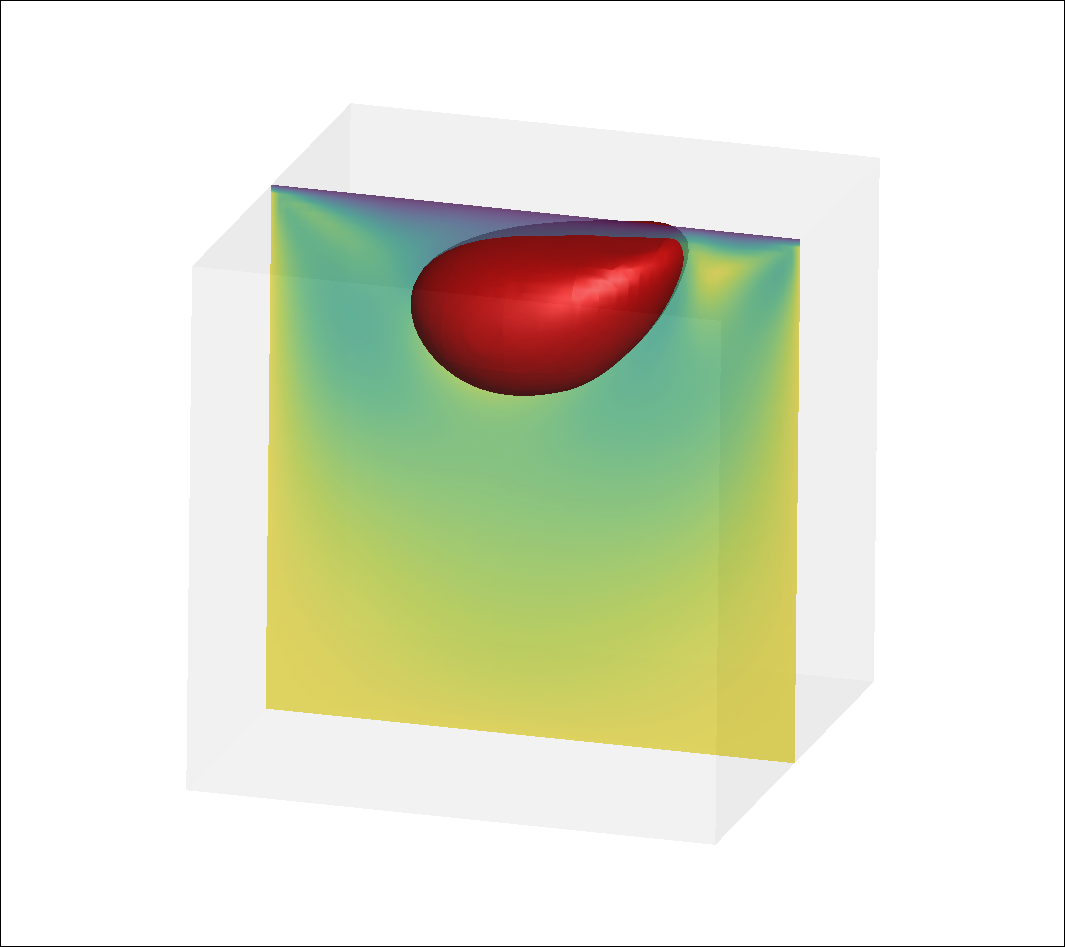}};
\end{tikzpicture}
%\put(-87,-2){{$x$}}
\caption{}
\label{subfig:VAL_disk_in_cavity_f}
\end{subfigure}
% ----------
\begin{subfigure}[t]{0.245\textwidth}
\centering
\begin{tikzpicture}
    \node[anchor=north west] at (0,0) {\includegraphics[width=1\linewidth,trim={60 50 125 100},clip]
{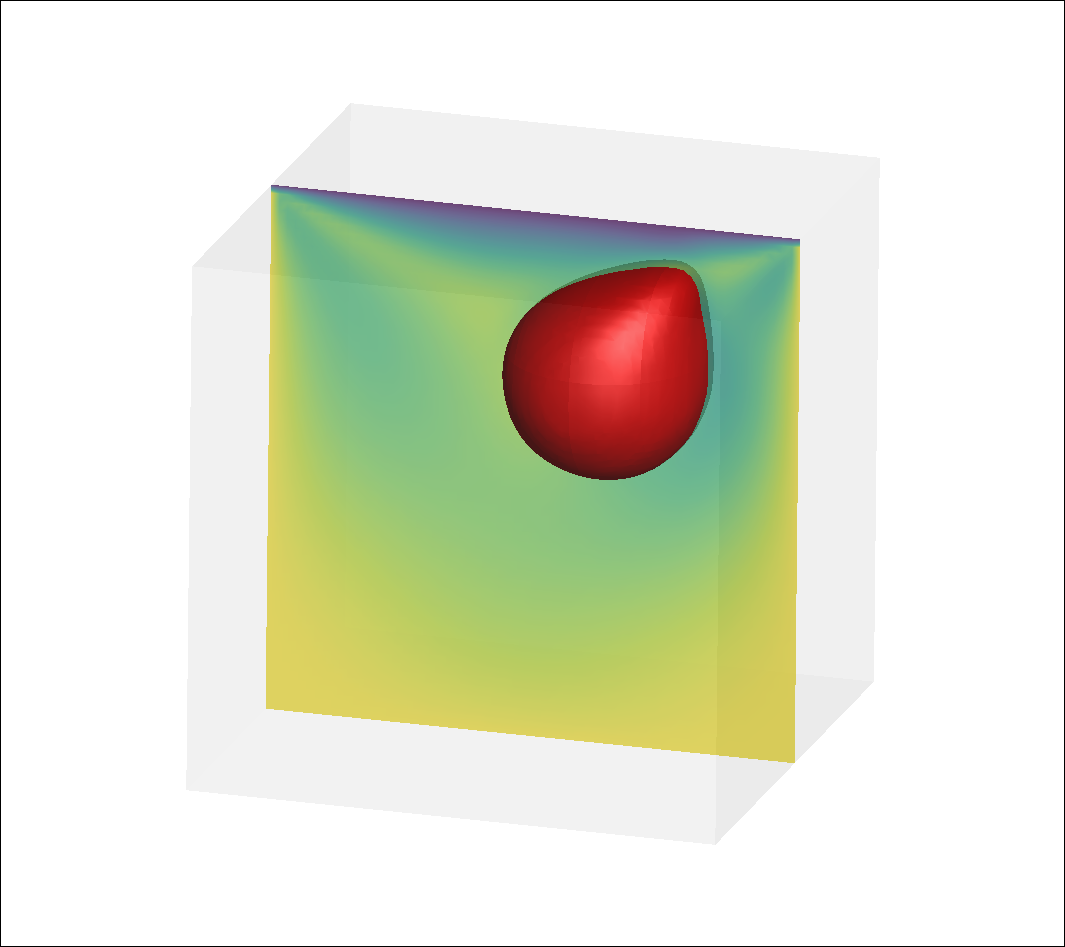}};
\end{tikzpicture}
%\put(-165,67.5){{$y$}}
%\put(-87,-2){{$x$}}
\caption{}
\label{subfig:VAL_disk_in_cavity_g}
\end{subfigure}
% ----------
\begin{subfigure}[t]{0.245\textwidth}
\centering
\begin{tikzpicture}
    \node[anchor=north west] at (0,0) {\includegraphics[width=1\linewidth,trim={60 50 125 100},clip]
{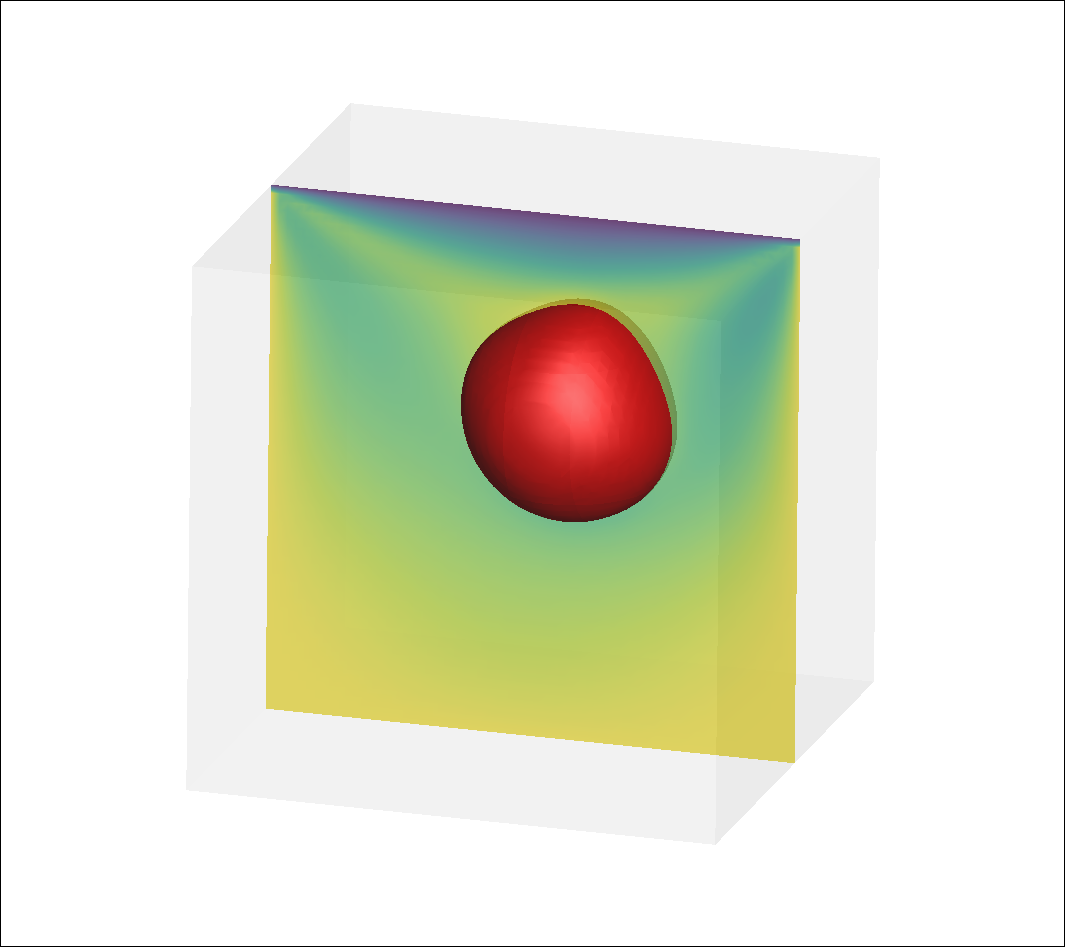}};
\end{tikzpicture}
%\put(-87,-2){{$x$}}
\caption{}
\label{subfig:VAL_disk_in_cavity_h}
\end{subfigure}
\caption{3D lid-driven cavity with a deformable sphere at different time instants (a-h) $t = 0, 1, 2, 3, 4, 5, 6, 7$. The vertical slice located at $z = 0.5$ shows flow speed contours and the red surface shows the solid shape ($\phis = 0.5$).\\\hspace{\textwidth}}
\label{fig:LDC_3Dsphere_snapshots}
\end{figure}

The solid sphere moves in a clockwise direction due to the circulating flowfield generated by the moving top wall and it experiences high stretching near the top lid before it slides along the lid. Despite high stretching, the VOF/PLIC method-based FSI framework does not generate any unphysical solid fragments. Figure \ref{fig:LDC_3Dsphere_snapshots} shows the deformed solid and fluid flow in the 3D cavity at multiple time instants $t = 0, 1, 2, 3, 4, 5, 6, 7$. The solid initially moves towards the top wall and undergoes significant stretching due to high fluid stresses. However, the lubrication effect \citep{skotheim2005soft} between the solid and the wall prevents it from touching the top wall. As time proceeds and it approaches a settled state, its deformation and movement reduce. These observations are consistent with Valizadeh et al. \cite{valizadeh2025monolithic}

% ======================================
% ======================================

\begin{figure}[h]
\centering
% ----------
\begin{tikzpicture}
    \node[anchor=north west] at (0,0) {\includegraphics[width=0.45\linewidth,trim={10 10 10 10},clip]
{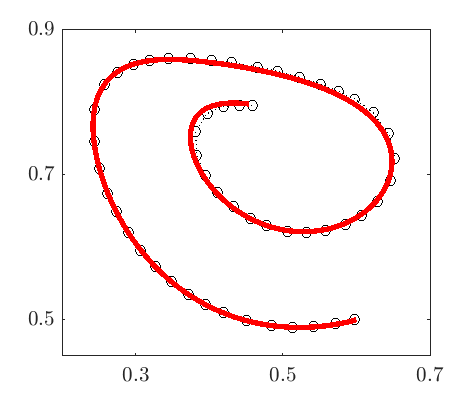}};
\end{tikzpicture}
\put(-228,93){{$y$}}
\put(-110,-2){{$x$}}
%\put(-92,94){{$-1$}}
\caption{ Deformable sphere centroid trajectory in 3D lid-driven cavity for $t \in [0,10]$. {\color{red}\fullthick} VOF/PLIC ($64 \times 64 \times 64$); {\color{black}{\emptycircle}} Valizadeh et al. \cite{valizadeh2025monolithic} ($h = 0.01$).\\\hspace{\textwidth}}
\label{fig:VAL_LDC_3Dsphere_centroid}
\end{figure}

To verify the results, we compute the 3D solid centroid trajectory as,
\begin{equation}
    \bm{x}_c(t) = \frac{\sum_{i=1}^{N_x}\sum_{j=1}^{N_y}\sum_{k=1}^{N_z}\bm{x}_{i,j,k}\phis_{i,j,k}(t)\Delta_x \Delta_y \Delta_z}{\sum_{i=1}^{N_x}\sum_{j=1}^{N_y}\sum_{k=1}^{N_z} \phis_{i,j,k}(t)\Delta_x \Delta_y \Delta_z}.
\end{equation}
Figure \ref{fig:VAL_LDC_3Dsphere_centroid} shows the solid centroid trajectory in the $x-y$ plane for $t \in [0,10]$. The solid centroid initially moves towards the top wall and spirals clockwise, consistent with the 3D snapshots. Throughout the time history, the trajectory agrees well with that of Valizadeh et al. \cite{valizadeh2025monolithic} obtained using phase-field modeling.

% ======================================
% ======================================

\subsection{Direct numerical simulation of turbulent channel flow with a deformable compliant wall}

% Problem setup

\begin{figure}[h]
\centering
% ----------
\begin{tikzpicture}
    \node[anchor=north west] at (0,0) {\includegraphics[width=0.5\linewidth,trim={0 0 0 0},clip]
{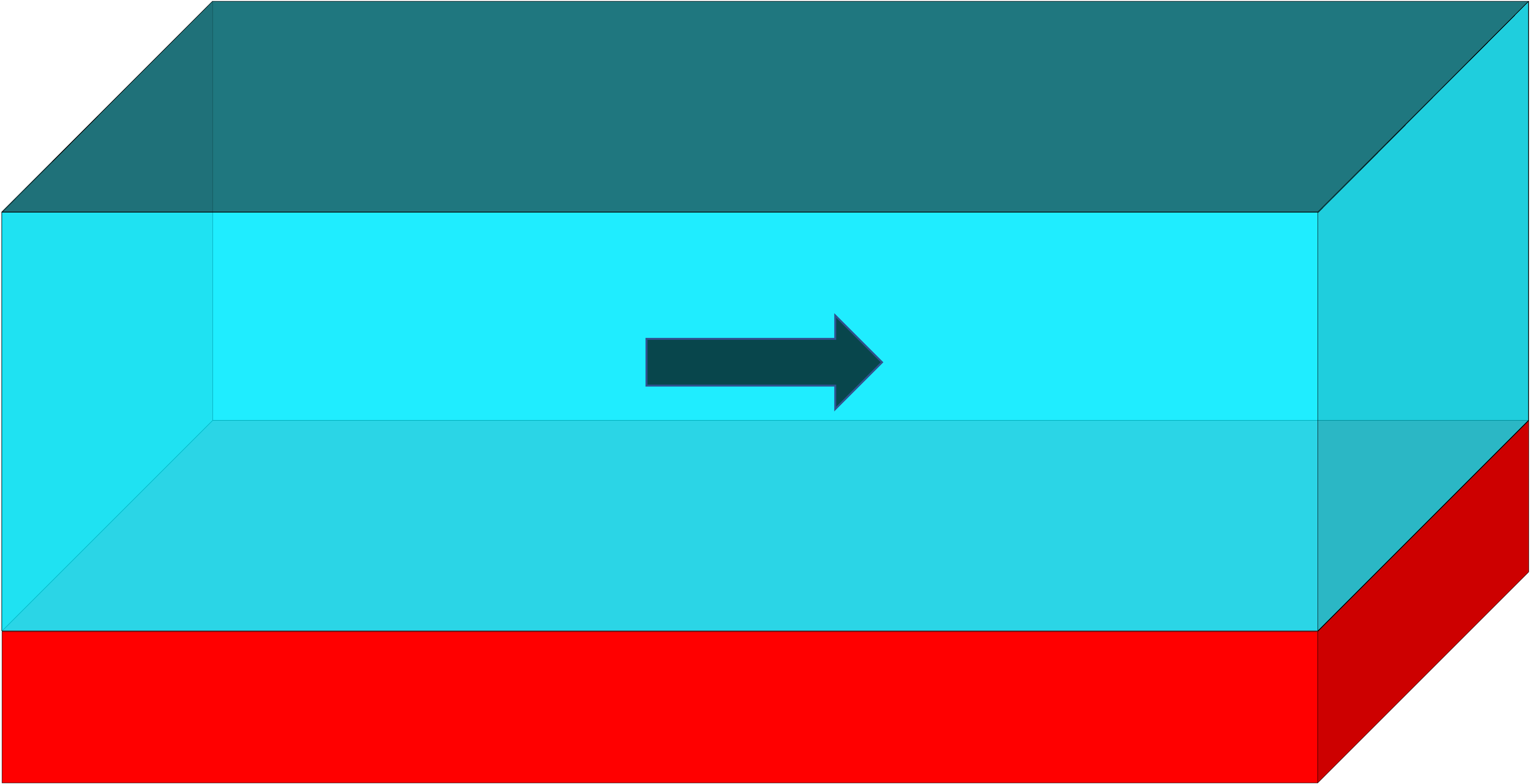}};
    %\draw[black,thick,->] (1.055,-2.883) -- (1.65,-2.962);
    \draw[black,thick,->] (1.35,-2.55) -- (1.35,-1.6);
    \draw[black,thick,->] (1.35,-2.55) -- (2.35,-2.55);
    \draw[black,thick,->] (1.35,-2.55) -- (0.834,-3.08);    
    %\draw[black,thick,->] (1.055,-2.883) -- (0.75,-3.215);
\end{tikzpicture}
\put(-195,56){{$x$}}
\put(-217,88){{$y$}}
\put(-233,43){{$z$}}
%\put(-165,67.5){{$y$}}
%\put(-88,-2){{$x$}}
\caption{Schematic of turbulent channel flow with a deformable compliant bottom wall and a rigid top wall.\\\hspace{\textwidth}}
\label{fig:tcf_compwall_schematic}
\end{figure}

We use the FSI solver to perform direct numerical simulation (DNS) of turbulent channel flow with a neo-Hookean compliant bottom wall and a rigid top wall. The streamwise, spanwise, and wall-normal directions are along $x$, $y$, and $z$ axes, respectively. Figure \ref{fig:tcf_compwall_schematic} shows a schematic of the problem. The domain size is $L_x \times L_y \times L_z = 2\pi\delta \times (2\delta + h^s) \times \pi\delta$, where $\delta$ is the half-height of the channel and $h^s$ is the compliant wall thickness. The flow field is initialized with a statistically stationary turbulent channel flow developed over rigid walls, occupying $x \in [0,L_x], y \in [0,2\delta], z \in [0,L_z]$ region, and the compliant wall has a flat interface initially with height $h^s$, occupying $x \in [0,L_x], y \in [-h^s,0), z \in [0,L_z]$ region. We impose no-slip conditions on the top and bottom walls, and consider periodicity in streamwise and spanwise directions.    

We consider a constant flow rate such that the bulk Reynolds number $Re_b = \rhof U_b\delta/\muf$ is equal to 2800, where $U_b$ is the fluid bulk velocity and the streamwise pressure gradient in the whole domain is adjusted to maintain constant $U_b$. For turbulent channel flow with rigid top and bottom walls, $Re_b = 2800$ corresponds to friction Reynolds number $Re_\tau = \rhof u_\tau \delta/\muf \approx 180$, where $u_\tau = \sqrt{\tau_w/\rhof}$ is the friction velocity and $\tau_w$ is the wall shear stress. To ensure two-way coupling, the parameters of the compliant wall are prescribed based on past works \citep{rosti2017numerical,rosti2020low,esteghamatian2022spatiotemporal}. The parameters in the fluid domain are $\delta=1$, $U_b=1$, $\rhof=1$, $\muf=Re_b^{-1}$, and in the solid domain are $h^s=0.5\delta=0.5$, $\rhos=1$, $\mus=Re_b^{-1}$, $G^s=0.5\rhof U_b^2=0.5$.   

The domain is discretized using uniform grid distribution in streamwise and spanwise directions. However, in the wall-normal direction, a non-uniform grid distribution is used in the fluid region away from the fluid-solid interface ($y \in (0.33,2]$), while a uniform distribution is used close to the interface and in the solid domain ($y \in [-0.5,0.33]$). The grid size is $N_x \times N_y \times N_z = 328 \times 312 \times 166$ and the grid resolution relative to the viscous length scale ($\delta_\nu = \muf/(\rhof u_\tau)$) based on turbulent flow over rigid walls is $\Delta_x/\delta_\nu=3.45$, $\Delta_z/\delta_\nu=3.41$, and $\Delta_ {y,min}/\delta_\nu=0.79$.    

\begin{figure}[h]
\centering
% ----------
\begin{tikzpicture}
    \node[anchor=north west] at (0,0) {\includegraphics[width=0.45\linewidth,trim={10 10 10 10},clip]
{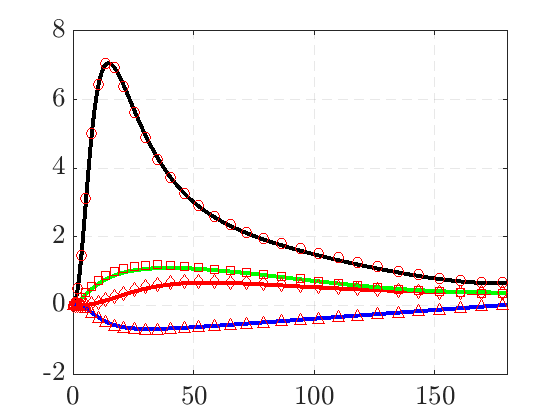}};
\end{tikzpicture}
\put(-235,90){{$\frac{\langle u_i^\prime u_j^\prime \rangle}{u_\tau^2}$}}
\put(-125,-2){{$y/\delta_\nu$}}
\put(-95,150){{{\color{black}\fullthick}}}
\put(-95,135){{{\color{red}\fullthick}}}
\put(-95,120){{{\color{green}\fullthick}}}
\put(-95,105){{{\color{blue}\fullthick}}}
\put(-75,150){{{\color{red}\emptycircle}}}
\put(-75,135){{{\color{red}\emptydiamond}}}
\put(-75,120){{{\color{red}\emptysquare}}}
\put(-75,105){{{\color{red}\emptytriangle}}}
\put(-65,150){$\langle u^\prime u^\prime \rangle/u_\tau^2$}
\put(-65,135){$\langle v^\prime v^\prime \rangle/u_\tau^2$}
\put(-65,120){$\langle w^\prime w^\prime \rangle/u_\tau^2$}
\put(-65,105){$\langle u^\prime v^\prime \rangle/u_\tau^2$}

%\put(-92,94){{$-1$}}
\caption{ Reynolds stress tensor components for turbulent channel flow with rigid top and bottom walls. Solid lines represent current results and symbols represent results of Moser et al. \citep{moser1999direct} for $Re_\tau \approx 180$. \\\hspace{\textwidth}}
\label{fig:RW_reystress}
\end{figure}

We first simulate the turbulent channel flow with rigid top and bottom walls until it reaches a statistically stationary state. Then, we replace the rigid bottom wall with a deformable compliant wall, where we analyze the system's behavior after it passes the initial transient phase. To verify the validity of turbulent channel flow constricted by rigid walls, we calculate Reynolds stress tensor defined as $\langle u_i^\prime u_j^\prime \rangle$, where $\langle\cdot \rangle$ represents spatiotemporal average (streamwise direction, spanwise direction, time) and $(\cdot)^\prime$ represents fluctuating component in the Reynolds decomposition. Figure \ref{fig:RW_reystress} compares the current Reynolds stresses at $Re_b = 2800$ ($Re_\tau \approx 180$) with that of Moser et al. \citep{moser1999direct} at $Re_\tau \approx 180$. The Reynolds stress components agree very well with the reference.           

% ======================================
% ======================================

\begin{figure}[h]
\centering
% ----------
\begin{subfigure}[t]{0.475\textwidth}
\centering
\begin{tikzpicture}
    \node[anchor=north west] at (0,0) {\includegraphics[width=1\linewidth,trim={10 100 10 100},clip]
{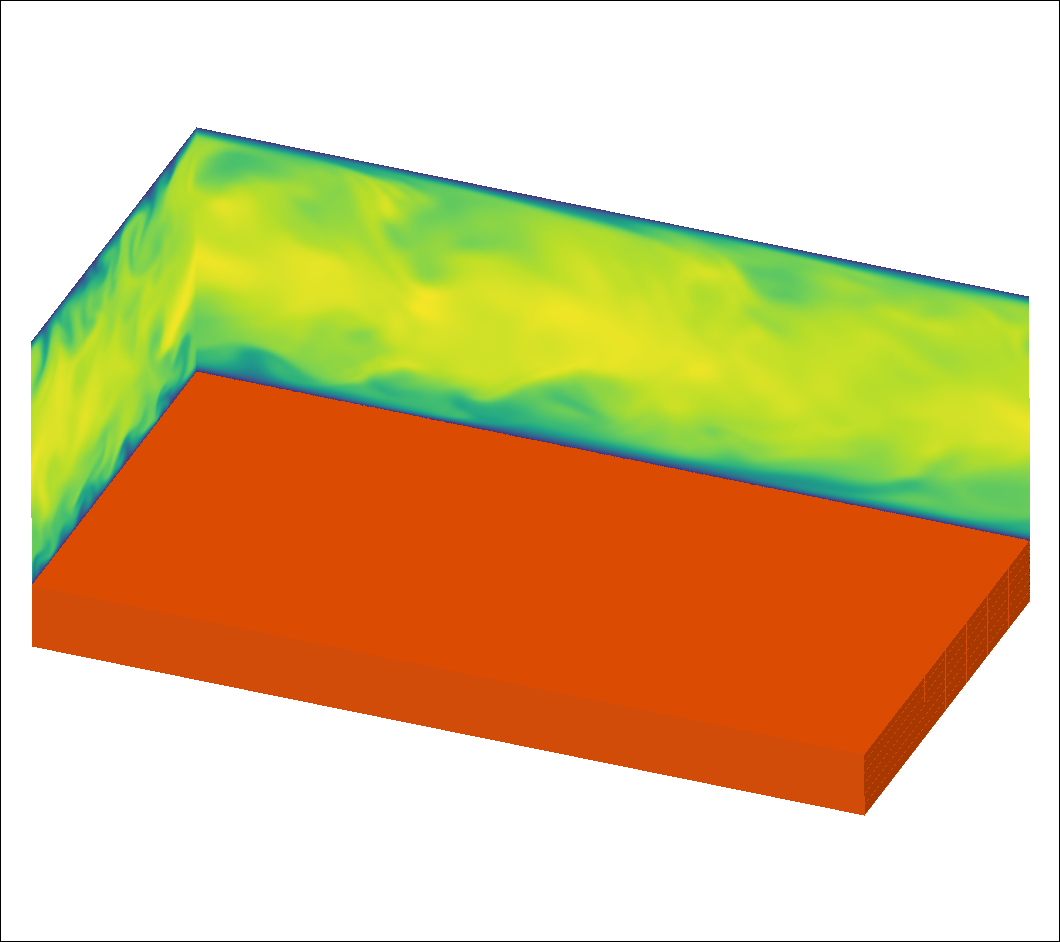}};
    \node[anchor=north west] at (4.5,0) {\includegraphics[width=0.2\textwidth,
        height=0.07\textwidth,trim={0 0 0 0},clip]
{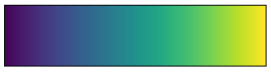}};
    \draw[black,thick,->] (1.62,-2.33) -- (1.62,-1.6);
    \draw[black,thick,->] (1.62,-2.33) -- (2.3,-2.46);
    \draw[black,thick,->] (1.62,-2.33) -- (1.25,-2.83);
\end{tikzpicture}
\put(-133,163){{$-0.2$}}
\put(-66,163){{$1.3$}}
\put(-99,176){{$u/U_b$}}
\put(-182,102){{$x$}}
\put(-198,129){{$y$}}
\put(-207,94){{$z$}}
\caption{}
\label{subfig:tcf_compwall_t0}
\end{subfigure}
% ----------
\begin{subfigure}[t]{0.475\textwidth}
\centering
\begin{tikzpicture}
    \node[anchor=north west] at (0,0) {\includegraphics[width=1\linewidth,trim={10 100 10 100},clip]
{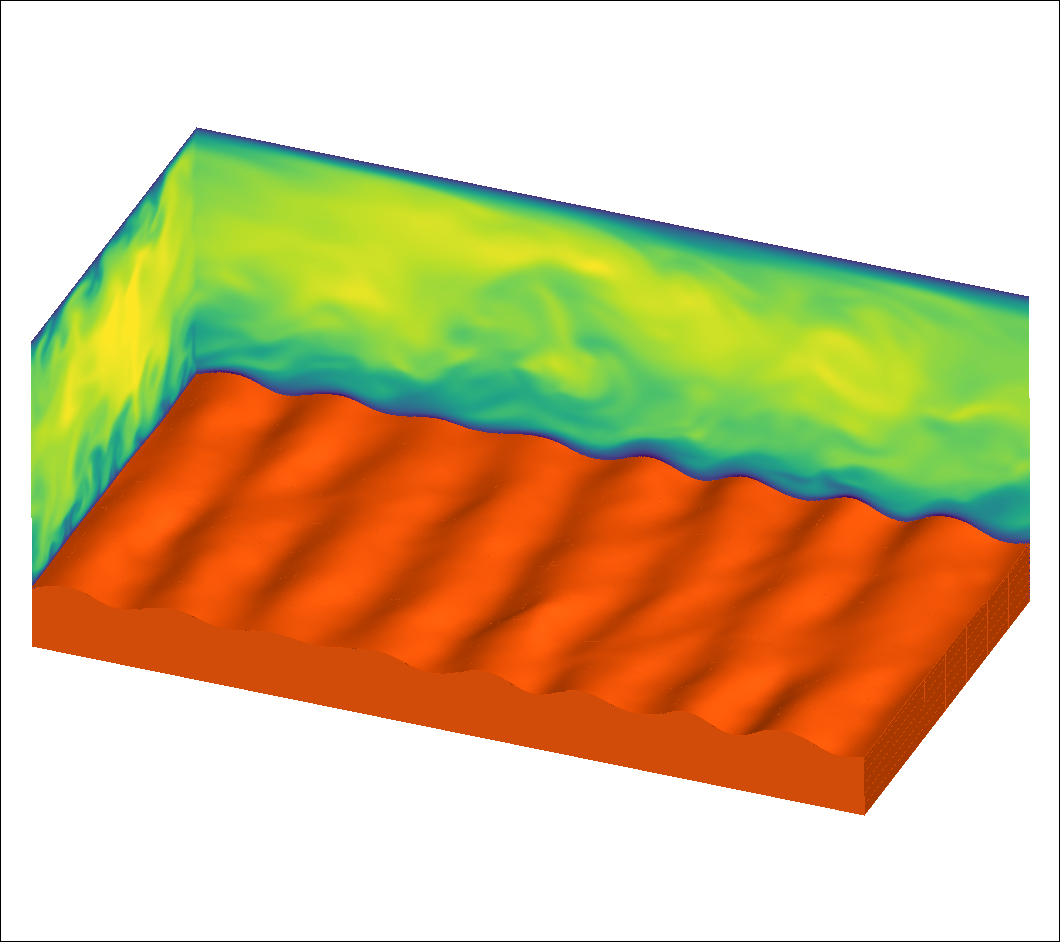}};
\end{tikzpicture}
\caption{}
\label{subfig:tcf_compwall_t200}
\end{subfigure}
\caption{Turbulent channel flow with (a) rigid bottom wall and (b) compliant bottom wall at $Re_b=2800$. Vertical slices show streamwise velocity contours and red shade shows the bottom wall.\\\hspace{\textwidth}}
\label{fig:tcf_compwall_3D_surf}
\end{figure}

Figure \ref{fig:tcf_compwall_3D_surf} compares instantaneous visualizations of turbulent channel flow over rigid and compliant bottom walls. The vertical slices show streamwise velocity, and the red shade shows the bottom wall. For the channel with a rigid bottom wall, the turbulent flow near the top and bottom rigid walls is qualitatively similar, and the rigid bottom wall has a flat surface. As the flow develops over the compliant wall, the fluid stresses deform the compliant surface, and its dynamic motion affects the near-wall turbulence and modifies the flow. The wavy undulations of the compliant surface are significant enough to alter the near-wall turbulent flow. As a result, turbulent flow near the compliant surface is no longer qualitatively similar to the rigid top wall. 

%------------------------------------------------------------------
%------------------------------------------------------------------

\begin{figure}[h]
\centering
% ----------
\begin{subfigure}[t]{0.475\textwidth}
\centering
\begin{tikzpicture}
    \node[anchor=north west] at (0,0) {\includegraphics[width=1\linewidth,trim={25 250 25 250},clip]
{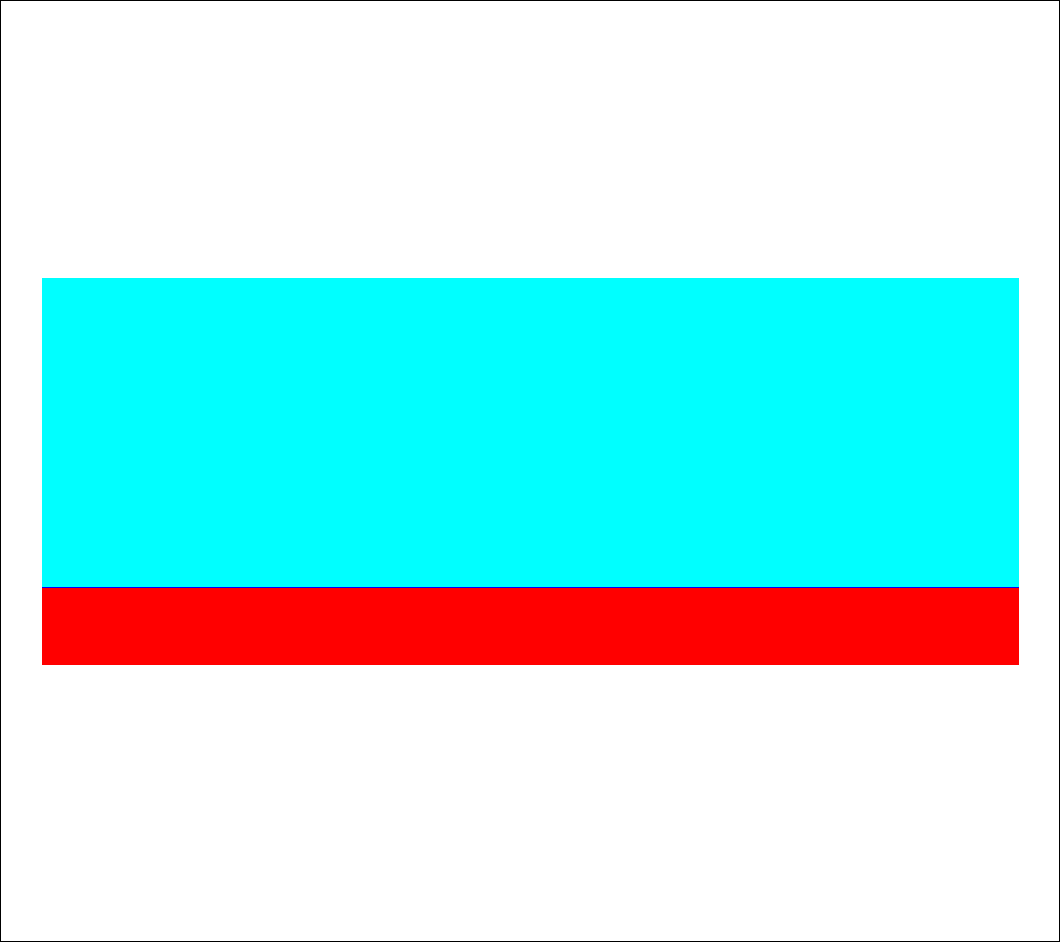}};
    \draw[black,thick,->] (0.265,-3.6) -- (0.265,-2.715);
    \draw[black,thick,->] (0.265,-3.6) -- (1.15,-3.6);
\end{tikzpicture}
\put(-215,2){{$x$}}
\put(-244,35){{$y$}}
\caption{}
\label{subfig:tcf_compwall_phis_0}
\end{subfigure}
% ----------
\begin{subfigure}[t]{0.475\textwidth}
\centering
\begin{tikzpicture}
    \node[anchor=north west] at (0,0) {\includegraphics[width=1\linewidth,trim={25 250 25 250},clip]
{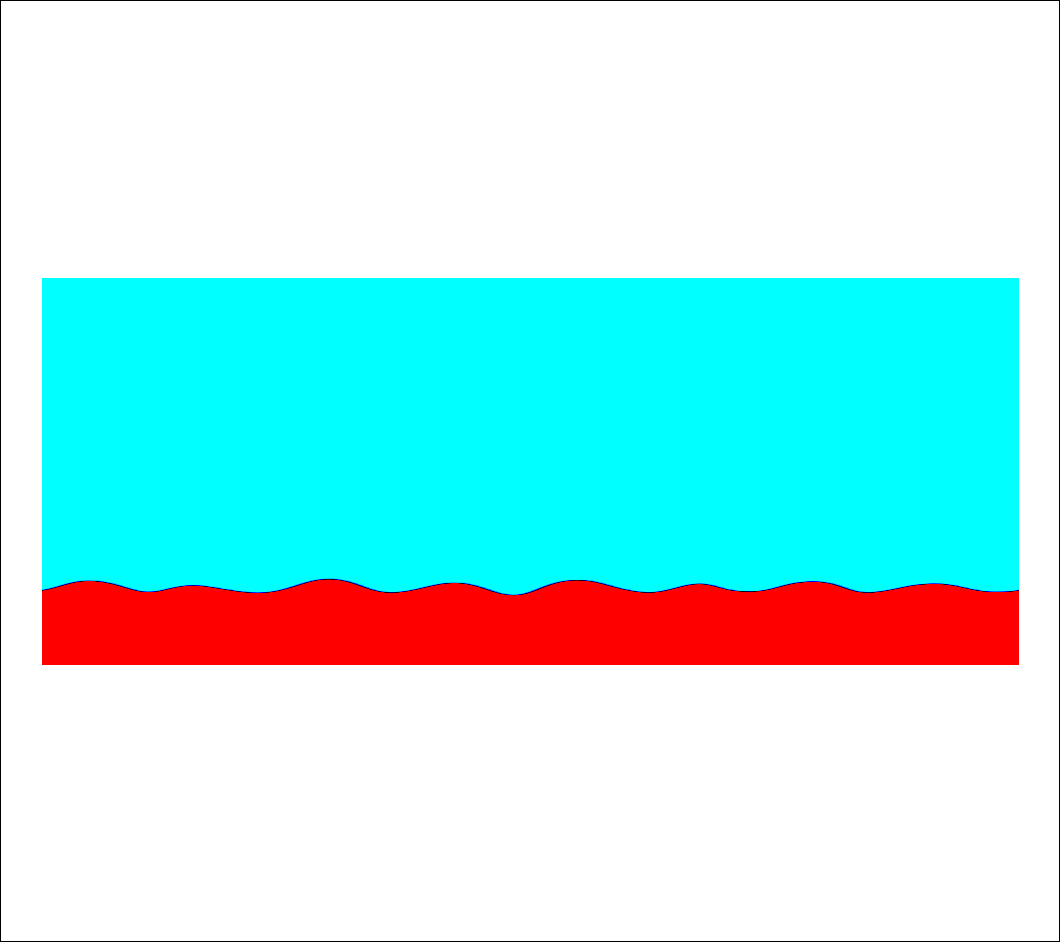}};
    \node[anchor=north west,rotate=90] at (8.36,-2.7) {\includegraphics[width=0.2\textwidth,
        height=0.07\textwidth,trim={0 0 0 0},clip]
    {figures/VOF_method_validations/LDC_compwall/Colormap_Doppler.png}};
\end{tikzpicture}
\put(-13,89){{$1$}}
\put(-14,30){{$0$}}
\put(-13,100){{$\phis$}}
\caption{}
\label{subfig:tcf_compwall_phis_200}
\end{subfigure}
\caption{Contours of $\phis$ on mid-spanwise plane of turbulent channel flow with compliant bottom wall at (a) $tU_b/\delta=0$ and (b) after the initial transient phase.\\\hspace{\textwidth}}
\label{fig:tcf_compwall_phis}
\end{figure}

The VOF/PLIC-based fluid-solid interface remains sharp and stable throughout time history, and despite turbulent interactions, we do not observe numerical artifacts such as solid fragments. Figure \ref{fig:tcf_compwall_phis} shows the contours of $\phis$ on a mid-spanwise plane for turbulent channel flow with a compliant bottom wall at $tU_b/\delta=0$ and after the initial transient phase.  

%------------------------------------------------------------------
%------------------------------------------------------------------

In past work involving such problems, fifth-order WENO scheme \citep{rosti2017numerical,rosti2020low} or LS \citep{esteghamatian2022spatiotemporal} methods were used to capture the fluid-solid interface. The WENO procedure requires high grid resolution to reduce interface diffusion due to the diffusive nature of the scheme. The LS method requires a reinitialization procedure to maintain stability, which can be problem-dependent. However, as discussed above, the VOF/PLIC-based FSI method maintains interface stability and sharpness, bypassing the requirement of any problem-dependent stabilization procedure or large grids.

\begin{figure}[h]
\centering
% ----------
\begin{subfigure}[t]{0.45\textwidth}
    \centering
    \begin{tikzpicture}
    \node[anchor=north west] at (0,0) {\includegraphics[width=1\linewidth,trim={10 10 10 0},clip]
    {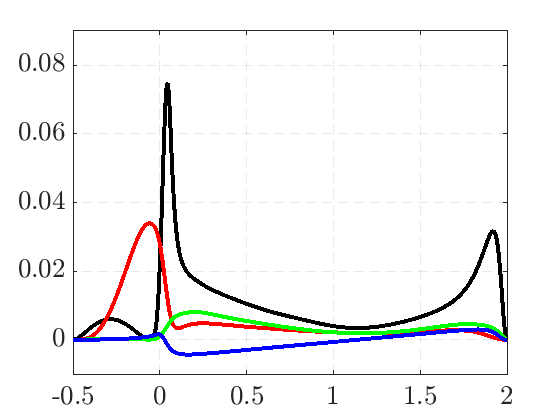}};
    \end{tikzpicture}
    \put(-252,90){{$\frac{ \langle u_i^\prime u_j^\prime  \rangle}{U_b^2}$}}
    \put(-115,-2){{$y/\delta$}}
    \put(-85,150){{{\color{black}\fullthick}}}
    \put(-85,135){{{\color{red}\fullthick}}}
    \put(-85,120){{{\color{green}\fullthick}}}
    \put(-85,105){{{\color{blue}\fullthick}}}
    \put(-65,150){$\langle u^\prime u^\prime \rangle/U_b^2$}
    \put(-65,135){$\langle v^\prime v^\prime \rangle/U_b^2$}
    \put(-65,120){$\langle w^\prime w^\prime \rangle/U_b^2$}
    \put(-65,105){$\langle u^\prime v^\prime \rangle/U_b^2$}
    
    \caption{}
    \label{fig:CW_reystress}
\end{subfigure}
% ----------
\begin{subfigure}[t]{0.45\textwidth}
    \centering
    \begin{tikzpicture}
    \node[anchor=north west] at (0,0) {\includegraphics[width=1\linewidth,trim={10 10 10 0},clip]
    {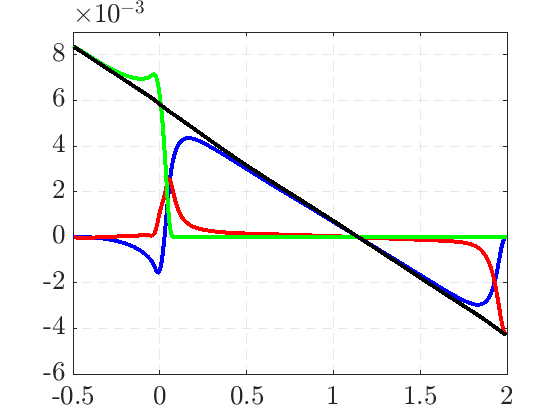}};
    \end{tikzpicture}
    \put(-224,90){{$\tau^i$}}
    \put(-115,-2){{$y/\delta$}}
    \put(-65,150){{{\color{black}\fullthick}}}
    \put(-65,135){{{\color{red}\fullthick}}}
    \put(-65,120){{{\color{green}\fullthick}}}
    \put(-65,105){{{\color{blue}\fullthick}}}
    \put(-45,150){$\tau^{total}$}
    \put(-45,135){$\tau^{V}$}
    \put(-45,120){$\tau^{H}$}
    \put(-45,105){$\tau^{R}$}    
    \caption{}
    \label{subfig:CW_totstress}
\end{subfigure}
\caption{ Stress distribution for turbulent channel flow with compliant bottom wall. (a) Reynolds stress tensor components and (b) the contribution of shear stress components to the total shear stress. \\\hspace{\textwidth}}
\label{fig:CW_stress}
\end{figure}

The wall compliance increases the Reynolds stress near the surface, and the shear stress component $\langle u^\prime v^\prime \rangle$ changes sign near the interface. Furthermore, the total shear stress follows a linear trend along the $y$ axis, and in most of the compliant wall region, the contribution of hyperelastic stress is the highest. The total shear stress is calculated as $\tau^{total} = \tau^{V} + \tau^{R} + \tau^{H}$, where $\tau^{V} = (\rho/\mu)\partial \langle u \rangle/\partial y$ is the viscous stress contribution, $\tau^{R} = -\rho \langle u^\prime v^\prime \rangle$ is the Reynolds stress contribution, and $\tau^H =  \rho G^s \langle \phis B_{xy} \rangle$ is the hyperelastic stress contribution. Figure \ref{fig:CW_reystress} shows the components of the Reynolds stress tensor, and figure \ref{subfig:CW_totstress} shows the total shear stress and its contributing components. The normal components ($ \langle u^\prime u^\prime \rangle, \langle v^\prime v^\prime \rangle, \langle w^\prime w^\prime \rangle$) of the Reynolds stress tensor are the highest close to the compliant surface (initially at $y=0$). The shear stress component ($\langle u^\prime v^\prime \rangle$) of the Reynolds stress tensor changes sign in the vicinity of the compliant surface and has the highest magnitude close to the surface. The total shear stress follows a linear trend along the $y$ axis, where its magnitude at the rigid support of the compliant wall ($y=-0.5$) is nearly double that at the rigid top wall ($y=2$). Furthermore,  in most of the compliant wall region, the contribution of hyperelastic stress ($\tau^H$) is the highest. These observations are consistent with the numerical work of Esteghamatian et al. \citep{esteghamatian2022spatiotemporal} that utilized the LS method to capture the interface. 

% ======================================
% ======================================

\begin{figure}[h]
\centering
% ----------
\begin{subfigure}[t]{0.475\textwidth}
\centering
\begin{tikzpicture}
    \node[anchor=north west] at (0,0) {\includegraphics[width=1\linewidth,trim={80 150 80 225},clip]
{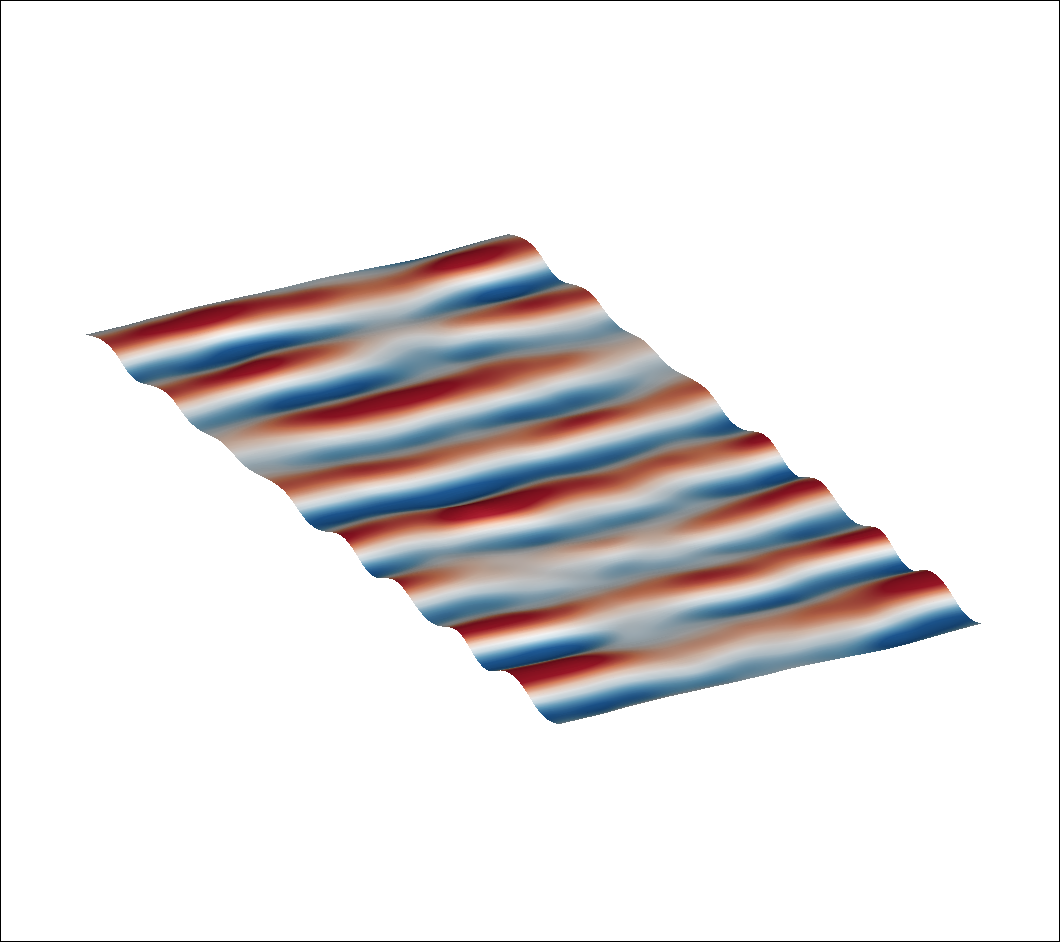}};
    \node[anchor=north west,rotate=90] at (0.5,-4.3) {\includegraphics[width=0.14\textwidth,
        height=0.07\textwidth,trim={0 0 0 0},clip]{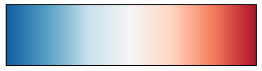}};
    \draw[black,thick,->] (4.2,-0.12) -- (4.9,-0.66);
    \draw[black,thick,->] (4.2,-0.12) -- (3.2,-0.35);
    %\draw[black,thick,->] (3.8,0) -- (4,0.7);    
    
    %\draw[black,thick,->] (1.52,-2.17) -- (2.15,-2.28);
\end{tikzpicture}
\put(-233,30){{$-0.05$}}
\put(-229,75){{$0.05$}}
\put(-228,85){{$d_y/\delta$}}
\put(-162,147){{$z$}}
\put(-105,135){{$x$}}
\caption{}
\label{subfig:tcf_compwall_def}
\end{subfigure}
% ----------
\begin{subfigure}[t]{0.475\textwidth}
\centering
\begin{tikzpicture}
    \node[anchor=north west] at (0,0) {\includegraphics[width=1\linewidth,trim={80 150 80 225},clip]
{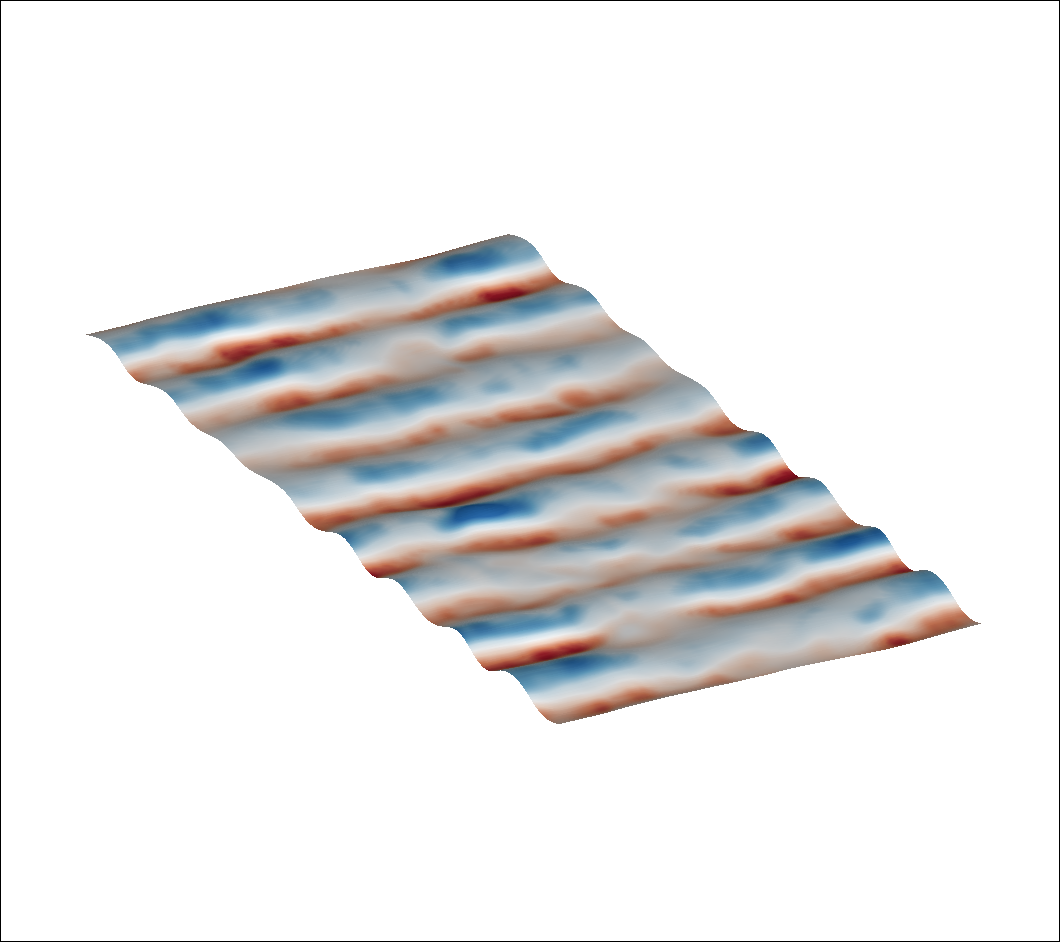}};
    \node[anchor=north west,rotate=90] at (0.5,-4.3) {\includegraphics[width=0.14\textwidth,
        height=0.07\textwidth,trim={0 0 0 0},clip]{figures/VOF_method_validations/tcf_compwall/deformation/diverging_blue_red.png}};
\end{tikzpicture}
\put(-233,30){{$-0.15$}}
\put(-229,75){{$0.15$}}
\put(-233,85){{$p/\rho U_b^2$}}
\caption{}
\label{subfig:tcf_compwall_press}
\end{subfigure}
\caption{Deformed compliant surface colored with contours of (a) wall-normal deformation and (b) pressure distribution.\\\hspace{\textwidth}}
\label{fig:tcf_compwall_def_press}
\end{figure}

The compliant surface exhibits spanwise aligned deformation patterns that propagate downstream, consistent with the past numerical \citep{rosti2017numerical,esteghamatian2022spatiotemporal} and experimental \citep{wang2020interaction} works. Figure \ref{subfig:tcf_compwall_def} shows the wall-normal deformation of the compliant surface, and figure \ref{subfig:tcf_compwall_press} shows the pressure distribution on the compliant surface. The deformation pattern consists of spanwise aligned alternating troughs and crests in the streamwise direction. Also, the low and high pressures are associated with surface crests and troughs, respectively.    

%------------------------------------------------------------------
%------------------------------------------------------------------

\begin{figure}[h]
\centering
% ----------
\begin{tikzpicture}
    \node[anchor=north west] at (0,0) {\includegraphics[width=0.6\linewidth,trim={50 50 50 400},clip]
{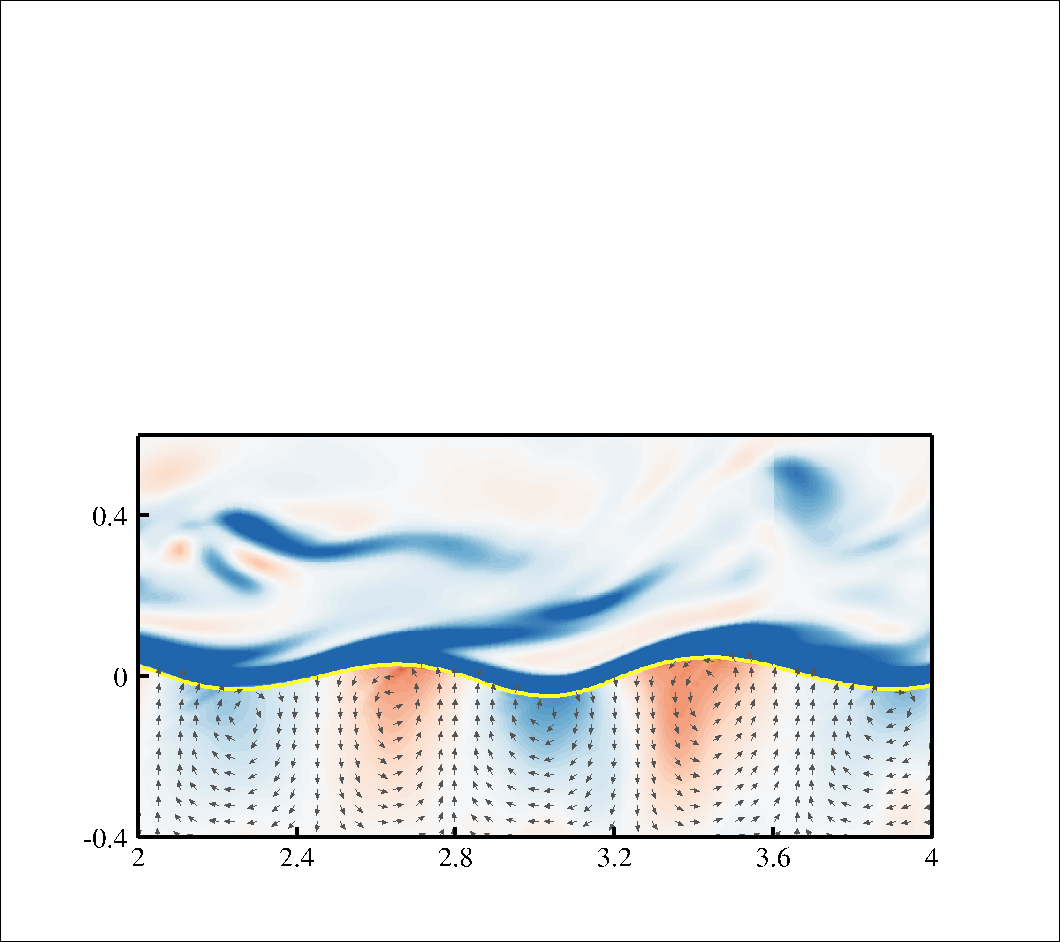}};
    \node[anchor=north west,rotate=90] at (10,-3.7) {\includegraphics[width=0.10\textwidth,
        height=0.03\textwidth,trim={0 0 0 0},clip]{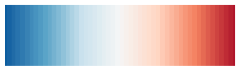}};
\end{tikzpicture}
\put(-294,77){{$y$}}
\put(-154,5){{$x$}}
\put(-20,51){{$-5$}}
\put(-17,113){{$5$}}
\put(-25,125){{$\omega_z\delta/U_b$}}
\caption{Spanwise vorticity field near the compliant surface on the mid-spanwise plane. The yellow curve indicates the surface and the arrows inside the compliant wall show the velocity vectors.\\\hspace{\textwidth}}
\label{fig:tcf_compwall_vorticity}
\end{figure}

Inside the compliant wall, the spanwise vorticity suggests the surface motion resembles Rayleigh waves. Figure \ref{fig:tcf_compwall_vorticity} shows the spanwise vorticity field near the fluid-solid interface on the mid-spanwise plane. Inside the solid, alternating positive and negative vortices are present, corresponding to the crest and trough of the surface, respectively. It indicates the existence of counter-rotating spanwise rolls, and the velocity vectors verify it. This observation is also consistent with the numerical work of Esteghamatian et al. \cite{esteghamatian2022spatiotemporal} 

% ---------------------------------------

% --------- Conclusions -----------------
\section{Conclusions}\label{sec:conclusions}

We have developed a monolithic framework to simulate three-dimensional fluid-structure interaction (FSI) problems involving incompressible flow and viscous hyperelastic solids on a fixed Cartesian grid. To simultaneously solve the governing equations of both fluid and solid subsystems, we use a one-continuum formulation, where the geometric VOF method \cite{hirt1981volume} is used to track the fluid-solid interface. Here, the piecewise linear interface calculation (PLIC) method \cite{debar1974fundamentals,youngs1982time,youngs1984interface,li1995calcul} is used to reconstruct the interface, and the Lagrangian Explicit (LE) method \cite{gueyffier1999volume,alame2020numerical,scardovelli2003interface} is used in the directionally split advection procedure. We consider linear and nonlinear Mooney-Rivlin materials \citep{sugiyama2011full,ii2011implicit} to capture the hyperelastic mechanical behavior of the solid, where the left Cauchy-Green deformation tensor ($\bm{B}$) \citep{sugiyama2011full} accounts for the solid deformation. We use the fifth-order weighted essentially non-oscillatory (WENO-Z) \citep{borges2008improved,jiang1996efficient} finite difference reconstruction procedure to treat the advection terms involved in the transport equation of $\bm{B}$ and a finite volume method (FVM)\cite{mahesh2004numerical} to solve the unified momentum conservation equations with incompressibility constraint. To the best of our knowledge, this is the first such 3D FSI framework developed using the geometric VOF (VOF/PLIC) interface capturing (IC) method and applied to assess turbulent FSI interactions.   

The performance of the Eulerian framework is evaluated using multiple benchmark problems \cite{zalesak1979fully,zhao2008fixed,sugiyama2011full,zhao2008fixed,ii2011implicit,valizadeh2025monolithic,mao20243d}, and our results agree well with the results reported in the previous literature. Reversibility tests are conducted to verify that the hyperelastic solid returns to its initial state after the stresses are released, eliminating any concerns about the reversibility of the solid on an Eulerian grid. Furthermore, the generality of the implementation on grids with non-unity cell aspect ratios was also tested. 

We have shown that the developed VOF/PLIC method-based framework maintains interface sharpness and stability throughout the time history. As a result, a high grid resolution is not required to compensate for interface diffusion, and no additional stabilization procedures are needed. The advantage of interface sharpness is demonstrated through one of the benchmark problems (circular disk in a lid-driven cavity), where our VOF/PLIC-based results on $128 \times 128$ grid agree well with the results of Sugiyama et al. \cite{sugiyama2011full} obtained on $1024 \times 1024$ Eulerian grid using the fifth-order WENO scheme to capture the interface.  

Despite the discontinuity of the interface and stress jumps across the interface, the FSI framework does not generate unphysical solid fragments even when the solid undergoes significant deformation or is susceptible to pinching. Furthermore, no unwanted oscillations at the solid surface are present at steady states. These are typical concerns when using the VOF/PLIC method to capture interfaces of systems involving different materials/phases.  
 
A direct numerical simulation (DNS) of turbulent channel flow with a deformable compliant bottom wall and a rigid top wall is performed, demonstrating the potential of the FSI framework to analyze complex turbulent interactions involving a wide range of length scales and time scales. The wall compliance increases the Reynolds stresses, and the compliant surface exhibits spanwise aligned deformation patterns that propagate downstream, consistent with past experimental and numerical works \citep{wang2020interaction,rosti2017numerical,esteghamatian2022spatiotemporal}. Inside the compliant wall, we verified the existence of counter-rotating spanwise rolls that suggest a similarity to Rayleigh wave motion, as observed in the numerical work of Esteghamatian et al. \cite{esteghamatian2022spatiotemporal}  

We have demonstrated that the geometric VOF-based FSI framework is robust in handling complex FSI problems involving turbulent interactions. Using such methods, one can bypass the requirements for high grid resolutions and stabilization procedures. Since solid shapes can be easily initialized using the VOF indicator function, one can use such a procedure to simulate the interaction of turbulent flows with complex geometries such as dynamic roughness elements and biofouling surface models.    

% ---------------------------------------

\bmsection*{Acknowledgments}
This work was supported by the United States Office of Naval Research (ONR) under ONR Grants N00014-17-1-2676 and N00014-21-1-2455 with Dr. Ki-Han Kim and Dr. Yin Lu Young as grant monitors. The computational resources were provided by the U.S. Army Engineer Research and Development Center (ERDC) Department of Defense (DoD) Supercomputing Resource Center (DSRC) through the DoD High Performance Computing Modernization Program (HPCMP), the U.S. Navy DSRC through the DoD HPCMP, and the Minnesota Supercomputing Institute (MSI) at the University of Minnesota. The authors thank Dr. Karim Alam{\'e} and Marc Plasseraud for helpful discussions. 

\bmsection*{Conflict of interest}

The authors declare that they have no known competing financial interests or personal relationships that could have appeared to influence the work reported in this paper.

% ---------------------------------------

%\bibliography{wileyNJD-AMA}
\bibliography{main}

%\appendix

\end{document}